\documentclass[usenatbib,useAMS,usegraphicx]{mn2e}

\usepackage{times}

\usepackage{xspace}
\usepackage[fleqn]{amsmath}

\providecommand{\eg    }{e.g.\xspace}%
\providecommand{\ie    }{i.e.\xspace}
\providecommand{\mrk   }{Mrk~421\xspace}%
\providecommand{\xray  }{X-ray\xspace}%
\providecommand{\xrays }{X-rays\xspace}%
\providecommand{\gray  }{$\gamma$-ray\xspace}%
\providecommand{\grays }{$\gamma$-rays\xspace}%

\providecommand{\me}{m_\mathrm{e}}%

\newcommand\araa{\textrm{ARA\&A}}%
\newcommand\apj{\textrm{ApJ}}%
\newcommand\apjl{\textrm{ApJ}}%
\newcommand\apjs{\textrm{ApJS}}%
\newcommand\apss{\textrm{Ap\&SS}}%
\newcommand\aap{\textrm{A\&A}}%
\newcommand\aaps{\textrm{A\&AS}}%
\newcommand\mnras{\textrm{MNRAS}}%
\newcommand\nat{\textrm{Nature}}%
\newcommand\pasp{\textrm{PASP}}%
\newcommand\pasa{\textrm{PASA}}%
\newcommand\physrep{\textrm{Phys.~Rep.}}%

\title[Time dependent simulations of \mrk]%
{Time dependent simulations of multiwavelength variability\\ of the blazar \mrk with a Monte Carlo multi-zone code}

\author[X. Chen et al.]{%
Xuhui~Chen,$^1$\thanks{gfossati@rice.edu, Xuhui.Chen@rice.edu}
Giovanni~Fossati,$^1$\footnotemark[1], 
Edison~P.~Liang$^1$
and Markus B\"ottcher$^2$ \\
$^1$ Department of Physics and Astronomy, Rice University, Houston, TX 77005 \\
$^2$ Astrophysical Institute, Department of Physics and Astronomy, Ohio University, Athens, OH 45701
}

\begin{document}

\date{Accepted 2011 June 9. Received 2011 June 8; in original form 2010 August 13}

\pagerange{\pageref{firstpage}--\pageref{lastpage}} \pubyear{2011}

\maketitle

\label{firstpage}

\begin{abstract}
We present a new time-dependent multi-zone radiative transfer code and its first application
to study the SSC emission of the blazar \mrk.
The code couples Fokker--Planck and Monte Carlo methods, in a 2 dimensional
(cylindrical) geometry. 
For the first time all the light travel time effects (LCTE) are fully considered,
along with a proper, full, self-consistent, treatment of Compton cooling, which
depends on them.
We study a set of simple scenarios where the variability is produced by
injection of relativistic electrons as a `shock front' crosses the emission region.
We consider emission from two components, with the second component either
being pre-existing and co-spatial and participating in the evolution of the 
active region (background), or being spatially separated and independent, 
only diluting the observed variability (foreground).
Temporal and spectral results of the simulation are compared to the
multiwavelength observations of \mrk in March 2001.
We find parameters that can adequately fit the observed SEDs and
multiwavelength light curves and correlations.
There remain however a few open issues, most notably:
i), The simulated data show a systematic soft intra-band \xray lag. 
ii), The quadratic correlation between the TeV gamma-ray and \xray flux
during the decay of the flare has not been reproduced.
These features turned out to be among those more affected by the spatial extent
and geometry of the source, \ie LCTE.
The difficulty of producing hard \xray lags is exacerbated by a bias towards
soft lags caused by the combination of energy dependent radiative cooling
time-scales and LCTE.  
About the second emission component, our results strongly favor the scenario
where it is co-spatial and it participates in the flare evolution, suggesting
that different phases of activity may occur in the same region.
The cases presented in this paper represent only an initial study,
and despite their limited scope they make a strong case for the need of
true time-dependent and multi-zone modeling.  
\end{abstract}

\begin{keywords}
galaxies: active --- 
galaxies: jets --- 
BL Lacertae objects: individual (\mrk) ---
gamma-rays: observations ---
X--rays: individual (\mrk) ---
radiation mechanisms: non-thermal
\end{keywords}

%%%%%%%%%%%%%%%%%%%%%%%%%%%%%%%%%%%%%%%%%%%%%%%%%%%%%%%%%%%%%%%%%%%%%%
\section{Introduction}
\label{sec:intro}

Blazars are the most extreme (known) class of AGN.
They are core-dominated, \textit{flat-radio-spectrum} radio-loud AGN.
Their properties are interpreted in terms of radiation from relativistic
jets pointing at us \citep{urry_padovani:1995:review}.
Because of relativistic beaming, jets greatly outshine their host
galaxies thus making blazars unique laboratories for exploring jet
structure, physics and origin.

Blazars emit strongly from radio through \gray energies.
Their spectral energy distribution (SED) comprises two major continuum,
non-thermal, components \citep{ulrich_maraschi_urry:1997:review,fossati_etal:1998:sequence}:
the first, peaking in the IR-optical-\xray range, is unambiguously
identified as synchrotron radiation of ultrarelativistic electrons.
The nature of the second component, sometimes extending to TeV energies, is
less clear and under debate.  
It is generally modeled as inverse Compton (IC)
scattering by the same electrons that produce the synchrotron emission. 
The seed photons can be synchrotron photons 
\citep[\textit{synchrotron self-Compton}, \textit{SSC}][]{mgc92_3c279,marscher_travis:1996} 
or external radiation fields such as emission directly from the
accretion disk, or the broad emission region (BLR), or a putative torus
present on a larger scale 
\citep[\textit{external Compton}, \textit{EC}, \eg][]{%
dermer_etal92, 
sbr94_ec, 
gg_madau:1996, 
blazejowski_etal:2000:compton_on_infrared,
sikora_etal:2009:constraining_emission_models}.
These models are generally referred to as leptonic models because the
particles and processes responsible for the emitted radiation are only
electrons and positrons.

A second class of models, hadronic models, consider the role played by
protons either by producing very high energy radiation directly via the
synchrotron mechanism, or by initiating a particle cascade leading to a second leptonic
population emitting a higher energy synchrotron component 
\citep{mannheim:1998:science_cosmic_rays,
rachen_correlated_x_tev,
sikora_madejski:2001:review,
arbeiter_etal:2005:ssc_and_pions,
levinson:2006:review,
boettcher:2007:emission_processes_review,
boettcher_reimer_marscher:2009:VHE_3c279}.

The frequency of the synchrotron peak ($\nu F_\nu$) has emerged as (one of)
the most important observational distinction across the blazar family
\citep[\eg][]{fossati_etal:1998:sequence}, leading to the classification of blazars as
`red' or `blue' according to their SED `color', \ie the location of the peak\footnote{%%%%
Blue and red SED objects are also called HBL and LBL, for High or Low peak.}. 
\cite{fossati_etal:1998:sequence} showed that blazar SEDs seem to change
systematically with luminosity; the most powerful objects are red, while
blue SEDs are associated with relatively weak sources, a result supported
by studies of high redshift blazars and of low power BL\,Lac objects 
\citep[see][]{fabian_etal:2001:gb1428,
fabian_etal:2001:pmn_j0525,
costamante_etal:2001:stretching_the_sequence,
gg_etal:2009:physical_properties_of_fermi_blazars}.

Another fundamental distinction among blazars concerns their `thermal' 
spectral properties, where they encompass a wide range of phenomenology,
ranging from objects with featureless optical spectra (BL\,Lac objects) 
to objects with quasar--like (broad) emission line spectra 
\citep[Flat Spectrum Radio Quasars, FSRQ][for a review]{urry_padovani:1995:review}.
This distinction is likely to have an impact on the mechanisms of
production of the \gray component in different types of blazars, with
BL\,Lacs being consistent with pure SSC, and FSRQ with EC.  
In fact in most FSRQs the prominence of the thermal components with
respect to the synchrotron emission suggests that EC must be dominant
over SSC.
The case for BL\,Lacs is more ambiguous.
Since particles in the active region (blob) in the jet would see the
external radiation greatly amplified by relativistic aberration with
respect to what we measure, the fact that we do not directly detect any
thermal component may not necessarily mean that in the jet rest frame
its intensity is not competitive with the internally produced
synchrotron radiation. 

However, the broad band emission of TeV detected BL\,Lac objects, like
\mrk, is well modeled with pure SSC and stringent upper limits can be
set on the contribution of EC to their SEDs
\citep{ghisellini_etal:1998:seds,
gg_etal:2009:physical_properties_of_fermi_blazars}.

%=====================================================================
\subsection{Variability}
\label{sec:intro:variability}

Rapid and large-amplitude variability is a defining observational
characteristic of blazars. 
It occurs over a wide range of time-scales and across the whole
electromagnetic spectrum \citep{ulrich_maraschi_urry:1997:review}. 
Flux variability is often accompanied by spectral changes, typically more
notable at energies around/above the peak of each SED component.
Multiwavelength correlated variability studies have been a major component
of investigation of blazars, but because of observational limitations so
far it has been focused on blue blazars.

Blue blazars / HBLs indeed constitute a particularly interesting subclass,
for their synchrotron emission peaks right in the \xray band, and the high
energy component reaches up to TeV \grays.
The \xray/TeV combination has been accessible observationally thanks to
ground based Cherenkov telescopes and the availability of several \xray
observatories.
Hence, the brightest HBLs have been studied extensively.
Simultaneous \xray/\gray observations showed that variations around
the two peaks are well correlated, providing us with diagnostics on the
physical conditions and processes in the emission region for HBLs.

Different models have been shown to successfully reproduce time-averaged or
snapshot spectral energy distributions of blazars.
So far, however, there has been remarkably little work taking advantage of the
information encoded in the observed time evolution of the SEDs by modeling it
directly, despite the tremendous growth and improvements on the observational
side, allowing in many cases to resolve SED on physically relevant
time-scales, fueled by several successful multiwavelength campaigns 
\citep[\eg, some of the most recent ones, for the brightest BL\,Lacs][]{%%%
sambruna00_mrk501_xray_tev,
ravasio_etal:2002:bllac,
takahashi_etal:2000:mkn421_1998,
fossati_etal:2000:mkn421_temporal,
fossati_etal:2000:mkn421_spectral,
krawczynski_etal:2004:1es1959,
blazejowski_etal:2005:multiwavelength_mrk421,
rebillot_etal:2006:multiwavelength_mrk421,
giebels_etal:2007:mkn421,
fossati_etal:2008:xray_tev,
aharonian_etal:2009:pks2155}. 
  
The spectral time evolution has been studied and characterized
by means of intra- and inter-band time lags, intensity correlation, and
hysteresis patterns in brightness--spectral shape space.
The main observed features unveiled by this type of analyses seem to be
well accounted for by attributing the \gray emission to SSC (in a one-zone
homogeneous blob model), and they emerge from the combination of
acceleration and cooling and depend on the relative duration of the related
time-scales \citep[\eg][]{takahashi_etal:1996:mkn421,
ulrich_maraschi_urry:1997:review,
kataoka:2000:phd_thesis}. 
A less empirical, more directly theoretical interpretation of this wealth
of data, requiring/exploiting the physical connection between series of
spectra, has remained relatively basic despite the clear richness of the
observed phenomenology 
\citep[\eg,][]{mastichiadis_kirk:1997:SSC_variability,
dermer:1998:spectral_variability,
li_kusunose:2000,
sikora_etal:2001:flares,
krawczynski_coppi_aharonian:2002:timedep,
boettcher_chiang:2002}.

%=====================================================================
\subsection{Light Crossing Time Effects}
\label{sec:intro:lcte}

One of the biggest challenges and limitations of the current models comes 
from the dealing with light crossing time effects (LCTE), usually treated 
in a simplified way, such as simply by introducing a photon escape
parameter \citep[\eg,][]{boettcher_chiang:2002}.
  
The observed variability on time-scales of hours indicates that light
crossing time effects \textit{within} the active region are very
important and must be dealt with.
There are two main aspects related to photon travel times that are important
for an accurate study.
The first, which we can call `external' \citep[following][]{katarzynski_etal:2008:pks2155}, 
is a purely geometric effect that pertains to the impact of the finite size
of the active region on the observed variability, namely the delayed
arrival time of the emission from different parts of the blob, and
consequent smearing of the intrinsic variability characteristics
\citep{protheroe:2002:factors_for_variability}.
It is relatively simple to implement.

The second effect, internal, pertains to the impact of these same delayed
times on the actual physical evolution of the variability (as opposed to
just our `perception' of it) due to the changing conditions inside the
active region.  
This effect constitutes the real challenge for proper multi-zone modeling.
In this respect the most important issue is that of the photon diffusion 
across the blob on the electrons' inverse Compton losses.
Proper accounting of this effect is significantly more complex and
computationally expensive, and traditionally neglected under the assumption
that electron cooling is dominated by synchrotron losses.
This is, however, a strong assumption, rarely valid, as suggested by the
observation that the luminosity of the synchrotron components is at best of
the same order as the IC component, more commonly lower.

It has long been realized that a simple one-zone homogeneous model is not
adequate to describe the temporal evolution of the blazar jets, and that
LCTE must be taken into account. 
\cite{mchardy_etal:2007:3c273_variability} suggested that the observed
delay between \xray and infrared variations in 3C\,273 could be related to
the time necessary for the soft (synchrotron) photon energy density to
build up as the they travel across the active region.

%=====================================================================
\subsection{Relevant Previous Work}
\label{sec:intro:previous_work}

Some progress has been made to develop multi-zone models, though with limited
success because the traditional analytical approach requires
significant assumptions, such as simple geometries or assumptions about the
relevance of different physical processes.
The inclusion of just the external LCTE is enough to yield new insights on SSC
light curves, such as on the way the interplay between cooling/acceleration
time-scales and source size affects the observed light curves as a function
of energy and combination of the various time-scales 
\citep{chiaberge_ghisellini:1999:timedep,
kataoka_etal:2000:crossingtimes_model,
katarzynski_etal:2008:pks2155}.
In all the cited cases the size of the active region and the duration of
the injection of fresh particles are related through $t'_\mathrm{inj} = R/c$,
where $R$ is the radius of a sphere or the length of a cubic region.  
The geometry is characterized by a single length-scale.
This kind of models, not accounting for internal LCTE and
non-locally-emitted radiation for IC emission, could yield correct
results for the evolution of the electron distribution if synchrotron
losses dominate, however even in this case their results for the evolution of
the IC component are not realistic because they ignore the contribution of
seed photons from other zones.

\citet{sokolov_marscher_mchardy:2004:SSC} and \citet{sokolov_marscher:2005:EC}
were the first to include the internal LCTE to calculate the IC spectrum,
for both SSC and external IC models. However, they did not properly account
for it when calculating IC energy losses.  
Their model is thus accurate only when synchrotron losses are dominant.
Observationally this corresponds (approximately) to cases where the peak of
the lower energy component of the SED (synchrotron) is significantly
brighter than that of the second peak (IC).  

\citet{graff_etal:2008:pipe} developed a model taking into account all the
LCTEs, but specialized to an elongated `pipe' geometry.
The geometry of the current implementation of their code is effectively
one-dimensional.  
The lack of an actual transverse dimension represents a significant
limitation when considering the LCTE, considering because of
relativistic aberration we are effectively observing a jet (also) from its side. 

\medskip
In this paper we introduce a more general and flexible code to simulate
blazar variability, addressing and overcoming most of the limitations
affecting previous efforts.

The general features and assumptions of the code are illustrated in section
\S\ref{sec:code}, followed by a comparison with the results of other
codes to test its robustness in \S\ref{sec:tests}. 
In \S\ref{sec:applications} we present results of a few test cases based on
the multiwavelength observations of \mrk in March 2001.
We conclude with a discussion of this first application and remarks about
future applications and developments in \S\ref{sec:conclusions}.

In order to keep the notation light, we will use primes for blob-frame
values sparingly, mostly to distinguish photon energies, luminosity and
times ($E$, $L$, $t$, $\tau$).  
We do not prime quantities that are usually not ambiguous because they are only
referred to in the blob-frame, such as magnetic field strength ($B$), source
size ($R$, $Z$), electron Lorentz factor ($\gamma$),
density ($n_\mathrm{e}$).  
Similarly, we do not use primes when the context is clear (for instance in the
discussion of the Fokker--Planck equation).

%-------------------------------------------------------
\begin{figure}
\centerline{%
\includegraphics[width=0.70\linewidth]{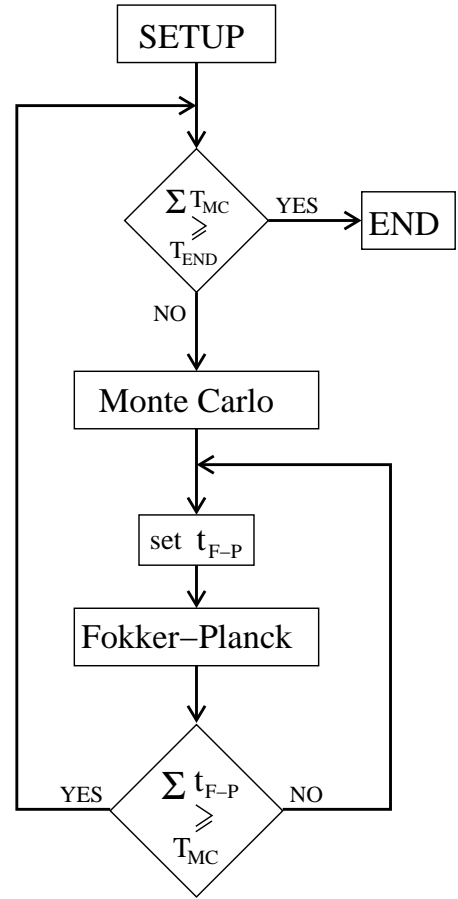}
}
\caption{%
Basic structure and work flow of the code. 
The Monte Carlo block handles photon emission/absorption processes (\eg,
synchrotron, IC, pair annihilation, self-absorption, escape).
The MC time-step is determined at setup and it does not change.
The Fokker-Planck block handles electron processes (\eg, injection, cooling,
pair production, escape).
The F-P time-step is adjusted at each iteration according to the current
physical conditions.
\label{fig:codeflow}
}
\end{figure}
%-------------------------------------------------------

%%%%%%%%%%%%%%%%%%%%%%%%%%%%%%%%%%%%%%%%%%%%%%%%%%%%%%%%%%%%%%%%%%%%%%
\section{The Monte Carlo / Fokker--Planck Code}
\label{sec:code}

Our code couples Fokker--Planck (F-P) and Monte Carlo (MC) methods in a 2 dimensional
(cylindrical) geometry. 
It is built on the MC radiative transfer code developed by Liang, B\"ottcher and
collaborators \citep{canfield_howard_liang:1987:MC_IC_relativistic_electrons,
boettcher_liang:2001:MC_code,
boettcher_jackson_liang:2003:MC_code}, 
parallelized by \citet{finke:2007:phd_thesis}.
The Monte Carlo method is ideal for multi-zone 2D/3D radiative transfer
problems.  Due to its tracking of the trajectory of every photon, LCTE are
automatically accounted for, regardless of the geometry. 
We modified the parent code significantly in several aspects, to make it
more general applicable, in particular to the
physical conditions of the active region in a blazar jet.

%=====================================================================
\subsection{Code Structure}
\label{sec:code:structure}

The code separates the handling of photon and electron evolution.
The electron evolution is governed by the Fokker--Planck equation, as
commonly done \citep[\eg][]{fabian_etal:1986:variability_in_compact_sources,
coppi92_tdep,
coppi_blandford_rees:1993:anisotropic_induced_compton,
kirk_rieger_mastichiadis:1998,
makino:1998:turku,
kataoka_etal:2000:crossingtimes_model,
chiaberge_ghisellini:1999:timedep}.
Photons are dealt with by the MC part of the code, which tracks photon
production and evolution by different mechanisms, including IC scattering with the
current electron population, \textit{and} propagation.
The code's basic structure and work flow is illustrated in Fig.~\ref{fig:codeflow}.
 
There are two main loop structures.
Since the evolution of the electron distribution is faster than that of the
photons, each MC cycle contains several F-P (electron) cycles.
Therefore the code has two main time-steps: a longer MC time-step ($\Delta
t'_\mathrm{MC}$), within which the F-P equation routine performs the
evolution of the electron spectrum on
shorter, variable length, time-steps ($\Delta t'_\mathrm{F-P}$).  
We describe them in more details in the next Sections.

%-------------------------------------------------------
\begin{figure}
\centerline{%
\includegraphics[width=0.99\linewidth,clip=]{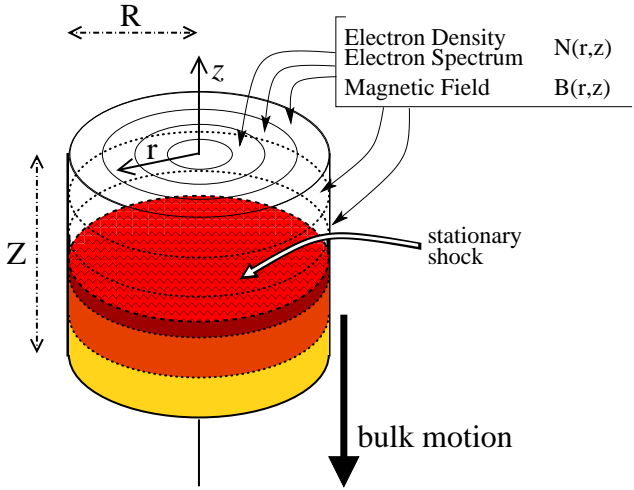}
}
\caption{%
The geometry of the blob model. The volume is divided into different zones
in $r$ and $z$ directions, each zone with its own electron distribution and
magnetic field.
We also schematically show the setup for the variability of the simulations
presented here.
The hatched layer represents the stationary shock (\S\ref{sec:code:changes:injection}).
The blob, simulation volume, is moving downward and crossing the shock front.
Zones that crossed the shock at earlier times have had some time to radiate
the newly injected energy and are plotted in lighter color shades. 
\label{fig:geometry_and_injection}
\label{fig:geometry}
\label{fig:injection}
}
\end{figure}
%-------------------------------------------------------

%=====================================================================
\subsection{Geometry}
\label{sec:code:geometry}

As illustrated in Fig.~\ref{fig:geometry}, the code is built with 2D
cylindrical geometry, with symmetry in the azimuthal direction.
The volume has radius $R$ and length $Z$, and it is divided evenly into
zones in the radial and vertical directions ($r$ and $z$ coordinates, 
$n_\mathrm{r}$, $n_\mathrm{z}$).
In all runs presented here $n_\mathrm{r}=9$ and $n_\mathrm{z}=30$.
The number of zones sets the resolution of the simulation for what
concerns spatial inhomogeneities in the physical properties, either
as directly set up or because of their different evolution (\eg radiation
energy density will always develop a radial profile, in turn inducing a
radial profile in the electron spectra).
In the scheme adopted for this work the number of zones is also related to
the duration of the Monte Carlo time-step (see \S\,\ref{sec:code:MC}).
For scenarios where the variability is produced by a perturbation crossing
the simulation volume moving along the $z$ axis, spatial/temporal resolution in
the $z$ direction is more important, hence we select a larger $n_\mathrm{z}$.
For the simulations presented here our choice yields a Monte Carlo time
resolution of $\Delta z/c/\delta \le 500$\,s in the observer's frame,
adequate to model the 2001 \mrk data.
It is possible to study faster variability, \eg $<5$ minutes as observed in 
PKS\,2155$-$304 \citep{aharonian_etal:2009:pks2155}, by increasing the
number of zones at the expense of increased computational time.
It is worth noting that as long as the choice of $n_\mathrm{r}$,
$n_\mathrm{z}$ ensures that each zone is small enough to sample properly  
the variations on the physically relevant time-scales, the results of the
simulations are substantially insensitive to these parameters.  

This geometrical setup is adequate for the case we want to study since the
assumption is that the active region is a slice of a collimated jet.
In principle, the code setup is flexible enough that slightly different
geometries could be simulated via the parameter settings in each zone.

Each zone has its own electron population (spectrum, density) and magnetic
field.  They can be setup individually and their time evolution is
independent from each other, except for the effect of mutual illumination.
Photons move freely among different zones, but the code assumes that
electrons stay in their given zone and do not travel across zones; 
the electron Larmor radius is very small for the energies and (tangled)
magnetic field strengths typical for the active region of a blazar jet, at
least those studied here.  
For instance, for $\gamma=10^6$ and $B=0.03$\,G, $r_\mathrm{L} = 5.7 \times
10^{10}$\,cm, to be compared with source size of the order of $10^{16}$\,cm.
The radiation emitted by the blob is registered in the form of a pseudo-photon
list (see~\S\ref{sec:code:MC:MC_particles}), with time, direction and
energy \citep[see also][]{Stern_etal:1995:MC_large_particle}.

All the calculations are done in the blob rest frame.
The transformation of all the quantities into the observer's frame is
performed afterwards.
The output is analyzed using a separate post processing code to produce 
SEDs and light curves. 
Since the product of the code is effectively a photon list, we have
significant freedom in the choices of bin sizes for time, energy, and
angle, mostly limited by statistics, much in the same way as for actual
observations.
Hence, we can tailor the simulation results to the characteristic of the
observations that we want to reproduce (\eg time binning, energy bands).

In all cases presented here, the observed spectrum is obtained by
integrating the beamed photons over a small solid angle centered around the
angle $\theta$ between jet axis and observer, assumed as customarily to be
$\theta = 1/\Gamma$, for which also $\delta = \Gamma$.
The typical width of the integration solid angle is $\Delta\cos(\theta)
\sim \mathrm{few} \times 10^{-4}$.

%=====================================================================
\subsection{The Monte Carlo section}
\label{sec:code:MC}

The MC part of the code uses the current electron distribution, as updated
in the F-P section of the code.
It includes all processes that involves changes in the radiation field,
such as Compton scattering and the production of new photons by various
radiative processes, the most important of which for our case is
synchrotron emission.   
Other notable processes are pair production and annihilation, and
synchrotron self-absorption.

The MC time-step is currently a user-set parameter, part of a run input setup.
It is adjusted depending on the geometry of the problem, \eg shorter
than the light crossing time of the smallest zone, and requirements
of physical accuracy, for instance with respect to the fact that during
each MC time-step the code does not change the electron distribution, 
which is evolved only during the F-P section of the code (\ie ensuring that
$\Delta t'_\mathrm{MC} < \tau'_\mathrm{cool}(\gamma)$ for the highest energy
electrons.)

%-----------------------------------------------------------
\subsubsection{Monte Carlo particles}
\label{sec:code:MC:MC_particles}

Since it is impossible to follow every individual photon a common technique
used in radiative transfer problems is to group them into packets,
pseudo-photons \citep[\eg][]{Abbott_Lucy:1985:MC,Stern_etal:1995:MC_large_particle},
to which we will refer as Monte Carlo particles. 
Every MC particle $k$ represents $n_\mathrm{k}$ photons with the same
energy, the same velocity vector, at the same position and time, carrying a
total energy of $E_\mathrm{k} = n_\mathrm{k}(\nu_\mathrm{k})\, h \nu_\mathrm{k}$.  
The $n_\mathrm{k}$ is also referred to as statistical weight of the MC particle.

The MC particles are born in the volume through emission processes,
primarily synchrotron radiation in our case. 
The luminosity of the newly radiated synchrotron contribution is computed
and converted into MC particles with a distribution
according to the probability given by their SED.

The position within a given zone, time within the current time-step, and
travel direction of the MC particle when it is generated are drawn randomly
from the appropriate probability distributions. 

At every time-step, each MC particle moves independently, with some
probability of being IC scattered. 
Absorption is handled as a decrease of the statistical weight of the MC particles. 

When a MC particle reaches the volume boundary, it is recorded with
the full information of the escape time, position, direction, and energy,
forming a list of emitted photons.

%=====================================================================
\subsection{The Fokker--Planck equation}
\label{sec:code:FP}

In each zone, the temporal evolution of the local electron population
is obtained by solving the Fokker--Planck equation:
\begin{equation}
\label{eq:FPeq}
\begin{split}
 \frac{\partial N (\gamma ,t)}{\partial t} & =
-\frac{\partial}{\partial \gamma}\bigg[N(\gamma,t)\dot{\gamma}(\gamma,t)\bigg] \\
 & +\frac{1}{2}\frac{\partial^2}{\partial \gamma^2}\bigg[N(\gamma,t)D(\gamma,t)\bigg]
+ Q(\gamma,t) - \frac{N(\gamma, t)}{t_\mathrm{esc}}
\end{split}
\end{equation}
$N(\gamma,t)$ is the electron spectrum, $\gamma$ is the random Lorentz
factor of electrons, $\dot{\gamma}$ is the total heating/cooling rate.
The IC cooling uses the time dependent radiation field calculated in the
Monte Carlo part of the code, with LCTEs accounted for, which is considered
constant for the duration of the F-P section of the code.
The full Klein--Nishina (K-N) scattering cross section is used (see
\S\ref{sec:code:changes:synchro_and_IC}).
$D(\gamma,t)$ is the dispersion coefficient which is not important
for the type of scenarios presented in this work, and it then set to zero. 
For generality, the $D(\gamma,t)$ term is still included in the solving of
the F-P equation. 
$Q(\gamma,t)$ is the electron injection term. 
Because, as noted in \S\,\ref{sec:code:geometry}, the electrons'
Larmor radius is much smaller than the size of the simulation zones, the
particle escape term is not considered\footnote{%%%%%
Except for the test runs discussed in Section~\S\ref{sec:tests:temporal_evolution}
for consistency with the model with which we are comparing the results.}.

The time-step of the F-P loop is adjusted automatically depending
on the rate of change (gain or loss) of energy of the particles to
ensure a physically meaningful solution.  It is constrained to be shorter
than one fourth of the MC time-step.

Rather than using the discretization scheme proposed by \citet{nayakshin_melia:1998}, 
as done in \citet{boettcher_jackson_liang:2003:MC_code}, we choose to adopt
the implicit difference scheme proposed by \citet{chang_cooper:1970}.
This scheme guarantees non-negative solutions, which in runs with the
original scheme resulted in wild oscillations of the electron distribution
at the high energy end \citep[for a discussion of this issue please refer to the
appendix of][]{chang_cooper:1970}.

The energy grid used for the electrons is logarithmic in kinetic energy
$x_j=\gamma_j-1$, with 200 mesh points from $x_\mathrm{min}=0.18$ to
$x_\mathrm{max}=3.1 \times 10^7$, \ie $x_j = 1.1\, x_{j-1}$.

After rewriting (\ref{eq:FPeq}) as
\begin{equation}
\label{eq:FP_rewrite}
\begin{split}
\frac{\partial N (\gamma ,t)}{\partial t} & =
     \frac{\partial}{\partial \gamma} \bigg[\bigg(-\dot{\gamma}(\gamma,t)+\frac{1}{2}\frac{\partial D(\gamma,t)}{\partial \gamma}\bigg)N(\gamma,t) \\
& + \frac{1}{2}D(\gamma,t)\frac{\partial N(\gamma,t)}{\partial \gamma}\bigg] 
  + Q(\gamma,t) - \frac{N(\gamma,t)}{t_\mathrm{esc}} 
\end{split}
\end{equation}
It is possible to discretize it as:
\begin{equation}
\label{eq:FP_discretized}
\begin{split}
 \frac{N_j^{n+1} - N_j^n}{\Delta t} &= \\
 \frac{1}{\Delta x_j} & \Bigg[\frac{1}{\Delta x_{j+1/2}}C_{j+1/2}\ w_{j+1/2}\frac{1}{1-e^{-w_{j+1/2}}}N_{j+1}^{n+1} \\
 -\, & \Bigg(\frac{1}{\Delta x_{j+1/2}}\ C_{j+1/2}\ W_{j+1/2}\ \\
 +\, &       \frac{1}{\Delta x_{j-1/2}}\ C_{j-1/2}\ w_{j-1/2}\ \frac{1}{1-e^{1-w_{j-1/2}}}\Bigg) N_j^{n+1} \\
 +\, &       \frac{1}{\Delta x_{j-1/2}}\ C_{j-1/2}\ W_{j-1/2}\ N_{j-1}^{n+1}\Bigg] \\
 +\, & Q_j^{n+1} - \frac{1}{t_\mathrm{esc}} N^{n+1}_j
\end{split}
\end{equation}
with
\begin{gather*} 
B_{j+1/2} = -\frac{1}{2}\big[\dot{\gamma}_j+\dot{\gamma}_{j+1}\big]+\frac{D_{j+1}-D_j}{2\Delta x_{j+1/2}}, \\
C_{j+1/2} = \frac{1}{4}(D_j+D_{j+1}), \\[12pt]
w_{j+1/2} = \Delta x_{j+1/2}B_{j+1/2}/C_{j+1/2}, \\[12pt]
W_{j+1/2} = w_{j+1/2}/(\exp(w_{j+1/2}) - 1). \\
\end{gather*}
Here the $j \pm 1/2$ subscripts refer to quantities computed as the average
values of the two adjacent grid points, such as 
\begin{gather*}
C_{j+1/2}=\frac{1}{2}(C_j+C_{j+1})
\intertext{An exception is that of}
\Delta x_{j}=\sqrt{\Delta x_{j+1/2} \Delta x_{j-1/2}} \\
\intertext{with}
\Delta x_{j+1/2}=x_{j+1}-x_j \\
\Delta x_{j-1/2}=x_j-x_{j-1}. 
\end{gather*}
In order to avoid infinity in our calculation, we set $D=10^{-40}s^{-1}$ instead of $D=0$.

The tridiagonal matrix formed by equation (\ref{eq:FP_discretized}) can be solved
using the standard algorithm in \cite{numerical_recipes92}.

%=====================================================================
\subsection{Synchrotron and inverse Compton}
\label{sec:code:radiative_processes}

The synchrotron spectrum is calculated adopting the single particle
emissivity averaged over an isotropic distribution of pitch angles
given by \citet{crusius_schlickeiser:1986:synchrotron} 
\citep{gg_guilbert_svensson:1988:boiler}:
\begin{equation}
\begin{split}
 P(\nu,\gamma) & =  \\
  \frac{3 \sqrt{3}}{\pi}\, & \frac{\sigma_\mathrm{T}\, c\, U_\mathrm{B}}{\nu_\mathrm{B}}\, y^2
  \left\{K_\frac{4}{3}(y) K_\frac{1}{3}(y) - \frac{3}{5}\,y \left[K^2_\frac{4}{3}(y) -
  K^2_\frac{1}{3}(y)\right]\right\}
 \label{eq:sync_sp}
\end{split}
\end{equation}
where $\sigma_\mathrm{T}$ is the Thomson cross section and
\begin{equation*}
   \gamma = \frac{E}{\me c^2}, \qquad
   \nu_\mathrm{B} = \frac{e B}{2\pi \me c}, \qquad
   y = \frac{\nu}{3 \gamma^2 \nu_\mathrm{B}}, 
\end{equation*}
with $E$ the total electron energy, and $K_{x}(y)$ is the modified
Bessel function of order $x$.

The total emitted synchrotron power and self-absorption coefficients are
calculated according to the formul\ae\ in \citet{rybicki_lightman}.
\begin{equation}
 \alpha_\nu = \frac{c^2}{8 \pi h\nu^3}\int dE\, P(\nu,E)\, E^2\,
              \left[\frac{N(E-h\nu)}{(E-h\nu)^2} - \frac{N(E)}{E^2}\right] 
\label{eq:sync_ab}
\end{equation}
where $P(\nu,E)$ is the synchrotron spectrum given above (\ref{eq:sync_sp}).

For the total Compton cross section, we used the angle-averaged cross section
given in \citet{coppi_blandford:1990}:
% After integration over the angle, this yields:
\begin{equation}
\label{eq:Compton_cross_section}
\begin{split}
 \sigma(\omega,\gamma) & =
 \frac{3 \sigma_\mathrm{T}}{32 \gamma^2 \beta \omega^2} \bigg[-\frac{x}{2}+\frac{1}{2(1+x)}+ \\
& \bigg(9+x+\frac{8}{x}\bigg)\ln(1+x)+4Li_2(-x)\bigg]
\Bigg|_{x=2\gamma(1-\beta)\omega}^{x=2\gamma(1+\beta)\omega}
\end{split}
\end{equation}
Where $Li_2(z)$ is the dilogarithm, which is evaluated numerically. 
To get the total cross section for a photon in an electron medium we need
to integrate over $\gamma$, weighted by the electron energy distribution.

%=====================================================================
\subsection{Other major changes}
\label{sec:code:changes}

Besides changing the numerical scheme to solve the F-P equation, we
implemented several other major changes in the code. 

%-------------------------------------------------
\subsubsection{Injection of electrons}
\label{sec:code:changes:injection}

The model of the electron injection process, as implemented currently,
involves a stationary shock perpendicular to the axis of the cylinder (jet)
(Fig.~\ref{fig:injection}).
Hence, in the frame of the blob the shock is traveling across the blob
with a speed equal to the bulk velocity of the blob $v_\mathrm{bulk} \sim c$.
This scenario is similar to the one discussed by \citet{chiaberge_ghisellini:1999:timedep}. 
The thickness of the shock is treated as negligible, in the sense that it
is considered active only in one zone at any given time, \ie it never
splits across a zone boundary.
However, during the time it takes to travel along a $Z$-zone, $\Delta z/c$,
particles are injected in the entire zone volume.  
From this point of view the `shock' thickness is not negligible.
Provided that the $\Delta z$ of each zone is small this approximation is
reasonable.  As noted, for the cases presented here, $\Delta z/c/\delta \le 500$\,s.
The total duration of injection is thus $t'_\mathrm{inj} = Z/v_\mathrm{bulk}$, 
and each slice of the simulation volume along the $z$ axis will eventually 
have an injected energy of $L'_\mathrm{inj} \Delta z/v_\mathrm{bulk}$,
where $\Delta z = Z/n_\mathrm{z}$ is the thickness of one slice.

Electron injection is included in the Fokker--Planck equation through the
term $Q(\gamma,t)$. 
The shock moves at the speed of $v_\mathrm{bulk}$ every F-P time-step. 
When the shock front is located in a given zone, electron injection is
active ($Q\not=0$), otherwise $Q=0$.
In the simulations presented here the injected electrons have a
power law distribution with a high energy exponential cutoff
\begin{displaymath}
Q(\gamma) = Q_0\, \left(\frac{\gamma}{\gamma_0}\right)^{-p}\, \mathrm{e}^{-\gamma/\gamma_\mathrm{max}}
\hfill \text{cm}^{-3}\, \text{s}^{-1}
\hfill  \gamma \ge \gamma_\mathrm{min}, 
\end{displaymath}
The value of the normalization $Q_0$ is controlled by the parameter $L'_\mathrm{inj}$.

Injection and acceleration time-scales and durations are in principle
independent from other parameters and could be set directly on the basis of 
a hypothesis on the details of physical processes underlying them.
In this work we are treating injection, and in turn the implied process for
accelerating the newly injected particles, phenomenologically, affording
ourselves the freedom to assume their spectrum and time-scales.

The underlying physical mechanism for the injection process is not specified. 
First order Fermi acceleration at a shock front or second order Fermi
acceleration by a plasma turbulence are two possible processes 
\citep[\eg][]{drury:1983:diffusive_shock_acceleration,
blandford_eichler:1987,
gaisser:1990:BOOK:CR_and_particle_physics,
protheroe:1996:cosmic_rays_acceleration,
kirk_rieger_mastichiadis:1998,
katarzynski_etal:2006:stochastic}.

%-------------------------------------------------
\subsubsection{Splitting of MC particles}
\label{sec:code:changes:splitting}

A major difficulty of using the Monte Carlo method to model broadband IC
emission, in the physical conditions typical of blazar jets, is the low
pseudo-photon statistics at high frequencies. 
Observations are affected by a very similar problem.

Blazar SEDs are approximately flat in ($\nu F_\nu$) over wide range of
energies.  
In blue blazars typical energies for the electrons responsible for the
($\nu F_\nu$) emission peaks, occurring in UV-X-ray and TeV bands, are
$\gamma \sim 10^{4}-10^{5}$.
When a photon (for us a MC particle) is scattered to the TeV
range, the energy of that MC particle will increase by about 9--11 orders of
magnitude depending on whether it was an \xray or optical photon,
and its `flux' will decrease by the same factor\footnote{%%%%%%%%
For constant statistical weight, the discretized spectrum would have $N_i
\sim N(\nu_i) (\Delta\nu)_i$ MC particles in each bin. 
Our grid of photon energies is equally spaced logarithmically, so we can
rewrite it as $N_i \sim N(\nu_i)\, \nu_i\, (\Delta \ln\nu)_i$, where 
$(\Delta \ln\nu)_i = \Delta\ln\nu$ is a constant.
Hence for a photon spectrum $N(\nu) \propto \nu^{-\Gamma}$, the relative
statistics of our discrete photon spectrum goes like $\nu_i^{-\Gamma+1}$.
For an approximately flat SED, \ie $\Gamma \simeq 2$, this goes like
$\nu_i^{-1}$.}, making the statistics of the high energy IC component very poor.

An additional challenge that we face is that the IC scattering probability
is very small. Under most reasonable conditions the active blob is very
optically thin.

In order to mitigate these problems, we introduced a method relying on the
splitting of MC particles.
The basic idea is that since every MC particle represents a packet of real
photons treated together, it is always possible to divide them into smaller
packets. 
If this splitting is applied in appropriate conditions, it is possible to
achieve a reasonable statistics on MC particles at high frequencies with
reasonable computer resources.  

We have implemented MC particle splitting in three different instances
within the context of the computation of IC scattering.
\begin{enumerate}

\item The first splitting is applied to every MC particle when it is considered
for IC scattering.  It is split into a large number of identical
subparticles (\eg $\sim10^{3}$).
The choice of this number depends on the trade off between improving the
statistics of the high energy photon spectrum and cost in terms of
computing resources (time and memory), and it was based on empirical testing.
Whether a particle is scattered or not is determined by comparing the
distance it would travel with a distance to the next scattering stochastically
determined from its mean free path.
Every MC subparticle draws a separate random number, and in turn has its
own probability of being scattered.  
All non-scattered MC subparticles are recombined into a MC particle, and
travel to a new position.  
The subparticles that do scatter (usually a small number) will be scattered
separately, to independent energies and directions (but see below).
This first splitting does not necessarily save a computational time, but
it decreases dramatically the memory allocation requirement to achieve the
desired statistics at the highest SED energies.
 
\item
The second instance of splitting is applied to MC (sub)particles that are
being scattered.
They are divided into another large number (\eg $\sim10^{3}$), and each of
these MC sub-subparticles will be scattered separately, to a frequency and
direction uncorrelated with those of the other particles.  
This splitting allows us to concentrate computation cycles on the rare
events of scatterings, which is what we are most interested in.

\item
Even with this second splitting, at highest energies the statistics of
the IC photon spectrum remains very poor.
To alleviate this problem, we implemented a third instance of MC particle
splitting.
It is triggered when one of the already twice-split MC particles is
scattered to very high frequency, above a threshold that is set a priori
and constant for each run, tailored to the characteristics of the studied SED.
This MC particle is split again, and each of its subparticles is
re-scattered from the original frequency. 
That scattering is accepted only when the scattered frequency is higher
than the preset threshold, otherwise it goes back and draws another random
number. 
This third splitting offers the benefit of avoiding the use of a much larger
number of subparticles in the second instance of splitting, and
subsequently avoiding the production of a very large number of MC particles
to be recorded in the computer memory.

\end{enumerate}
Splitting causes the number of MC particles to grow during the simulation.
Nevertheless, the advantage over directly setting up the simulation with
more MC particles is significant both in terms of number of MC particles
and more importantly because the new MC particles are created where they
are most needed, thus increasing greatly the efficiency of the code.
In typical runs the increase in the number of MC particles due to the
splitting is modest, of the order of 10--20 per cent of the number of newly
emitted synchrotron photons at each MC step.

%-------------------------------------------------
\subsubsection{Arbitrary electron energy distribution}
\label{sec:code:changes:arbitrary_electrons}
\label{sec:code:changes:synchro_and_IC}

Although the F-P equation can calculate the time dependent evolution of the
electrons with arbitrary spectrum, earlier versions of the code forced the
decomposition of the electron population into a low-energy thermal
population plus a high-energy power law tail. 
The emissivity of cyclotron, non-thermal synchrotron and thermal
bremsstrahlung radiation processes were calculated on the basis of this
decomposition. 
The calculations of the synchrotron self-absorption coefficient and the total
scattering cross section of a photon in the medium were dependent on this
`thermal plus power law' approximation as well. 
In order to make the code more general, and in particular more suitable for
blazar simulations, in which there is usually a dominant non-thermal
lepton population, we have entirely rewritten the relevant sections of the
code.
The code now calculates all physical quantities using the actual electron
spectrum, as updated by solving the F-P equation 
(see \S\,\ref{sec:code:radiative_processes})

%=====================================================================
\subsection{Deactivated Features}
\label{sec:code:deactivated_features}

Some features of the code have been deactivated in this study. 
Among these are the cyclotron and bremsstrahlung emission and Coulomb
scattering of electrons with protons, all considered not important in
blazar jets.
Others are turned off because they are not the focus of this paper; 
these include external sources of photons, 
which will be subject of future investigations.

%-------------------------------------------------------
\begin{figure*}
\centerline{%
\hfill
\includegraphics[width=0.49\linewidth]{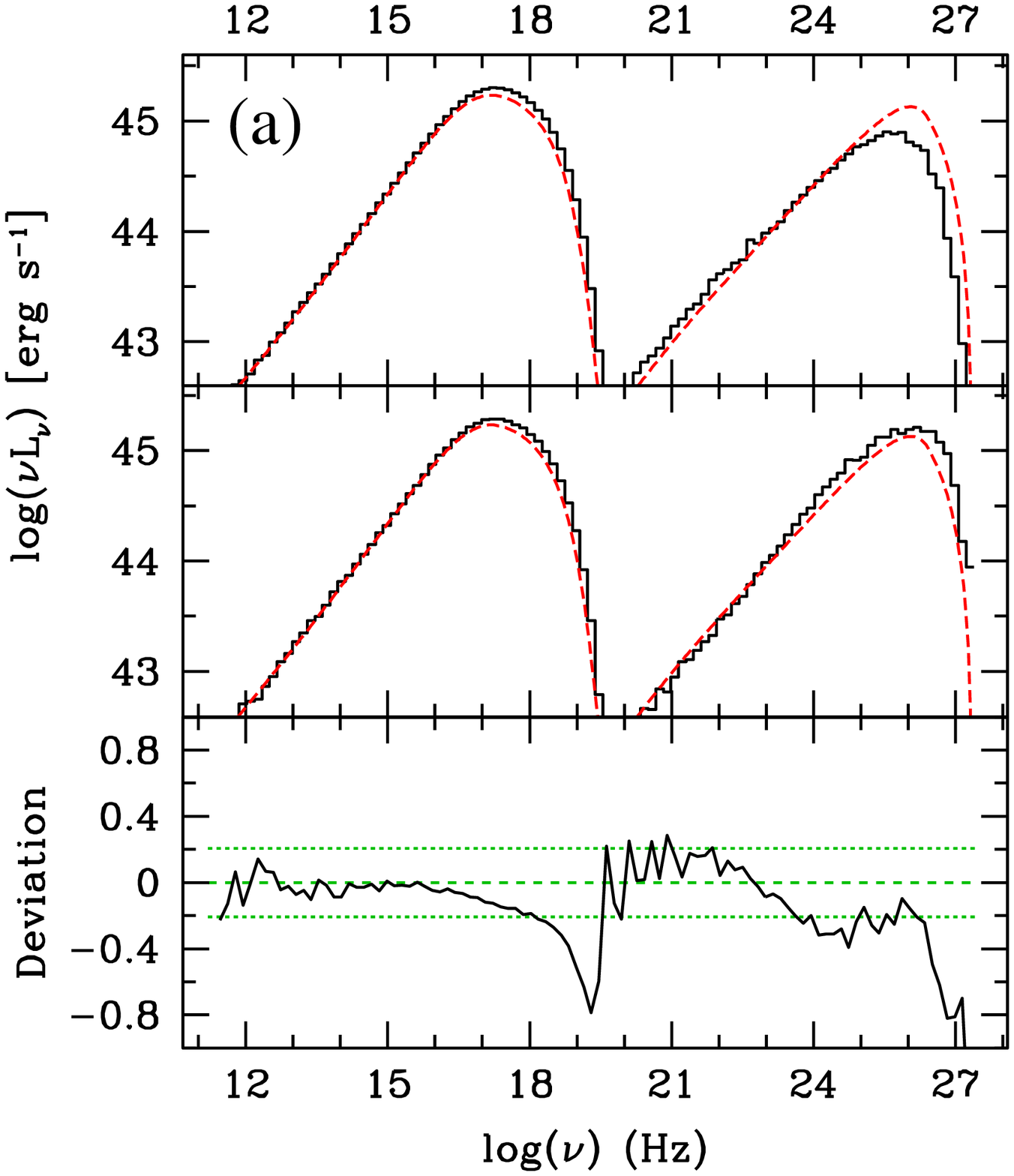}
\hfill
\includegraphics[width=0.49\linewidth]{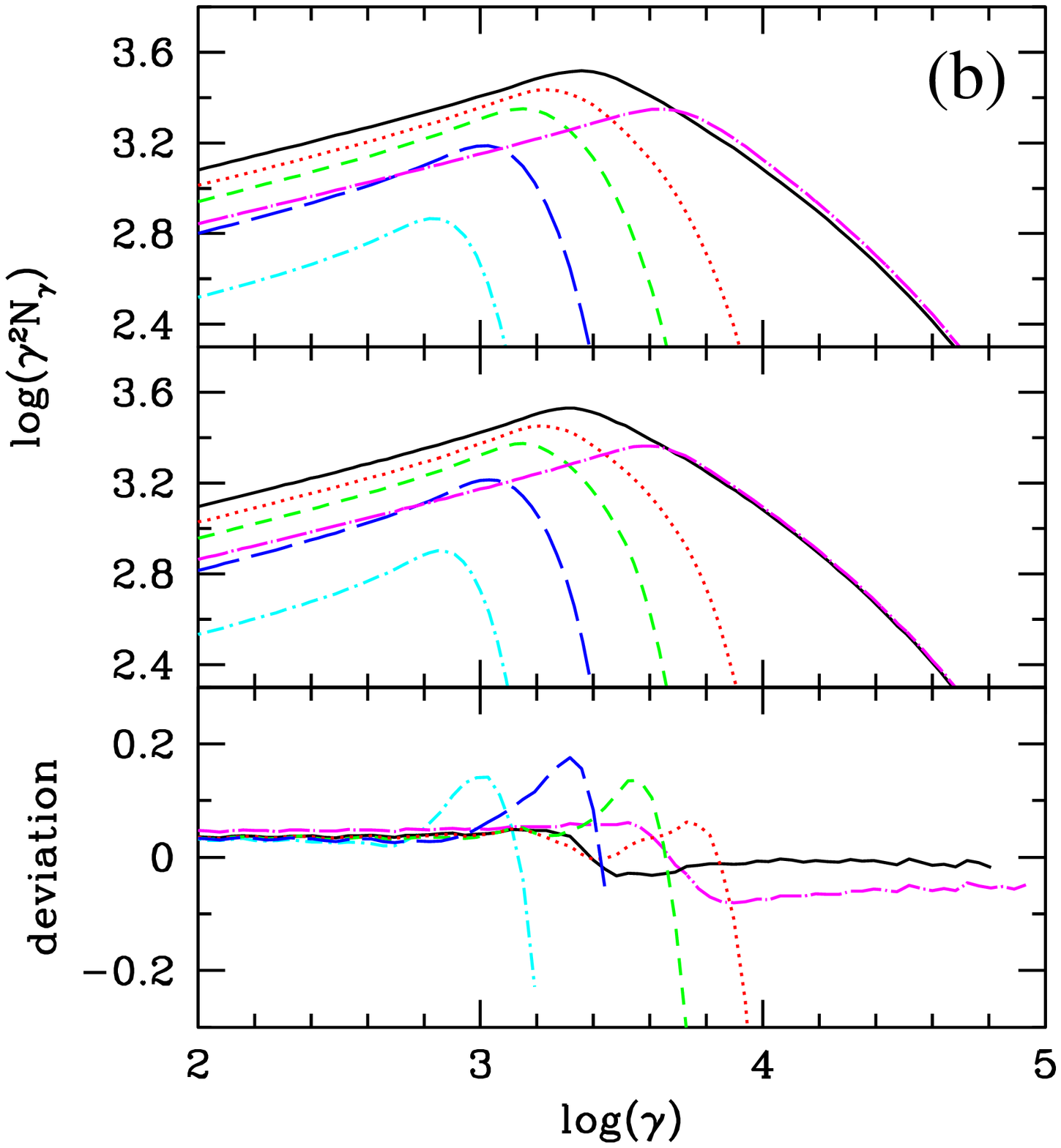}
\hfill
}
\caption{%
(a) Results of the steady state homogeneous model simulation.
The smooth red lines are the SEDs from \citet{fossati_etal:2008:xray_tev}.
The black histogram is the SED produced by our MC/F-P model run with the
closest possible source setup.
In the top panel we used the full K-N cross section, while in the middle
we used a step function approximation as done in \citet{fossati_etal:2008:xray_tev}.
The bottom panel shows the fractional deviation, $\Delta y/\langle y\rangle$, 
between these two latter SEDs.
Positive sign corresponds to the one-zone spectrum having a larger value.
\label{fig:test_steady_state}
%%%%%
(b) Results of the time-dependent homogeneous model simulation.
Electron distributions, as $\gamma^2 N(\gamma)$, for different
times for our MC/F-P code (top), or the one-zone code of
\citet{chiaberge_ghisellini:1999:timedep} (middle).
In different colors and line types we plot the spectra at the
following times (in units of $R/c$): 
0.5 (magenta, long dash dot),
1 (black, solid), 
1.25 (red, dotted), 
1.5 (green, sort dash), 
2 (blue, long dash),
3 (cyan, dot dash).
In the bottom panel we show the fractional deviation, $\Delta y/\langle
y\rangle$, between the spectra for the two codes. 
Lines are truncated at the energy where $\gamma^2 N_\gamma$ drops a
factor of 30 below its peak value.
\label{fig:test_pl_injection}
}
\end{figure*}
%-------------------------------------------------------

%%%%%%%%%%%%%%%%%%%%%%%%%%%%%%%%%%%%%%%%%%%%%%%%%%%%%%%%%%%%%%%%%%%%%%
\section{Test Runs} 
\label{sec:tests}

In order to test the reliability and robustness of the code, we compared the
results of our code with those of other authors using different codes, 
for cases where the codes' capabilities are comparable. 
We first compare the results with a non time-dependent code to test the MC
radiative part of the code. 
Then we compare the electron evolution with a time dependent code, in a
single-zone homogeneous case.  
Generally the results match very well.

%=====================================================================
\subsection{Steady state SED of homogeneous models}
\label{sec:tests:quasi_equilibrium}

To test how the code handles the radiative processes, we try to reproduce
the theoretical SED shown in \citet{fossati_etal:2008:xray_tev}, 
for the non-extreme parameter choice (solid line in their fig.~10a).
That SED was computed with a single-zone homogeneous SSC model. 
The electrons are assumed to be continuously injected and reach a steady state
\citep[\eg][]{ghisellini_etal:1998:seds}. 
For this test, we take the equilibrium electron distribution calculated in
the homogeneous code as our input electron distribution, and turn off the
F-P evolution of the electrons. 
We cut the volume into several identical zones just to make use of the
parallel structure of the code.
Since the single-zone model uses a spherical geometry, while our MC model
uses a cylindrical geometry, we choose to use the same radius
($R=10^{16}$\,cm), but with the height of the cylinder $Z = 4/3 R$, in
order for the two models to have the same volume.
The produced SED is shown in Fig.~\ref{fig:test_steady_state}a in the top
panel as black histogram, directly compared with that of
\citet{fossati_etal:2008:xray_tev}.
In general the two SEDs match well, except for a slight discrepancy around
the peak of the IC component. 
This arises from the fact that the single zone model uses a step function
to approximate the K-N cross section, while our code
implements the full K-N cross section.

We then also tested our code with the step function approximation. 
The result in shown in Fig.~\ref{fig:test_steady_state}a, middle panel.
The overall shapes of the SEDs match better. 
The total luminosity seems a little higher in the MC model. 
However, it is worth noting that although we are matching the volume, 
the geometry is different in the two codes and this has a small effect on
the IC component.
Moreover, in order to achieve a reasonable statistics the emitted photons
are integrated over a finite solid angle, \ie a range of angles $\theta$
(photon direction angle with respect to the jet axis, in the observer frame).  
Hence for a given bulk Lorentz factor $\Gamma_\mathrm{bulk}=26$, we are 
effectively integrating over a range of Doppler factors $\delta \simeq 23.5
\sim 28.7$, not exactly $\delta=26$ as for the comparison model. 

%=====================================================================
\subsection{Temporal evolution (one zone model)}
\label{sec:tests:temporal_evolution}

The other important aspect of our MC/F-P code, the Fokker--Planck evolution
of the electrons, was tested by comparing the code with the one-zone time
dependent homogeneous SSC code by \citet{chiaberge_ghisellini:1999:timedep}.
We set the number of zones to one, and used a power law injection 
with the same parameters they used 
($B=1$\,G, 
$\gamma_\mathrm{min}=1$, 
$\gamma_\mathrm{max}=10^5$, 
$p=1.7$, 
$L'_\mathrm{inj} = 3.69 \times 10^{41}$\,erg/s,
$t'_\mathrm{esc} = 1.5 R/c$, 
$t'_\mathrm{inj} = R/c$), except that our geometry is a
cylinder with $R = 1.1547 \times 10^{16}$\,cm, $Z= 10^{16}$\,cm, while they
use a sphere with $R = 10^{16}$cm. 

The electron spectra at different times are shown in
Fig.~\ref{fig:test_pl_injection}b; the upper panel shows the one produced
by the MC code, the middle panel the one produced by the one-zone code, 
while the bottom panel shows the deviation. 
The two spectra match reasonably well, giving us confidence that our code
handles the evolution of the electron distribution correctly.

%%%%%%%%%%%%%%%%%%%%%%%%%%%%%%%%%%%%%%%%%%%%%%%%%%%%%%%%%%%%%%%%%%%%%%
\section{Application to Mrk 421} 
\label{sec:applications}

\mrk is the archetypical `blue' blazar, the most luminous and best
monitored object in the UV, \xray and TeV bands.
It was the first extragalactic source detected at TeV energies
\citep{punch_etal:1992:mkn421_tev_discovery}.
As such it has been the target of multiple multiwavelength campaigns
with excellent simultaneous coverage by \xray and TeV telescopes.
\citep{takahashi_etal:2000:mkn421_1998,
maraschi_etal:1999:letter_sax_mkn421,
fossati_etal:2000:mkn421_temporal,
fossati_etal:2000:mkn421_spectral,
krawczynski_etal:2001:mrk421_x_tev,
blazejowski_etal:2005:multiwavelength_mrk421,
rebillot_etal:2006:multiwavelength_mrk421,
giebels_etal:2007:mkn421,
fossati_etal:2008:xray_tev,
donnarumma_etal:2009:mrk421_2008}.

For a first application of our code we focused on one of best flares ever observed, 
occurred on 2001 March 19 \citep{fossati_etal:2008:xray_tev}.
It was a well defined, isolated, outburst that was observed both in the \xray
and \gray bands from its onset through its peak and decay.
It uniquely comprised several rare favorable features, namely absence
of data gaps (except \textit{RossiXTE}'s short orbital gaps), excellent
TeV coverage by the HEGRA and Whipple telescopes and large amplitude
variation in both \xray and \gray bands.

%=====================================================================
\subsection{Observational constraints and goals}
\label{sec:applications:goals}

We aim to reproduce several observational features.
Some of them can be regarded as constraints on the setup of a
baseline model, as they provide guidance on the general properties and
parameter values yielding an acceptable fit to the SEDs
(\eg \citealp{tavecchio_maraschi_gg:1998:tev},
\citealp{bednarek_protheroe:1997:SSC_constraints}, and 
\citealp{fossati_etal:2008:xray_tev} for an example specific for the
observations studied here).
In this respect, we have five fundamental observables we want to match:
\begin{itemize}

\item The peak frequencies of the synchrotron and IC components,
$\nu_\mathrm{p,S}$, $\nu_\mathrm{p,IC}$, which for \mrk are observed
in the \xray and \gray bands.

\item The peak luminosity and the relative strengths of the two SED
components, $\nu L_\mathrm{p,S}$, $\nu L_\mathrm{p,IC}$.

\item The variability timescale ($t_\mathrm{var}$).
Combined with an hypothesis on the Doppler factor it provides
a constraint on the size of the blob.
For \mrk in \xray and \gray it is typically of the order of tens of kiloseconds.

\end{itemize}

Besides giving an indication about the size of the active region, the latter
can be different for different energy bands and in turn its energy dependence
can provide additional constraints on the model parameters and source geometry.
  
There is then a set of observational features whose explanation remains to
a large extent an open question.
They represent the ultimate goal of our work and the driver for the
development of a time dependent multi-zone model.
\begin{enumerate}
 
\item 
The quasi-symmetry of flare light curves, showing similar rising and
falling time-scales, both in \xrays and \grays.
The symmetry seems to be a quite common feature at several wavelengths.
It would seem to support the interpretation that the flare evolution is
governed by the geometry of the active region 
\citep{chiaberge_ghisellini:1999:timedep,
kataoka_etal:2000:crossingtimes_model}.
However, this could be true only if all other (energy dependent)
time-scales are shorter than the blob crossing time, or relevant
geometric time-scale, or only for emission at energies for which this
is true.

\item 
The characteristics of the multiwavelength correlated intensity variations.
The flare amplitude is generally larger in \grays than in the \xray
band, and flux variations show a quadratic (or higher order) relationship that
holds during both the rise and the decay phases of the flare.
This behavior was observed in \mrk on 2001 March 19, and also for other
`clean' flares, including for other blue blazars \citep[\eg][reporting
a cubic variation for PKS\,2155$-$304]{aharonian_etal:2009:pks2155}.

\item 
The existence and length of inter-band (\xray vs. \gray) and intra-band (soft
\xray vs. hard \xray) time lags, often with changing sign from flare to
flare (see references given above for \mrk).
In the isolated flare of 2001 March 19, \citet{fossati_etal:2008:xray_tev} report 
a possible lag of about 2 kilo-seconds of the TeV flux with respect to a
soft \xray band (2--4\,keV), whereas TeV and harder \xrays (9--15\,keV)
were consistent with no lag.  
In turn an \xray intraband lag was detected.

\item
The fact that even during large outbursts the optical flux changes 
little.
This may constrain the characteristics of the particle injection,
such as their spectrum (shape and density) and energy span.
On the other hand, the time dependent spectral behavior of blazars has
led people to speculate that there is more than one component
contributing to the blazar emission \citep{fossati_etal:2000:mkn421_spectral,
krawczynski_etal:2004:1es1959,
blazejowski_etal:2005:multiwavelength_mrk421,
ushio_etal:2009:mrk421_in_2006_with_suzaku}.
It is not clear if this additional zone is co-spatial with the zone
undergoing the flare or it is far enough elsewhere along the jet that
the two do not interfere with each other and evolve independently.

\item SED shape, and its time variations, particularly around the two peaks. 
For \mrk we mostly focus on the \xray and TeV \gray spectra.

\end{enumerate}
These features have been observed in several instances for \mrk, mostly
cleanly in the case of the 2001 March 19 flare, and the other well
studied TeV detected blue blazars.
For the brightest blue blazars there is an extensive database of
multiwavelength observations and studies of time resolved spectral
variability.  The phenomenology is richer and more complex than 
the few items just introduced, on which we focus.
In this respect, one of the most interesting findings is the
observation of a correlation between luminosity and position of the
peak of the synchrotron component \citep[\eg][]{tavecchio_etal:2001:mkn501,
fossati_etal:2000:mkn421_spectral,
tanihata_etal:2004:mrk421_evolution,
tramacere_massaro_cavaliere:2007:synchrotron_signatures}.

In this work we are mostly aiming at illustrating the capabilities of our
code with respect to investigating the above observational findings, by
presenting the results of simulations of three simple scenarios.

%===============================================================================
\begin{table*}
\begin{minipage}{\textwidth}
\caption{Summary of model parameters}
\label{tab:model_parameters}
\begin{tabular}{@{}l @{~~~~}c cc cc @{~~~~}c cccc @{~~~~}c ccc}
%-------------------------------------------------------------------------------
\hline
%%%%%%%%%%%%%%%%%%%%%%%%%%%%%%%%%%%%%%%%
Case && 
\multicolumn{4}{c}{general source parameters}  &&
\multicolumn{4}{c}{back-/fore-ground component} &&
\multicolumn{3}{c}{injected component} \\[6pt]
%%%%%%%%%%%%%%%%%%%%%%%%%%%%%%%%%%%%%%%%
  &&
$R$ &
$Z$ &
$\Gamma$ &
$B$ &
  &
$\gamma_\mathrm{min}$ &
$\gamma_\mathrm{b}$ &
$\gamma_\mathrm{max}$ &
$n_\mathrm{e}$ &
  &
$\gamma_\mathrm{min}$ &
$\gamma_\mathrm{max}$ &
$L'_\mathrm{inj}$ \\
%%%%%%%%%%%%%%%%%%%%%%%%%%%%%%%%%%%%%%%%
  &&
$10^{16}$ &
$10^{16}$ &
 &
 &
 &
 &
 &
 &
 &
 &
 &
 &
$10^{40}$  \\
%%%%%%%%%%%%%%%%%%%%%%%%%%%%%%%%%%%%%%%%
  &&
cm & 
cm & 
 &
G & 
 &
 &
 &
 &
cm$^{-3}$ &
 &
 &
 &
erg s$^{-1}$ \\
\hline
%%%%%%%%%%%%%%%%%%%%%%%%%%%%%%%%%%%%%%%%
% case #1: Background Initial energy: 2.17e46
%          injected number density  : 1.72e-2 (This is the injected number density, not the injection rate.)
% case #2: Background Initial energy: 1.52e44 
%          Foreground Initial energy: 2.31e46
%          injected number density  : 2.67e0 
% case #3: Background Initial energy: 2.38e46
%          injected number density  : 2.41e-2
%%%%%%%%%%%%%%%%%%%%%%%%%%%%%%%%%%%%%%%%
{1: with `background'} &&
 1.0 & 1.33 & 33  & 0.1 &&
 50 & $2 \times 10^{4}$ & $2   \times 10^{5}$ & 4.0 &&
      $50$              & $1.9 \times 10^{5}$ & 5.5 \\ %%& $5   \times 10^{5}$ \\%%[0.2in]
%%%%%
{2: with `foreground'} && 
 1.0 & 1.33 & 33 & 0.08 &&
 50 & $1 \times 10^{4}$ & $1   \times 10^{5}$ & 6.0 &&
      $50$              & $1.9 \times 10^{5}$ & 6.0 \\ %%& $5   \times 10^{5}$ \\%%[0.2in]
%%%%%
{3: better TeV spectrum} &&
 1.5 & 2.0 & 46 & 0.035 &&
 50 & $2 \times 10^{4}$ & $2   \times 10^{5}$ & 1.56 &&
      $50$              & $1.9 \times 10^{5}$ & 3.2 \\ %% & $8  \times 10^{5}$  \\%%[0.2in]
%%%%%
\hline
\end{tabular}
\\
% \medskip
All background or foreground components electron spectra are broken
power laws (with exponential cutoff), with spectral indices $p_1=1.5$,
$p_2=2.5$ ($=p_1+1$).  
The injected power law has spectral index $p=1.5$ in all cases.
\end{minipage}
\end{table*}
%===============================================================================

%=====================================================================
\subsection{On model parameters}
\label{sec:applications:parameter_choice}

This code affords us a great freedom.  In particular, we could setup each
zone with different initial conditions.  
However, for the scenarios presented in this work we took a conservative
approach and setup each zone with identical values for the usual set of
physical parameters.

Our homogeneous (at least initially) SSC model is defined by the following
quantities (see also Table~\ref{tab:model_parameters}): 
\begin{itemize}
\item source size/geometry ($R$, $Z$, or aspect ratio), 
\item Lorentz factor ($\Gamma$), 
\item magnetic field strength ($B$),
\item various parameters describing the electron spectrum, 
\eg four for an injected power law: $\gamma_\mathrm{min}$,
$\gamma_\mathrm{max}$, $p$, $L'_\mathrm{inj}$.
For a broken power they would be six because there would be a spectral
break $\gamma_\mathrm{b}$ and two spectral indices ($p_1$, $p_2$) instead of one.
\end{itemize}

With simple considerations we can reduce the number of model parameters to
constrain from 8 (or 10) to 5 ($B$, $\Gamma$, $R$, $\gamma_\mathrm{max}$,
$L'_\mathrm{inj}$) and as illustrated in the previous section we have $5+$
fundamental observables to do it.

The source aspect ratio can be at least qualitatively constrained by the
profile of the flare light curve, for in first approximation extreme
geometries would yield fairly distinctive flare shapes due to LCTE.
For this work we adopted a conservative, stocky, volume aspect ratio
$R/Z = 3/4$, \ie width:length = 3:2. 

Among the electron spectrum parameters, $\gamma_\mathrm{min}$ and $p$ (or
$p_1$) can be set with reasonable confidence based on considerations on the
SED shape and variability (or lack thereof) at frequencies below the
synchrotron peak.
The precise value of $\gamma_\mathrm{min}$ is however not well constrained by
observations.
The emission by electrons at $\gamma \le 10^3$ would be below the optical
band, where there is not much simultaneous coverage, and emission by much
lower energy electrons would fall in a band (\ie $\nu \le 10^{11}$\,Hz)
where observations suggest that the SED is dominated by radiation from
other regions of the jet \citep[\eg][]{kellerman_pauliny:1981:compact_radio_sources}.
Moreover, cooling time-scales for electrons at those energies are long
compared with the typical duration high-density multiwavelength campaigns
(see eq.\,\ref{eq:tau_sync}), making it difficult to set a constraint on
$\gamma_\mathrm{min}$ based on variability.  
Higher values of $\gamma_\mathrm{min}$ affect the synchrotron emission in the
optical band and in turn the IC component, mainly in the GeV band, and
therefore we can assess their viability with current and future observations.
Given that during the 2001 campaign \citep{fossati_etal:2008:xray_tev} there
seemed to be a modest level of variability in the optical band, $\Delta
m_\mathrm{V} \simeq 0.4$, though not directly from observations simultaneous
with the March 19 flare, we simulated scenarios where the injected electron
population has a relatively low $\gamma_\mathrm{min}=50$
(Table~\ref{tab:model_parameters}).
We choose to truncate the electron distribution at this value also
because the number of low energy electrons grows rapidly, thus increasing
significantly the computational time without adding much to the investigation
presented here; as noted, emission from lower energy electrons would not be
detectable, and they would not significantly alter the properties of the
emission and its variability observed in blue blazars.  
This is of course an assumption that is valid for this work but that should
be revisited, for instance for the study of red blazars.

The spectral indices of the injected electron distributions $p$ or $p_1$ 
($p_2=p_1+1$ as expected for a cooling break) are mostly constrained by the 
shape of the synchrotron SED at energies below the optical range.
The preferred value for $p,p_1=1.5$ constitutes a somewhat hard spectrum but
it is consistent with values discussed by several particle acceleration
studies.
In particular, stochastic (2$^\mathrm{nd}$-order Fermi acceleration) and
acceleration at relativistic shear layers have been suggested to produce
very hard ($p < 2$) particle spectra 
\citep{virtanen_vainio:2005:particle_acceleration,
stawarz_ostrowski:2002:acceleration,
rieger_duffy:2004:shear_acceleration,
rieger_duffy:2006:shear_acceleration}.

In the context of this discussion, it is worth adding that we
considered scenarios with and without a pre-existing population of
relativistic electrons or an external `diluting' SED (see 
\S\,\ref{sec:applications:standard_case} and
\S\,\ref{sec:applications:no_bkg_electrons} for details), and their
characteristics could be regarded as an additional degree of freedom of our
modeling. 
In this respect, however, while a particular choice of values has some
effect on the best parameters for the component responsible for the
variability, its effect is fairly limited and the most important point 
is the existence or not of such secondary component (see
\S\,\ref{sec:discussion}).

Next, we illustrate some general arguments and estimates for values of the
fundamental physical parameters.
We then present and discuss the results obtained with the set of parameters
that we deemed more successful, and in turn `fit' the SEDs and light
curves of the 2001 March 19 outburst testing several different parameter
combinations, including the possible dilution by emission from a different
region of the jet not involved in the flare, and the presence of a
pre-existing electron population in the region that becomes active.

%=====================================================================
\subsection{Estimates of active region parameters from observables}
\label{sec:applications:parameter_estimates}

Key parameters in the modeling of blazars with the SSC model include the
Lorentz factor, the magnetic field strength, the size of the volume, and the
energy of the electrons that are responsible for the synchrotron peak of the
SED, $\gamma_\mathrm{p}$.
This latter is associated with a break in the electron distribution or its
maximum, depending on the spectral index.
We use the observational results of \citet{fossati_etal:2008:xray_tev}
as the benchmark of our analysis.  
There are several observed features that constrain the value of these
parameters (see previous section).
Additionally, independent estimates of the relativistic beaming parameters
of blazars, from observed superluminal motion as well population
statistics, yield Lorentz factors of the order of tens
\citep{urry_padovani:1995:review}.  
As we mentioned before, we make the common assumption to be observing the
source at the angle $\theta = 1/\Gamma$, hence $\delta = \Gamma$
\citep[see][for a deeper statistical analysis, showing that the most likely
combination is $\Gamma\sin\theta\simeq0.7$]{cohen_etal:2007:beaming_and_intrinsic_properties}.

The observed peak of the synchrotron component (at energy $E_\mathrm{p,S}$)
results from the combination of electrons' $\gamma$, $B$ and $\delta$.
Assuming mono-energetic emission the relationship is
$E_\mathrm{p,S} = \nu_\mathrm{B}\, \gamma^2_\mathrm{p}\, \delta$.
For \mrk $E_\mathrm{p,S} \simeq 1$\,keV. 
Parameterized\footnote{%%%%
Because the redshift of Mrk\,421 is small, $z=0.031$, for simplicity we left
out factors $(1+z)$.} on fiducial values for these three parameters the
$\gamma_\mathrm{p}$ of the electrons emitting at the synchrotron peak is:
\begin{equation}
\gamma_\mathrm{p} \simeq 1.7 \times 10^{5}\; 
        \left(\frac{B}{0.1\ \mathrm{G}}\right)^{\!\!-1/2} 
	\left(\frac{\delta}{30}\right)^{\!\!-1/2}
        \left(\frac{E_\mathrm{p,S}}{1\ \mathrm{keV}}\right)^{\!\!1/2} 
\end{equation}
If the IC component peak resulted from scattering of photons of the
synchrotron peak in Thomson regime we could directly infer the energy of
the electrons emitting at both SED peaks as
\begin{equation}
\gamma_\mathrm{p} \simeq \left(\frac{3\, E_\mathrm{p,IC}}{4\, E_\mathrm{p,S}}\right)^{\!\!1/2} \!\!\simeq 
        2.7 \times 10^{4} 
        \left(\frac{E_\mathrm{p,IC}}{1\ \mathrm{TeV}}\right)^{\!\!1/2} 
        \left(\frac{E_\mathrm{p,S}}{1\ \mathrm{keV}}\right)^{\!\!-1/2} 
\end{equation}
with $E_\mathrm{p,IC}$ is the peak energy of the IC component.
However, as discussed by \citet[][see Fig.~11 therein]{fossati_etal:2008:xray_tev},
the SED shape and variability time-scale observed in \mrk in 2001 
favor parameters such that the scattering between electrons at
$\gamma_\mathrm{p}$ and synchrotron peak photons at $E_\mathrm{p,S}$ would
happen in the K-N regime \citep[see also][]{tavecchio_maraschi_gg:1998:tev,bednarek_protheroe:1999:mrk501_constraints}.
In this case the IC peak energy would be largely independent of
$E_\mathrm{p,S}$ and the expression would instead be: 
\begin{equation}
\gamma_\mathrm{p} \simeq \frac{E_\mathrm{p,IC}}{\delta\, \me c^2} \;\simeq\; 6.5 \times 10^{4} 
        \left(\frac{E_\mathrm{p,IC}}{1\ \mathrm{TeV}}\right)\, 
        \left(\frac{\delta}{30}\right)^{\!\!-1} \, .
\end{equation}

Requiring that the condition for Thomson regime, $\gamma x' \leq 3/4$ (where
$x' = E'_\mathrm{target}/(\me c^2)$), holds true for $E'_\mathrm{target} = E'_\mathrm{p,S}$ 
and $\gamma=\gamma_\mathrm{p}$, one can derive a rough estimate of what
($B$, $\delta$) combination would be necessary to push into the Thomson
regime the scattering between $\gamma_\mathrm{p}$ and its own synchrotron
photons, emitted at $E_\mathrm{p,S}$.
%REV2% , \ie to make the IC peak the exact SSC match of the synchrotron one.
\begin{equation}
\left(\frac{B}{0.1\ \mathrm{G}}\right) \left(\frac{\delta}{30}\right)^{\!\!3} \ge 220 \,
        \mathbf{\left(\frac{E_\mathrm{p,S}}{1\ \mathrm{keV}}\right)^{\!\!3}}
\end{equation}
As shown by \citet{fossati_etal:2008:xray_tev}, it is indeed possible to
achieve an acceptable SED fit with high $B$ and $\delta$. 
However, while this kind of model matches equally well a static SED, its
smaller blob size and extreme Lorentz contraction make it implausible when
compared with more dynamic observational findings, beginning with the
variability time-scales.

The rest frame synchrotron cooling time can be expressed as function of
electron energy and the magnetic field:
\begin{equation}
\tau'_\mathrm{cool,S} = \frac{7.7\times10^8}{\gamma B^2} ~~\mathrm{s} 
\label{eq:tau_sync}
\end{equation}
or, more directly related to observables, in terms of observed photon energies:
\begin{equation}
\tau'_\mathrm{cool,S} = 4.6\! \times\! 10^5\, 
        \left(\frac{B}{0.1\ \mathrm{G}}\right)^{\!\!-3/2} 
	\left(\frac{\delta}{30}\right)^{\!\!1/2}
        \left(\frac{E_\mathrm{S}}{1\ \mathrm{keV}}\right)^{\!\!-1/2} ~\mathrm{s}
\label{eq:tau_sync_parameterized}
\end{equation}
or, in the observer's frame,
\begin{equation}
\tau_\mathrm{cool,S} = 15.1\,
        \left(\frac{B}{0.1\ \mathrm{G}}\right)^{\!\!-3/2} 
	\left(\frac{\delta}{30}\right)^{\!\!1/2}
        \left(\frac{E_\mathrm{S}}{1\ \mathrm{keV}}\right)^{\!\!-1/2} ~\mathrm{ks}
\label{eq:tau_sync_obs_parameterized}
\end{equation}
showing its dependence on the inverse square root of the energy of the
observed photons.

A general constraint among the observed variability time-scale and source
size and time-scale of the acceleration, injection or cooling process is:
\begin{equation}
t_\mathrm{var} \ge \frac{1}{\delta}\,
\max\left(\tau'_\mathrm{cool}, \tau'_\mathrm{acc}, t'_\mathrm{inj}, \frac{R}{c}, \frac{Z}{c}\right)
\end{equation}
As noted in Section\,\ref{sec:code:changes:injection}, in this work we take 
a simplified approach, whereby we do not specify the acceleration
mechanisms underlying the particle injection, and we choose to link the
injection time-scale to the geometry of the source, namely its dimension
along the jet axis, $Z$.  
Hence we have a simplified relationship with the observed variability, and
considering that the 2001 March 19 flare has a flux doubling and halving
time of the order of $10^4$\,s, we have approximately
\begin{equation}
\max\left(\tau'_\mathrm{cool}, \frac{R}{c}, \frac{Z}{c}\right) \simeq 
\max\left(\tau'_\mathrm{cool}, \frac{R}{c}\right) \simeq 10^4\, \delta \quad\mathrm{s}
\end{equation}
Please note that this constraint could actually vary with the observed band
because some time-scales are likely to be energy dependent.

If the IC cooling rate is similar to the synchrotron cooling rate
$\tau'_\mathrm{cool} \sim \tau'_\mathrm{cool,S}/2$. 
The condition $\tau'_\mathrm{cool} < R/c$ translates into
\begin{equation}
E_\mathrm{S} > 0.46\; 
    \left(\frac{R}{10^{16}\ \mathrm{cm}}\right)^{\!\!-2}\,
    \left(\frac{B}{0.1\ \mathrm{G}}\right)^{\!\!-3}\, 
    \left(\frac{\delta}{30}\right) \quad\mathrm{keV}
\end{equation}

From the constraints and relationships illustrated above we infer that a
good starting point to model the SED of \mrk is a combination of 
$R \sim 10^{16}$\, cm,
$B\sim 0.1$\, G, 
$\Gamma \simeq \delta \sim 30$,
$\gamma_\mathrm{p} \sim 10^{5}$.

Because of computational limitations we did not perform an actual fit to
identify the best set of parameters values reproducing the SED and the
flare evolution properties.
We explored a limited range of values for $R$, $B$, $\Gamma$ around the
values obtained from the above analysis, and focused on changes of the
maximum electron energy $\gamma_\mathrm{max}$ and the injected
power $L'_\mathrm{inj}$.  

We ran a large number of short simulations aimed at sampling a reasonable
range of values around our initial guesses and evaluated them mostly on the
basis of their matching the \xray spectra and variability. 
In a second stage we honed in on the best cases and adjusted the parameters
by means of full length simulations\footnote{%%%%
A typical run takes around 24 hours on eight Xeon 2.83\,GHz CPU cores,
using up to 16 GB of memory.  
As currently implemented the code does not scale well with the number
of CPUs, only gaining a factor of three in speed by going to 96 CPUs.
The bottleneck is mostly due to the longer computational time required by
the zones with larger volume because it scales with the number of photons
contained in each zone.}.

%-------------------------------------------------------
\begin{figure*}
\centerline{%
\hfill
\includegraphics[width=0.49\linewidth]{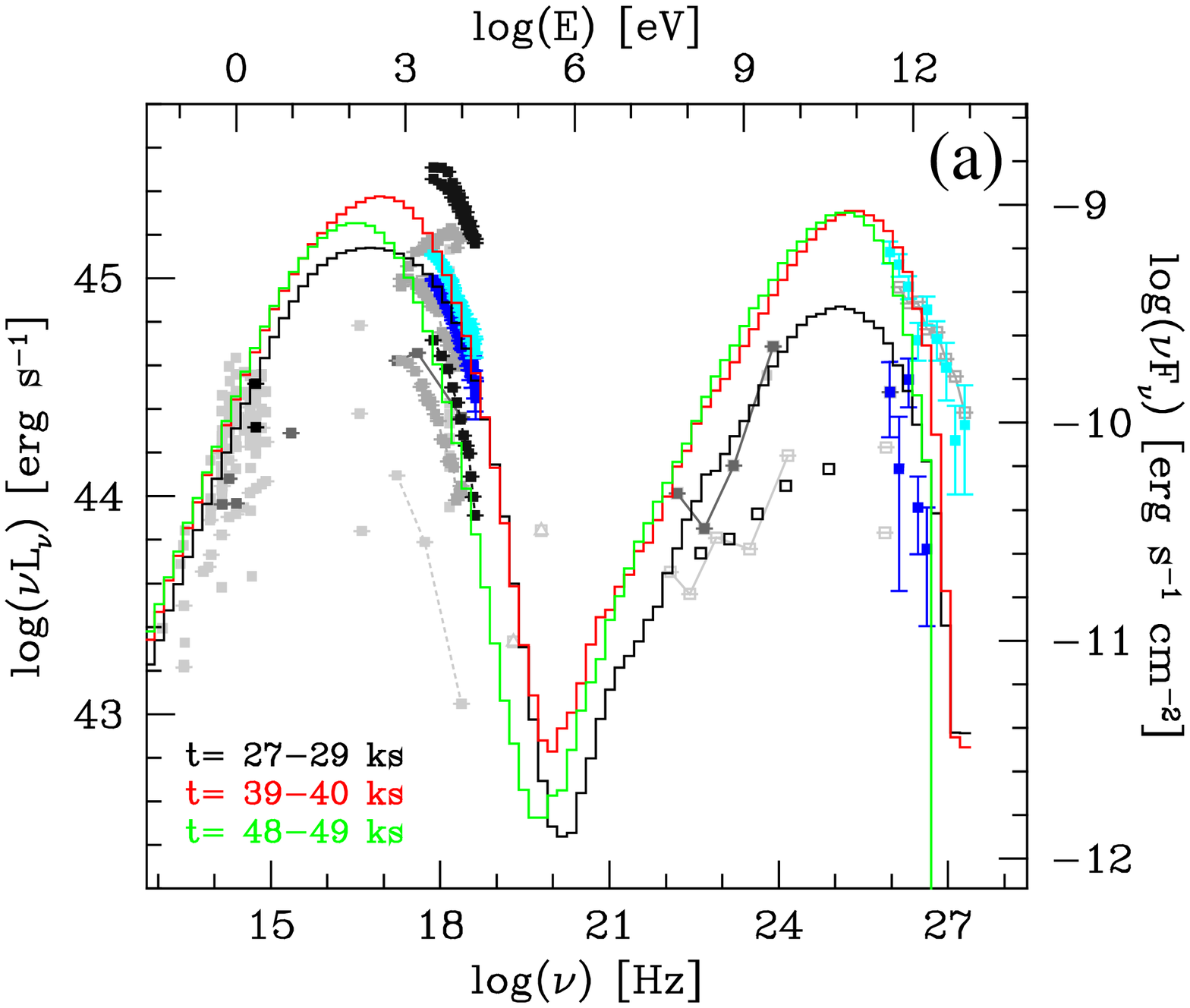}
\hfill
\includegraphics[width=0.49\linewidth]{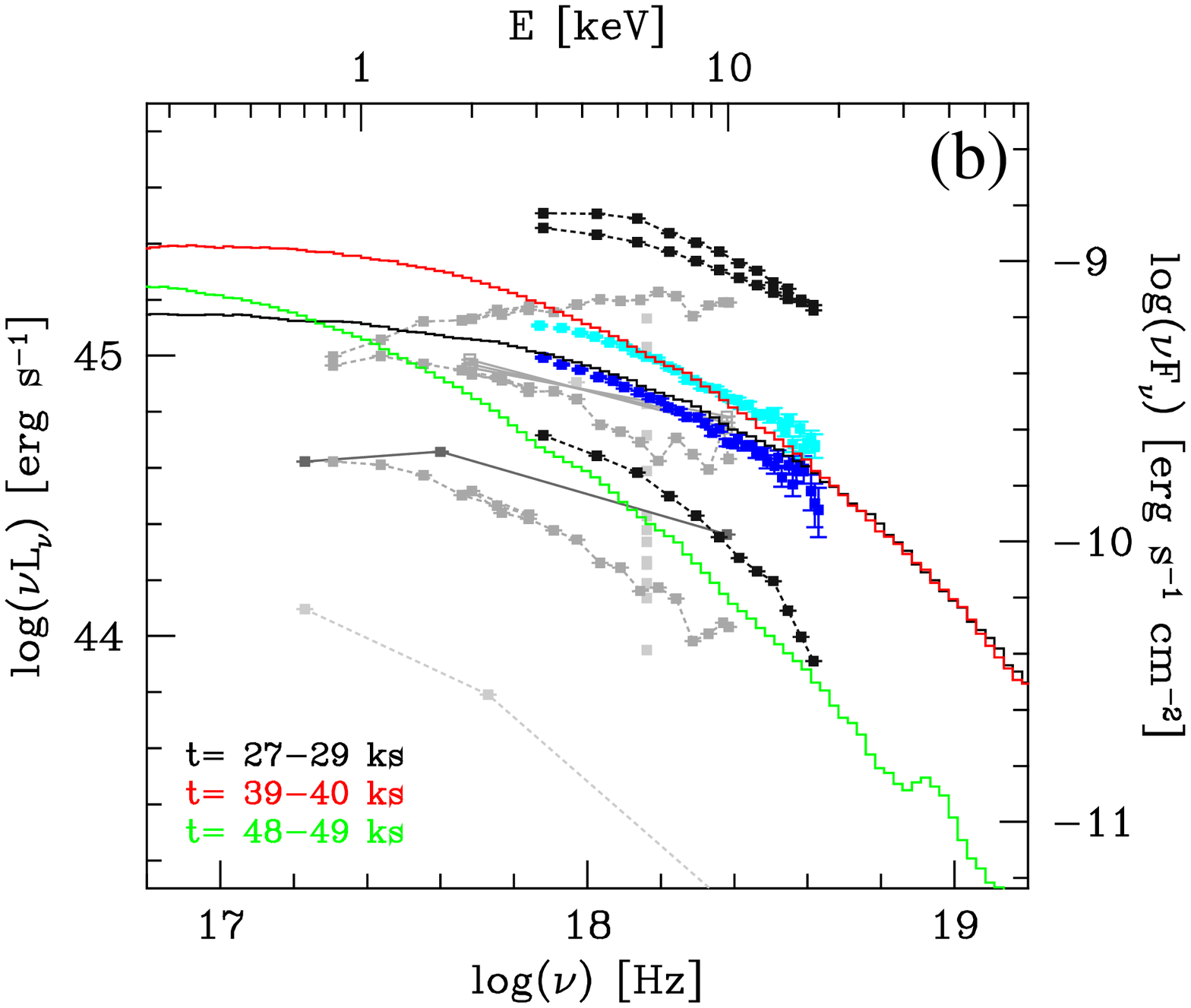}
\hfill
}
\centerline{%
\hfill
\includegraphics[width=0.45\linewidth]{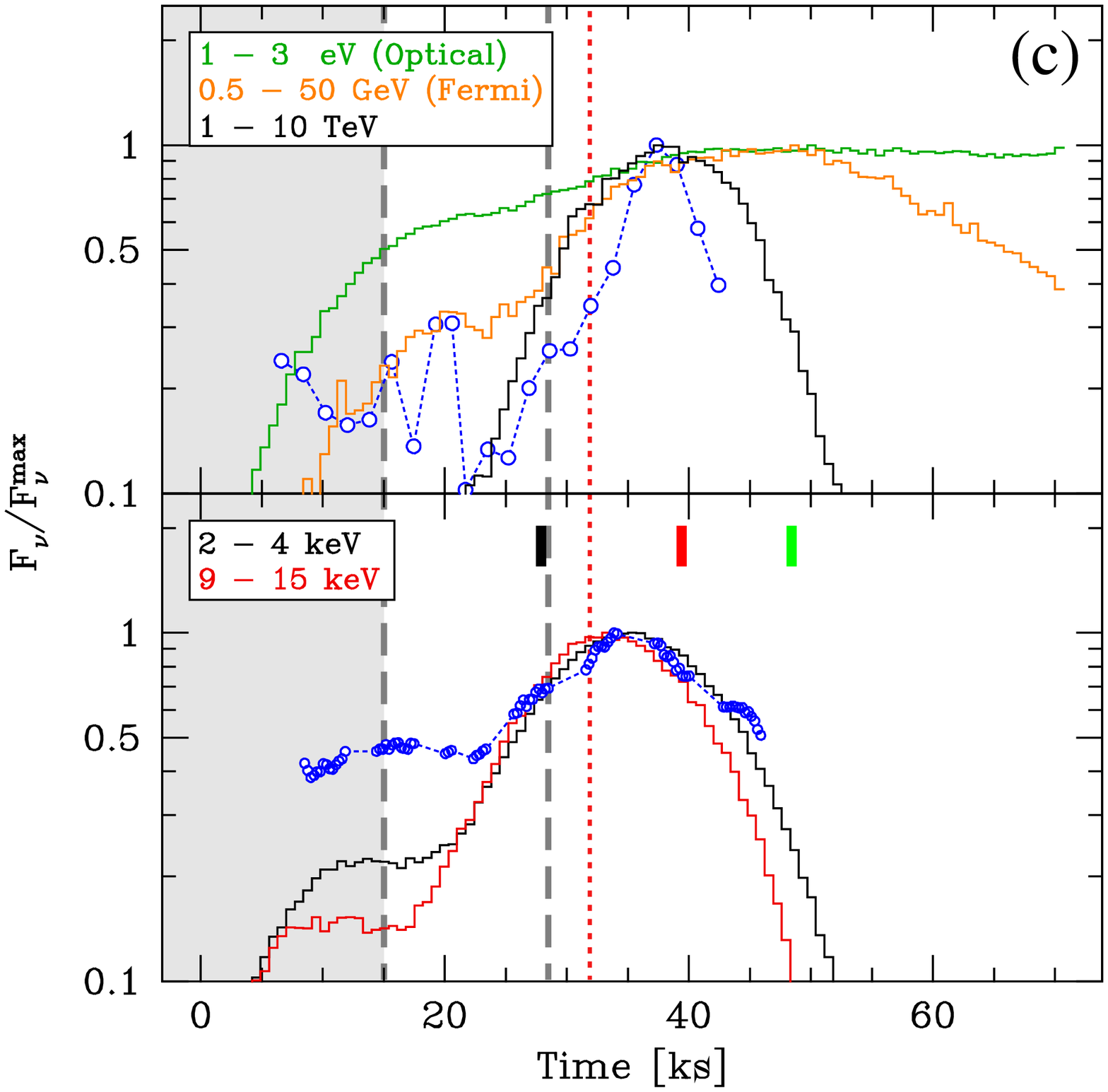}
\hfill
\includegraphics[width=0.45\linewidth]{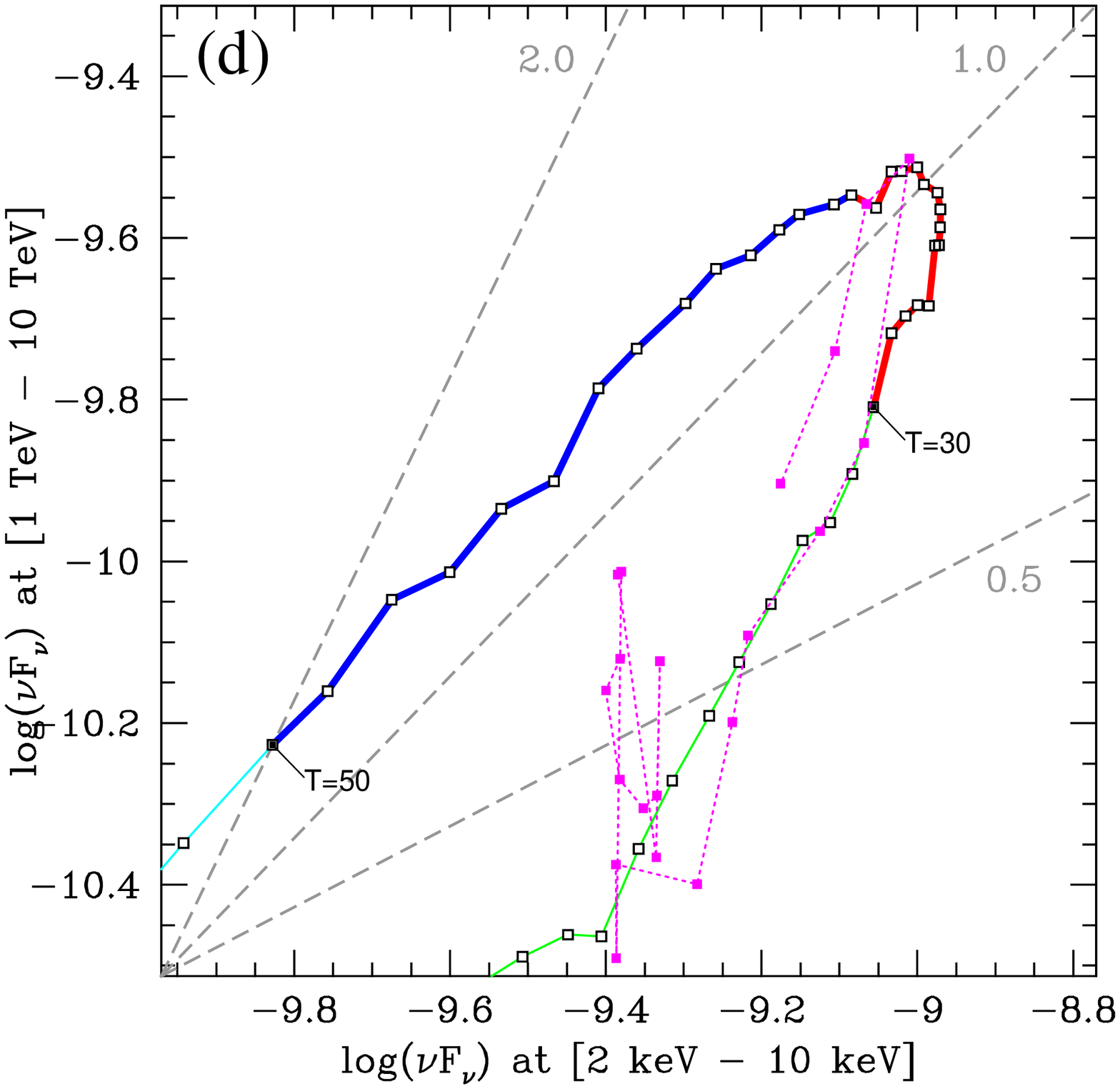}
\hfill
}
\caption{%
Summary of results of the first case, with pre-existing background electron
population. All quoted times are in the observer's frame.
(a) Broad band SEDs for three representative times (see labels) during the
simulation of the flare.
Observed spectra for high and low states during the 2001 March 19 flare in
\xray and TeV \grays are plotted in cyan and blue.  
Other historical data points in grey or black are the same as in
\citet{fossati_etal:2008:xray_tev}.   
The black empty squares in the GeV \gray band are the \textit{Fermi}/LAT 
1-year averages in five bands.
\label{fig:standard_seds}
(b) Zoom centered on the \xray band.
\label{fig:standard_seds_zoom}
(c) Light curves at 5 different energy bands, on 700\,seconds bins.
Light curves are normalized to their peak values.
In the bottom panel we show two \xray energy bands, and in blue the
\textit{RossiXTE}/PCA 2--4\,keV data.
The top panel comprises the simulated TeV band light curve, to be compared
with the blue data points, as well as light curves in optical and a band
representative of the \textit{Fermi}/LAT bandpass.
\label{fig:standard_lc}
The grey shaded area is intended to highlight the initial section of
the light curve which is not meaningful because it corresponds to the
interval during which the pre-existing electron population is being
`prepared'.
The long dashed vertical grey lines mark the injection period.
The dotted red line marks the time corresponding to the largest
cross-section of the active region along planes of equal observed
times (see text, \S\,\ref{sec:applications:geometric_effects}).
The colored short thick segments mark the times corresponding to the SEDs
plotted in (a) and (b).
(d) The flux vs. flux plot for \xray and \grays.
Simulations data have been smoothed with a one-hour width boxcar filter.
Colors highlight different 10\,ks time intervals. 
As labeled, red starts at t=30\,ks.
For the red and blue lines we used a thicker trait to highlight the
central time interval of the outburst.
The magenta points connected by a dotted line show the observed correlation of
TeV flux in Crab units vs. \xray in count rate for the 2001 March 19 flare.
Because of the different units they are plotted at an arbitrary position.
\label{fig:standard_ff}
}
\end{figure*}
%-------------------------------------------------------

%-------------------------------------------------------
\begin{figure*}
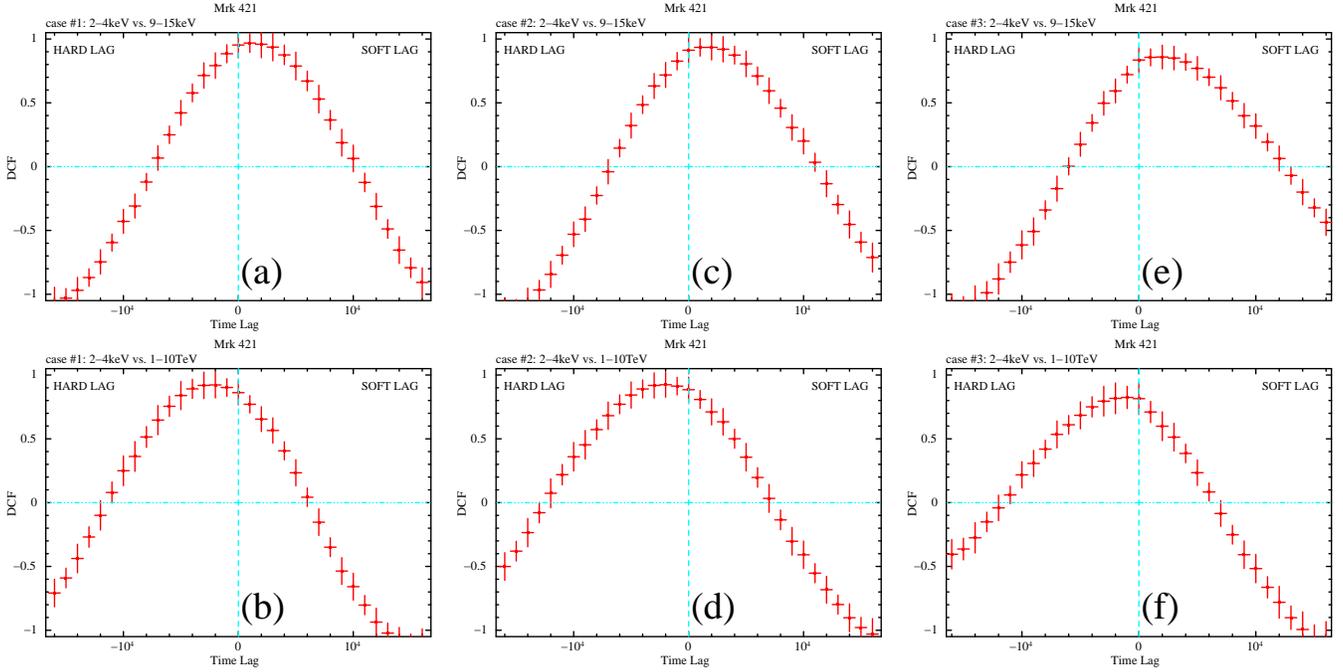

\centerline{%
\includegraphics[angle=270,scale=0.24]{f05a.ps}
\hfill
\includegraphics[angle=270,scale=0.24]{f05c.ps}
\hfill
\includegraphics[angle=270,scale=0.24]{f05e.ps}
}
\centerline{%
\includegraphics[angle=270,scale=0.24]{f05b.ps}
\hfill
\includegraphics[angle=270,scale=0.24]{f05d.ps}
\hfill
\includegraphics[angle=270,scale=0.24]{f05f.ps}
}
\caption{%
Discrete cross correlations analysis.
Our simulated cases run left to right, 
(a, b) for the `background' case, 
(c, d) for the `foreground' case, 
(e, f) for the `better TeV spectrum' case.
Top panels show the intra-band \xray correlation between \xray ($2-4$\,keV) 
and \xray ($9-15$\,keV), 
and bottom panels the inter-band correlation between soft(er) \xray
($2-4$\,keV) and TeV \grays.
\label{fig:dcf}
}
\end{figure*}
%-------------------------------------------------------

%=====================================================================
\subsection{Case 1: injection in a blob with a pre-existing (background) electron population}
\label{sec:applications:standard_case}

In all cases presented in this paper, the outburst is attributed to the
injection into our active volume of a new population of higher energy
electrons, with fixed injected spectrum (power law with exponential cutoff).

In this first scenario the blob is not empty, but it is filled with a
`background' population of electrons, homogeneous throughout the volume.  
%REV% These electrons serve as a quasi-steady (slowly evolving) component in the
These electrons serve as a slowly evolving component in the electron
distribution and in turn the SED, which can be regarded as the remnants of
a previous phase of activity.  
They participate fully in the time evolution of the blob, cooling and
emitting radiation.

The overall best case has the following parameters:
$R=10^{16}$\,cm, 
$Z=4/3 \times 10^{16}$\,cm, 
$B=0.1$\,G, 
$\Gamma=33$.
Parameters for this and all following cases are summarized in
Table~\ref{tab:model_parameters}.

At $t=0$ the electron spectrum for the `background' population is
a broken power law distribution: 
\begin{align*}
 N(\gamma) & = N_\mathrm{b}\left(\frac{\gamma}{\gamma_\mathrm{b}}\right)^{-p_1} 
	   & \text{cm}^{-3} &
           & \gamma_\mathrm{min} < \gamma < \gamma_\mathrm{b} \\
 N(\gamma) & = N_\mathrm{b}\left(\frac{\gamma}{\gamma_\mathrm{b}}\right)^{-p_2}\, 
                           e^{-\gamma/\gamma_\mathrm{max}}
	   & \text{cm}^{-3} &
           & \gamma_\mathrm{b} \leq \gamma 
\end{align*}
The spectral indices are $p_1=1.5$ and $p_2=2.5$. 
The break is at $\gamma_\mathrm{b} = 2 \times 10^4$, the high-energy cut-off at
$\gamma_\mathrm{max} = 2 \times 10^5$. 
The number density of this `background' population is $n_\mathrm{e}=4$\,cm$^\mathrm{-3}$. 
Their total energy content is $2.2 \times 10^{46}$\,ergs.

By the time when the new flare begins, \ie the shock begins to cross the
region and inject electrons, this pre-existing population has cooled to
a $\gamma_\mathrm{max}$ of a few times $10^4$, yielding a synchrotron peak
at around 50\,eV.
In the observer's frame, the cooling time-scale for the peak of the
`background' component is of the order of 1 day and we could think of it as
due to the aging of the electron spectrum from a previous active phase
occurred a few days earlier.  
In most recent long observing campaigns \mrk exhibited flares on about this
time-scale \citep[\eg][]{takahashi_etal:2000:mkn421_1998}.

The injection of electrons begins at $t'_\mathrm{start,inj} = 5 \times 10^5$\,s, 
with a power law distribution (\S\ref{sec:code:changes:injection}).
The parameters of the injected spectrum are: 
$p=1.5$, 
$\gamma_\mathrm{min}= 50$,
$\gamma_\mathrm{max}= 1.9 \times 10^5$, 
$L'_\mathrm{inj} = 5.5 \times 10^{40}$\,erg/s.

The emitted --beamed-- photons are integrated over the angle of
$0.99944<\cos(\theta)<0.99964$, which corresponds to a Doppler factor of
$27<\delta<42$.

%-------------------------------------------------------
\begin{figure*}
\centerline{%
\hfill
\includegraphics[width=0.49\linewidth]{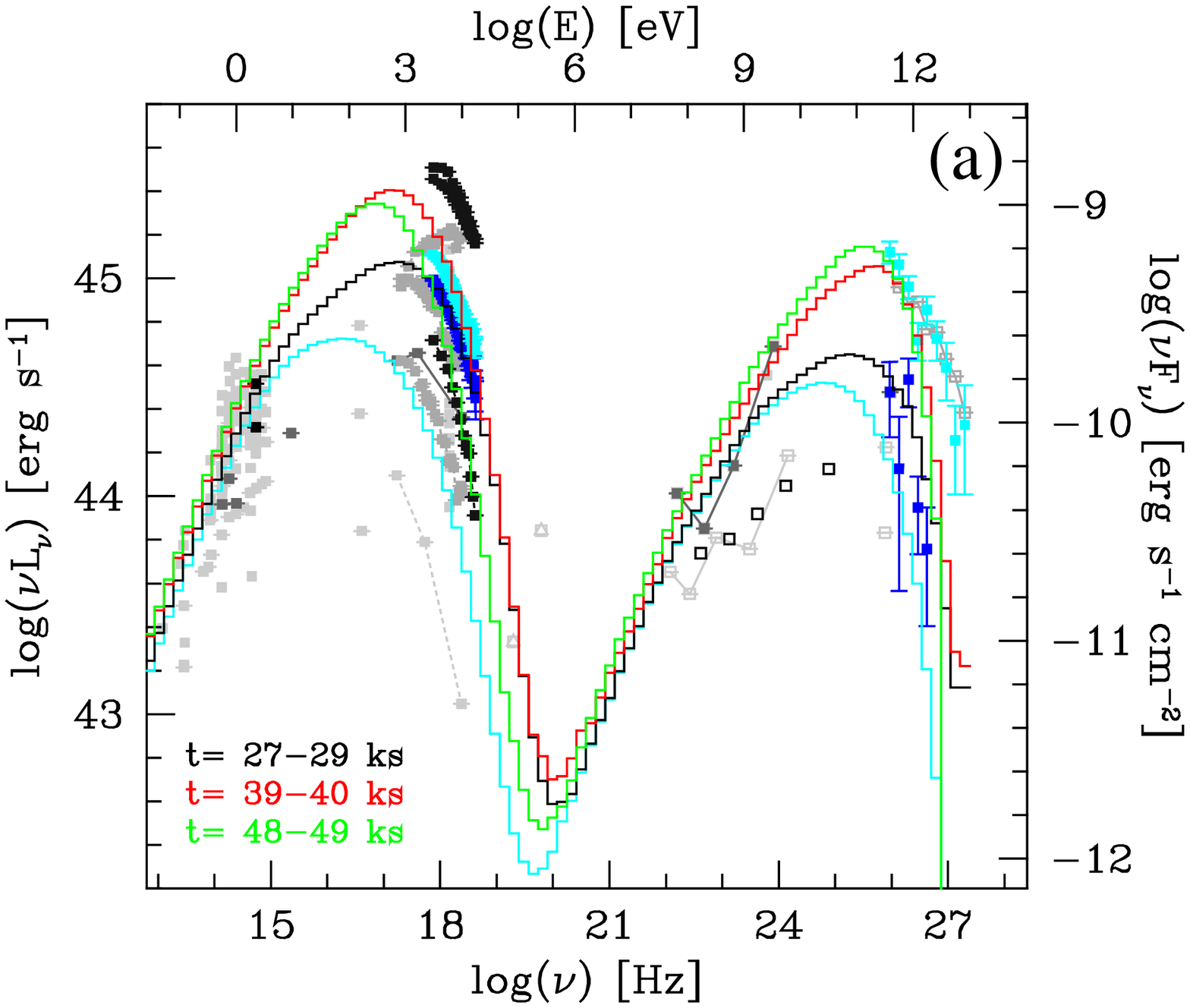}
\hfill
\includegraphics[width=0.49\linewidth]{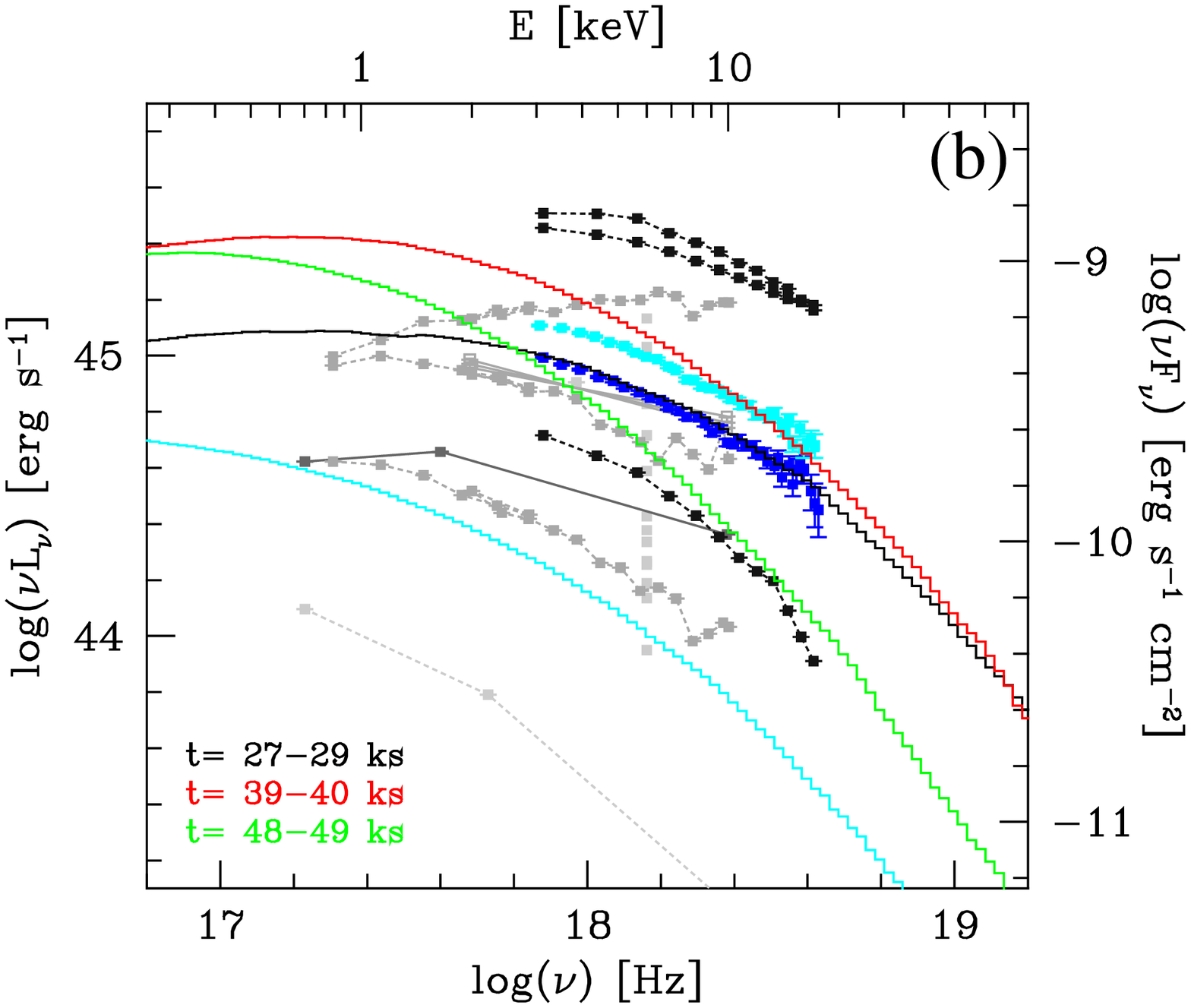}
\hfill
}
\centerline{%
\hfill
\includegraphics[width=0.45\linewidth]{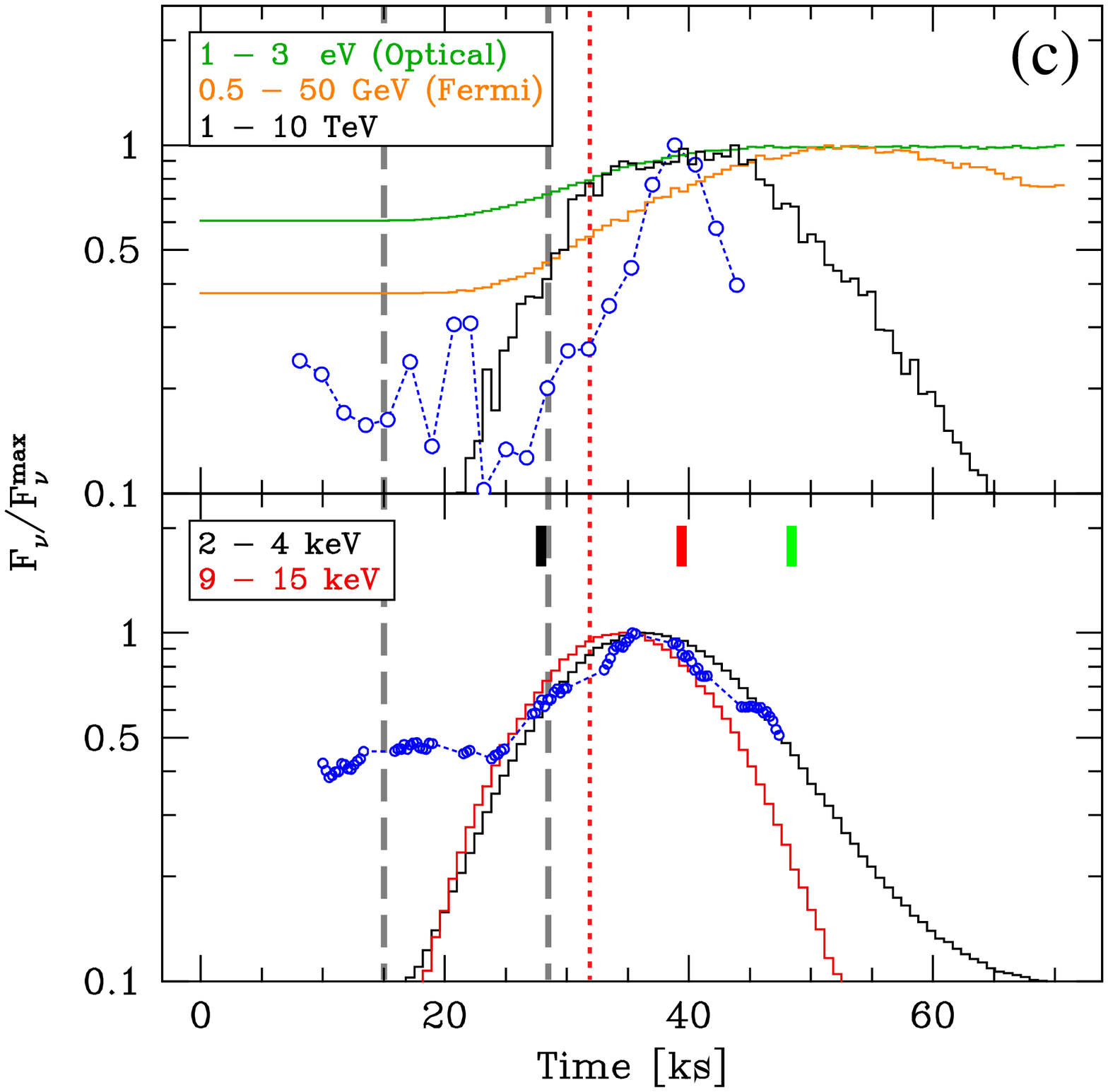}
\hfill
\includegraphics[width=0.45\linewidth]{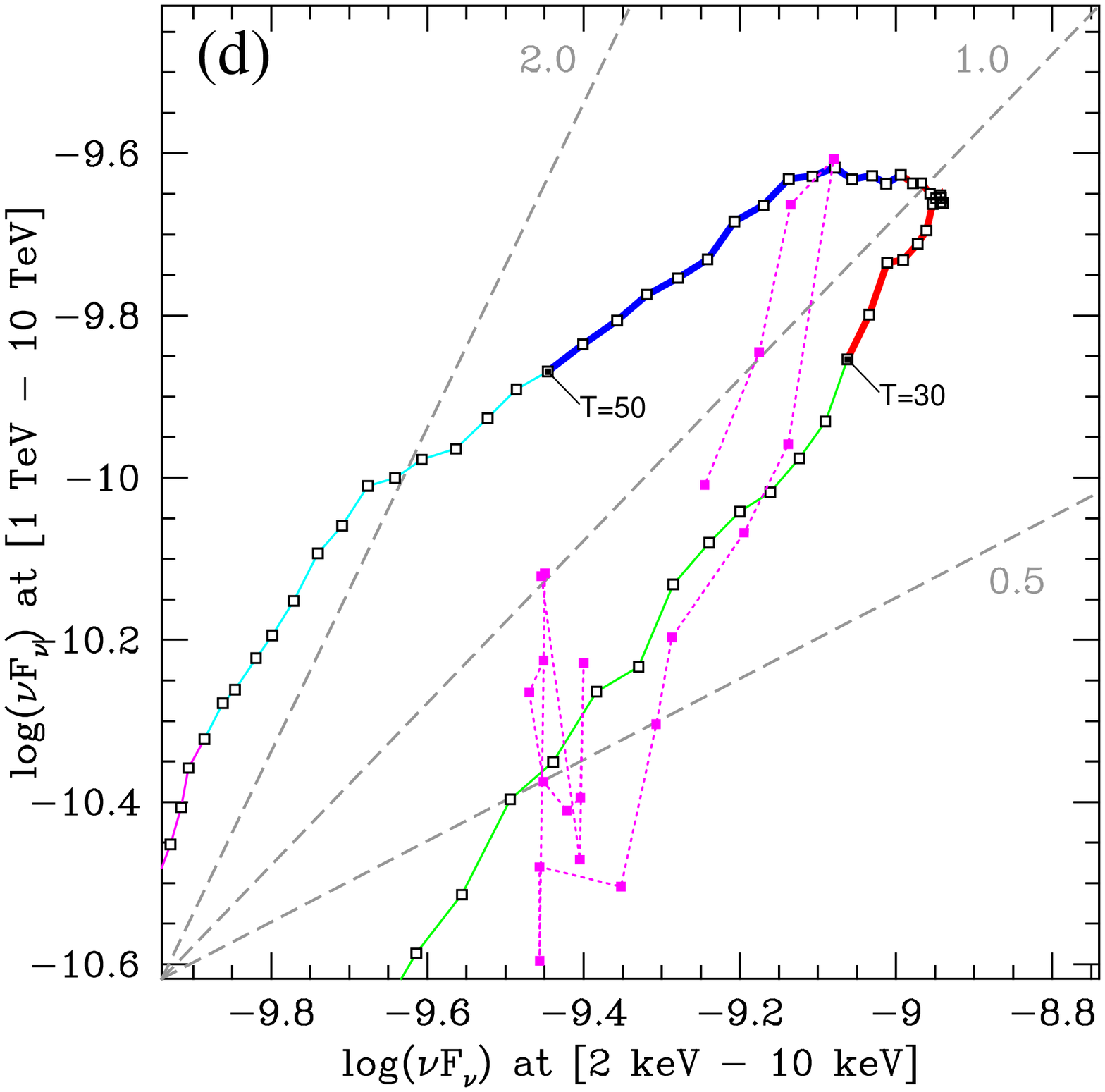}
\hfill
}
\caption{%
Summary of results for the second case (empty blob, with foreground emission).
All panels, colors, symbols are the same used in Fig.\,\ref{fig:standard_seds},
with the addition in a) and b) of a cyan SED representing the `foreground' component.
In this case there is no preliminary phase to prepare the active region.
\label{fig:empty_fg_seds}
\label{fig:empty_fg_seds_zoom}
\label{fig:empty_fg_lc}
\label{fig:empty_fg_ff}
}
\end{figure*}
%-------------------------------------------------------

%-------------------------------------------------
\subsubsection{Results}
\label{sec:applications:standard_case:results}

In Fig.~\ref{fig:standard_seds}a-d we show a summary of the main
comparisons with observations, as SEDs, light curves and flux-flux correlation.
The broadband SEDs at 3 different times are shown in
Fig.~\ref{fig:standard_seds}a, with \xray and TeV \gray spectra for 2001
March 19 and historical multiwavelength (from radio to TeV) data points.
Corresponding SEDs zoomed around the \xray band are shown in
Fig.~\ref{fig:standard_seds_zoom}b.
Light curves for 5 relevant energy bands are plotted in Fig.~\ref{fig:standard_lc}c,
while the fluxes in the \xray and TeV \gray bands 
are plotted against each other in Fig.~\ref{fig:standard_ff}d.
About the light curves, it is important to note that the evolution during
the first 15 kilo-seconds ($=t'_\mathrm{start,inj}/\delta$) of
simulation (highlighted with grey shading) simply reflects the initial
setup of the pre-existing background electron population, reaching its
(approximately) steady radiative state as the blob fills with the
radiation from all the zones, and radiation begins to escape.

In Figures~\ref{fig:dcf}a,b we show the Discrete Correlation Function 
\citep[DCF][]{edelson_krolik:1988:DCF} computed between the light curves in two
\xray bands (2$-$4 and 9$-$15\,keV) and \xray and TeV (2$-$4\,keV and $>1$\,TeV).
They are shown for illustrative purposes, and no extensive statistical analysis
has been performed to assess the uncertainty on the lag value.

In the framework of the observational issues outlined in
Section~\ref{sec:applications}, we note that:

\begin{enumerate}
\item  %% SYMMETRY
The flare light curves are approximately symmetric for both the \xray bands as 
well as for the TeV \grays.
The GeV \gray light curve asymmetry reflects the relative duration of the
light crossing times and of the cooling time-scales of the electrons emitting
the seed photons and doing the IC scattering.
For photons emitted by electrons (and seed photons) varying on a
time-scale shorter than the geometric one this latter dominates the
flare profile, hence making it symmetric.  For bands whose emission
processes are characterized by physical time-scales longer than the
geometric ones, the --slower-- cooling decay profile emerges.

\item  %% FLUX-FLUX CORRELATION
The amplitude correlation between \xray and \gray fluxes is
a quadratic, \ie $F_\gamma/F_{\gamma,0} \sim \left(F_\mathrm{X}/F_\mathrm{X,0}\right)^\eta$ 
with $\eta \simeq 2$, during the rise of the flare.
Shortly after the peak the trend flattens, becoming linear.
At this point the March 19 light curves were still showing a quadratic
correlation, which lasted until the end of the TeV (Whipple) observational
coverage (see magenta points in Fig.~\ref{fig:standard_ff}).

\item  %% TIME LAGS
A soft lag is clearly discernible between the different \xray bands,
while a similarly short hard lag is present between the \gray and
the softer \xray band (see also Figures~\ref{fig:dcf}a,b).
In the 2001 March 19 flare a short hard lag was observed in both cases 
\citep{fossati_etal:2008:xray_tev}.
We will discuss a possible important factor responsible for the soft
intra-band \xray lags and the role played by geometry and LCTE later, in
Section~\ref{sec:applications:geometric_effects}.

\item  %% OPTICAL STEADY LEVEL
Since the active region was previously filled by a population of electrons
emitting a lower luminosity slowly varying SED this scenario easily accounts
for the modest variability in the optical band.

\item  %% SPECTRAL SHAPE
While matching the observed \xray spectra is relatively easy, for the 
\gray spectrum we encountered the usual challenge: the observed \gray
spectrum is harder than what predicted by simulations 
\citep{fossati_etal:2000:mkn421_spectral,
blazejowski_etal:2005:multiwavelength_mrk421}.

\end{enumerate}

In order to investigate these points in more details, we explored two
alternative scenarios, which we discuss in turn below.

%=====================================================================
\subsection{Case 2: injection in empty blob, with emission diluted by a
separate steady component (foreground)}
\label{sec:applications:no_bkg_electrons}

With a similar setup we tested a scenario in which there is no
background electron population pre-existing in the blob. 
The steady broader band emission observed in the optical band is
attributed to a component from a different region in the jet, which we
will call `foreground' component.
We assume that there is no interaction between the two regions.
The `foreground' component is combined a posteriori with the radiation
from the flaring blob, simply by adding it as a steady SED to the
emission from the time dependent simulation.
A very important difference with respect to the previous case is that
photons from this component do not contribute to the IC emission by the
freshly injected electron population.

The volume size, geometry and Doppler factor of the active blob are the same
as for the previous case. 
Because of the lack of extra local seed photons for the IC emission,
in order to match the SED, in particular to boost the IC component with
respect to the synchrotron one, it is necessary to decrease the magnetic
field strength.
The injection of electrons begins at 
$t'_\mathrm{start,inj} = 5 \times 10^5$\,s. 
The injected distribution has a spectrum with
$p=1.5$, 
$\gamma_\mathrm{min}= 50$,
$\gamma_\mathrm{max}= 1.9 \times 10^5$, 
$L'_\mathrm{inj}=6 \times 10^{40}$\,erg/s.

The foreground component is simulated with the same code, run separately.
For convenience, its electrons are assumed to be in similar
geometric and magnetic environment to the active region.
They have a broken power law distribution, with spectral indices 
$p_1 = 1.5$ and $p_2 = 2.5$.
The break is at $\gamma_\mathrm{b} = 10^4$, 
the high-energy cut-off at $\gamma_\mathrm{max} = 10^5$. 
The electron density is $n_\mathrm{e}=6$\,cm$^\mathrm{-3}$
(total energy content is $2.3 \times 10^{46}$\,ergs).
These parameters for the putative `foreground' emission are such that its
time evolution is modest on the time-scales in which we are interested here.

%-------------------------------------------------------
\begin{figure*}
\centerline{%
\hfill
\includegraphics[width=0.49\linewidth]{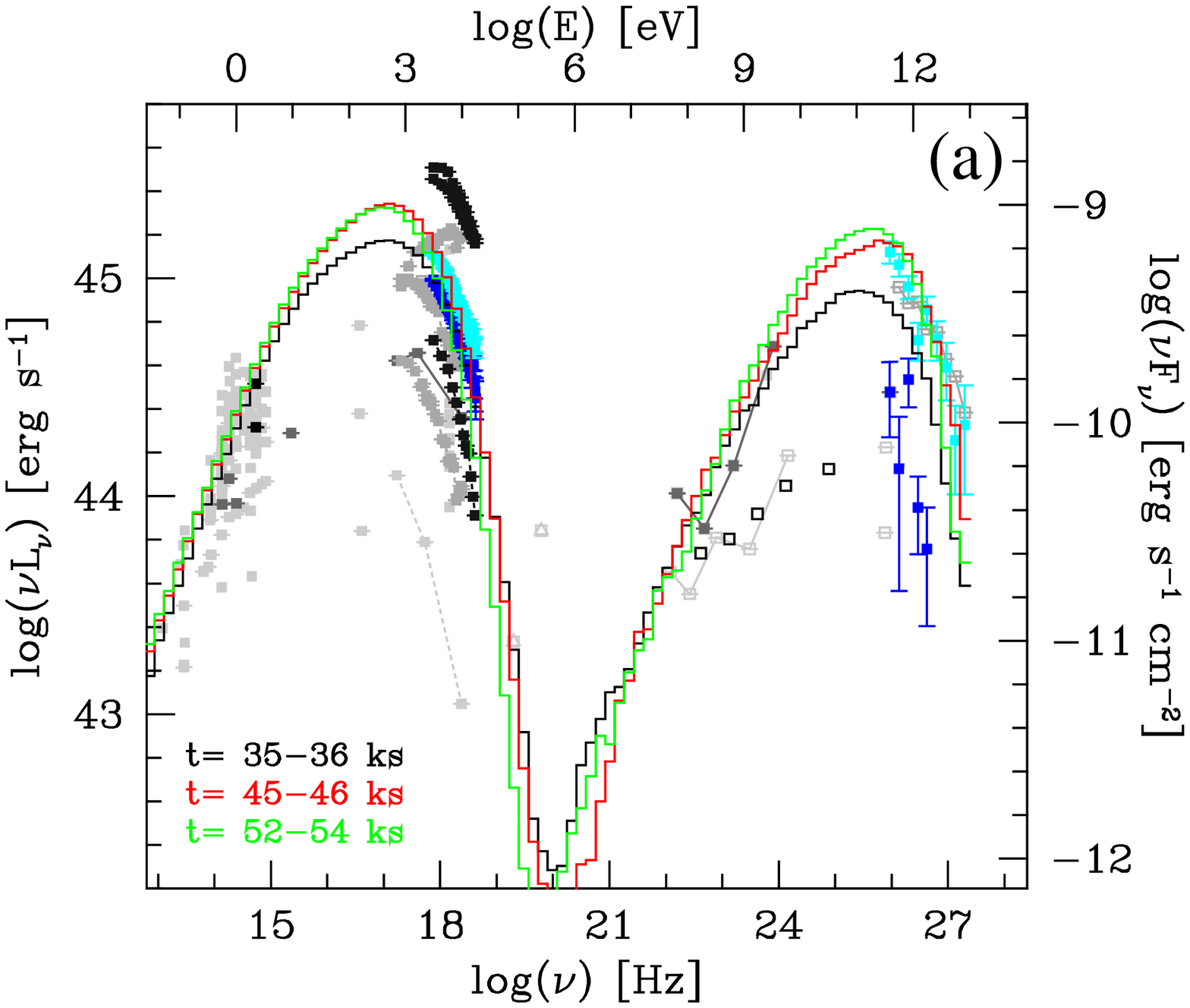}
\hfill
\includegraphics[width=0.49\linewidth]{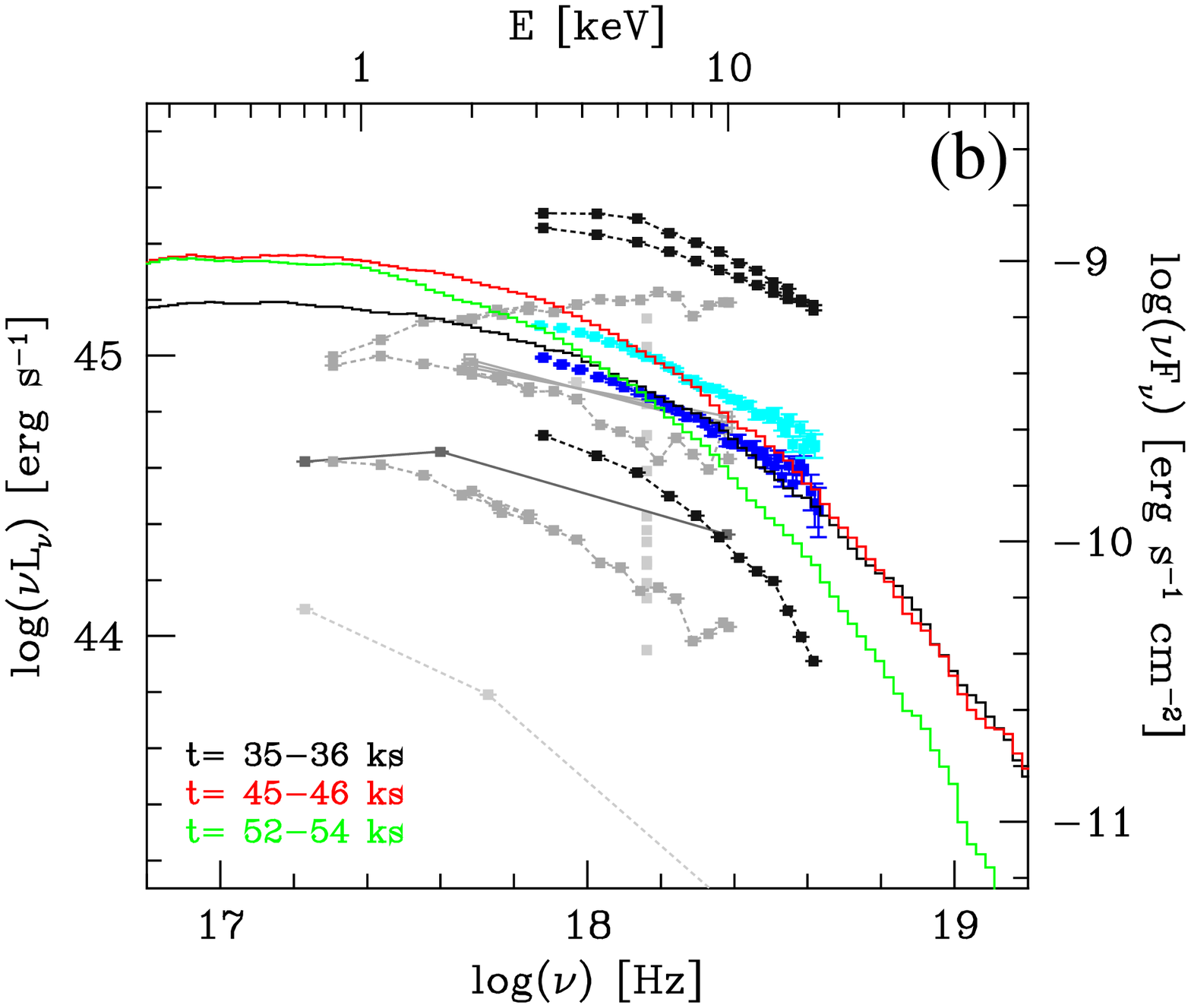}
\hfill
}
\centerline{%
\hfill
\includegraphics[width=0.45\linewidth]{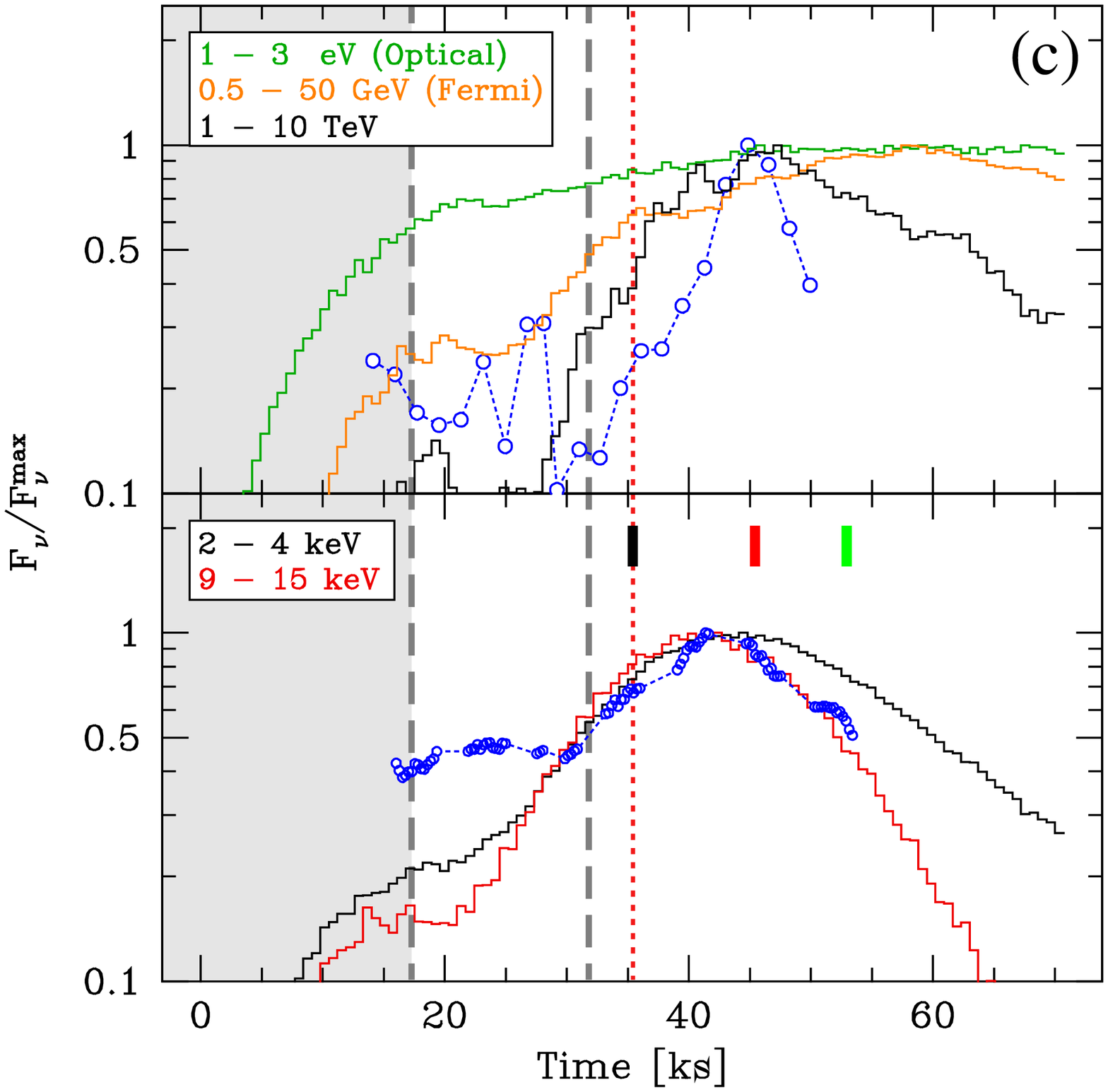}
\hfill
\includegraphics[width=0.45\linewidth]{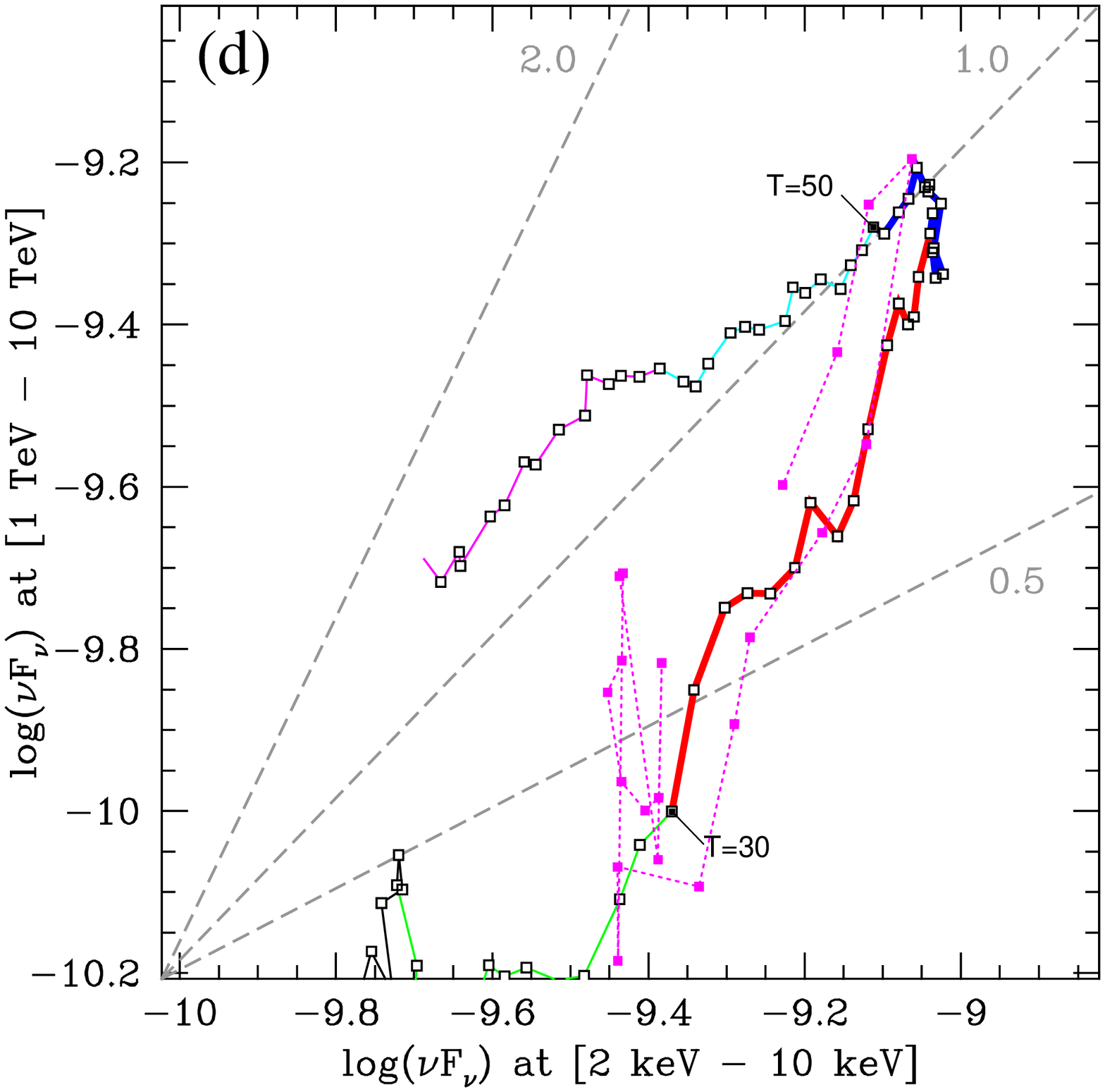}
\hfill
}
\caption{%
Summary of results for the case~3 (blob with pre-existing background electron
population, with parameters adjusted to better match the TeV spectrum.)
All panels, colors, symbols are the same used in Fig.\,\ref{fig:standard_seds}.
\label{fig:bettersed_seds}
\label{fig:bettersed_lc}
\label{fig:bettersed_ff}
}
\end{figure*}
%-------------------------------------------------------

%-------------------------------------------------
\subsubsection{Results}
\label{sec:applications:empty_fg:results}

The resulting SEDs, light curves, and the \xray vs. TeV flux-flux
correlation are shown in Fig.~\ref{fig:empty_fg_seds}a-d, the DCFs 
in Fig.~\ref{fig:dcf}c,d.
This scenario does not reproduce the main features of the reference
observations better than the first one.

\begin{enumerate}
\item  %% SYMMETRY
The flare is asymmetric in TeV \grays and in the softer \xray band.
It remains symmetric for harder \xrays.

The relative length of cooling and geometric time-scales is again
an important factor, cleanly shown by the \xray bands.
In TeV \grays however this is compounded by the effect of LCTEs.
The first steeper rise (up to $t\!=\!30$\,ks) of TeV flux is driven by the
increase of electrons as they are injected in the blob by the moving shock
combined with the fact that we see a larger and larger fraction of the blob
volume, modulated by external LCTE (see \S\,\ref{sec:applications:geometric_effects}).
This is also signaled by the fact that the knee occurs at around the time when
the observer would see the largest section of the blob (red dotted line in
Fig.~\ref{fig:empty_fg_lc}c).
The slow rising, flat-top, phase ($30\!-\!45$\,ks) of the TeV light curve is
due to the increase of seed photons available at each location within the
blob due to diffusion from the rest of the blob, delayed by internal LCTE. 
It's a slow rise also because the high energy electrons responsible for
most of the IC scattering to the TeV band are already cooling rapidly.
At some point the radiation energy density in each location in the blob will
stop increasing because enough time has passed for photons to diffuse
throughout the blob. After that time the evolution is simply
determined by particle cooling and external LCTE. 
Because the electrons emitting the bulk of the observed TeV flux 
have a cooling time larger than $R/c$, in this case the TeV flare decay
shape is determined by cooling rather than LCTE.

\item  %% FLUX-FLUX CORRELATION
As in case~1, the flux-flux amplitude correlation is reproduced only partially.
The trend is almost quadratic during the rising phase of the flare, and it
turns to sub-linear on the decaying phase, after a short horizontal shift
corresponding to the flat top of the TeV light curve.

\item  %% TIME LAGS
The path of the flux--flux diagram signals the presence of a time lag
between the soft \xray and the TeV \gray emission, which is shown
in the DCF (Fig.~\ref{fig:dcf}d). 
The TeV \gray lags the 2-4 keV soft \xray by about 2 kilo-seconds, comparable
to the observation of the March 19 flare as in the first scenario.
Also similar to case~1 is the soft lag between the two \xray bands,
opposite to what observed on 2001 March 19.

\item  %% OPTICAL STEADY LEVEL
For what concerns the optical band, since we designed also this second
scenario to address directly its minimal variability, it is not
surprising that the light curve exhibits only a modest variation.

\item  %% SPECTRAL SHAPE

Finally, also in this scenario we have not been able to produce TeV
\gray spectra as hard as the observations and in the end we limited
ourselves to matching the flux level at around 1\,TeV.

\end{enumerate}

Some of the differences with respect to the first case are ultimately due 
to the weaker magnetic field making synchrotron cooling time-scales
$\sim60$\% longer: for the highest energy electrons emitting in \xray and
TeV $\tau'_\mathrm{cool}$ becomes longer than $R/c$.
It is worth emphasizing that the decrease of $B$ is dictated by
observational constraints, namely the relative luminosity of the
synchrotron and IC components and the need to compensate for the 
absence of the additional source of seed photons for IC scattering provided
in case~1 by the co-located `background' component.
This is in fact a good example of how the model is globally constrained.

%=====================================================================
\subsection{Case 3: with pre-existing electron population, adjusted to
better match the TeV spectrum}
\label{sec:applications:better_sed}

As we pointed out, in the previous two cases, the simulated SED in the
TeV \gray range is softer than the observed spectra. 
To try to improve the match of the TeV part of the SED, we considered a 
modified version of the first scenario.
We increased the Lorentz factor ($\Gamma\simeq46$) and decreased the
magnetic field strength ($B=0.035$\,G), the goal being to move the
inverse Compton peak to higher energy while leaving the synchrotron
peak approximately unchanged.
The parameters are: 
$R=1.5 \times 10^{16}$\,cm, 
$Z=2 \times 10^{16}$\,cm, 
$B=0.035$\,G, 
$\Gamma=46$.
At $t=0$ the `background' electrons have the same broken power law
distribution as in the first case, but with lower number density,
$n_\mathrm{e}=1.56$\,cm$^\mathrm{-3}$.
The volume is slightly larger yielding a total energy content of 
$2.9 \times 10^{46}$\,ergs.

The injected electrons have a power law distribution with 
$p=1.5$, 
$\gamma_\mathrm{min}= 50$,
$\gamma_\mathrm{max}= 1.9 \times 10^5$, 
$L'_\mathrm{inj}= 3.2 \times 10^{40}$\,erg/s.
Injections starts in this case at $t'_\mathrm{start,inj} = 8 \times 10^5$\,s. 

Results are integrated over $0.99971< \cos(\theta) <0.99981$, which
corresponds to a Doppler factor range $37<\delta<57$.  
%-------------------------------------------------
\subsubsection{Results}
\label{sec:applications:better_sed:results}

SEDs, light curves, and \xray vs. TeV flux-flux correlation are shown in
Fig.~\ref{fig:bettersed_seds}, and DCFs in Fig.~\ref{fig:dcf}e,f.
The itemized summary of the main reference observations does not
show improvements beyond the slightly higher VHE SED peak.

\begin{enumerate}
\item  %% SYMMETRY
Because of the larger Doppler factor the electron emitting at the SED
peaks have lower energy, which combined with a weaker magnetic field
results in longer cooling time-scales (see
eq.~\ref{eq:tau_sync_parameterized}), in turn exceeding the source
crossing time. 
This has the effect of increasing the asymmetry of the light curves in
bands whose emission involves lower energy electrons and/or photons.
The soft \xray and TeV \gray light curves indeed have a slowly decaying
tail.

\item  %% FLUX-FLUX CORRELATION
Once again during the rising phase of the flare the \gray-\xray
correlation is approximately quadratic, until the peak of the \xray
light curve.  
After the TeV flare peak the correlation is approximately linear, as
expected when the variation in both bands is driven only by the cooling of
the (same) electrons, because the IC seed photons are emitted by particles
with longer cooling time-scale.

\item  %% TIME LAGS
The results concerning time-lags are equivalent to those of the other
scenarios, perhaps with a hint of a smaller lag between TeV and softer \xray.
More extensive analysis would be necessary to quantify this possibility.

\item  %% SPECTRAL SHAPE
The slight shift of the IC peak to higher energy enables a better match
with the observed spectra, although the actual spectral indices of the
simulated SEDs remain softer than the observed values.

\end{enumerate}

Further increases of the Doppler factor can still produce good SEDs, as
long as we concurrently reduce the size of the volume.
However, the light crossing time would rapidly become smaller than the
observed flare duration, and it would have minimal impact on the
observed phenomenology.
Therefore the observed flare shapes must represent the true
acceleration and cooling of the electrons, and the symmetry of the
light curves must be caused by similar heating (or injection) and
cooling time-scales.

%-------------------------------------------------
\subsection{Geometric Effects on Light Curves}
\label{sec:applications:geometric_effects}

There are complex geometry-related effects that have an impact not only
on the shape of the observed light curve (\eg its symmetry), but can also
leave an imprint on other observables such as time lags and
energy-dependent flare shape.  
Depending on how the particle injection and acceleration processes are
distributed spatially, differences in physical time-scales for particles of
different energy effectively may add a further geometric effect by inducing
inhomogeneities (\eg stratification) in the source 
\citep[see also][]{chiaberge_ghisellini:1999:timedep,sokolov_marscher_mchardy:2004:SSC}.

We would like to illustrate with an extremely simple toy-model some
aspects of the role of the geometry of the emitting region, and its
interplay with some of the intrinsic physical time-scales, responsible 
for the fact that the peaks of the simulated \xray light curves did not
correspond to either the time when the shock exits the active region
and injection is not present anywhere anymore, or to the time
corresponding to the largest cross-section of the cylindrical volume
along planes of equal observed times.  
In Figures~\ref{fig:standard_lc}c,~\ref{fig:empty_fg_lc}c,~\ref{fig:bettersed_lc}c, 
these two times are marked as the second dashed grey line and the
red-dotted line, respectively.
Moreover, the shift changes with the light curve energy band as noticeable
in the case of \xray light curves.

To illustrate how time shifts are caused by the different size of the
observable regions filled with electrons contributing most of the emission
at those frequencies, we consider a purely geometrical model solely based
on the `appearance' of slices of different thickness through a cylinder.

\begin{figure}
\centerline{%
\includegraphics[width=0.98\linewidth]{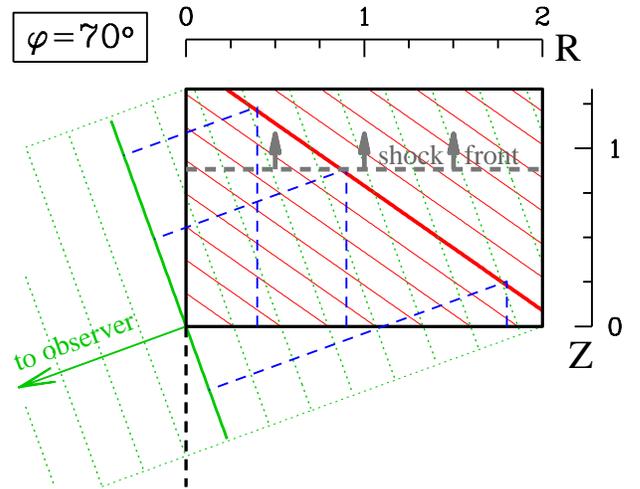}
}
\caption{%
Diagram illustrating the geometry of the toy-model 
(see text, \S\,\ref{sec:applications:geometric_effects}).
The black rectangle represents a side 2D view of the cylinder, with
$R=1$ and $Z=4/3$ (same aspect ratio of our simulations).
The grey dashed line represents the `shock front' traveling at the speed of
light through the cylinder, along its axis of symmetry, in the direction
marked by the arrows.
The green arrow represents the direction to the observer, while the
green lines (solid and dotted) are planes perpendicular to the
line of sight. In this example the viewing angle is $\varphi=70\degr$.
The red lines represent the loci of points whose photons reach the
observer simultaneously, taking into account the `shock` travel
distance/time until of their activation and the light travel time to
the observer since that moment.  
The blue dashed lines illustrate this by showing three paths of equal
length from the beginning of the `flare'.
The red loci form an angle $90\degr-\varphi/2$ with the line of sight
or, equivalently, an angle $\varphi/2$ with the front face of the volume.
  %%%%%
\label{fig:toy_model_cartoon}
}
\end{figure}

Like in our real blob model, a shock is traveling along the axis
of symmetry of the cylinder turning `on' a thin local slice.   
Each point of this slice stays `on' for a limited time, $\tau_\mathrm{on}$.
We do not consider a variation of brightness with time, just an on/off state.
We build light curves where `flux' is simply the size of the volume that is
seen `on' by the observer at any given time.
The `on' volume visible at each time from the observer point of view is
computed taking into account light travel times and it cuts through the
cylinder along planes yielding constant arrival time to the observer.
For an observer viewing the cylinder at an angle $\varphi$ with
respect to its axis, the loci of points whose photons he sees
simultaneously are planes with an inclination $\varphi/2$ with respect to
the face of the cylinder ($90\degr-\varphi/2$ with respect to the cylinder
axis).
Figure~\ref{fig:toy_model_cartoon} shows a 2D schematic of the geometry
of the problem.
  
%-------------------------------------------------------
%-------------------------------------------------------

%-------------------------------------------------------
\begin{figure*}
\centerline{%
\hfill
\includegraphics[height=0.35\linewidth,bb= 40 184 550 692,clip=]{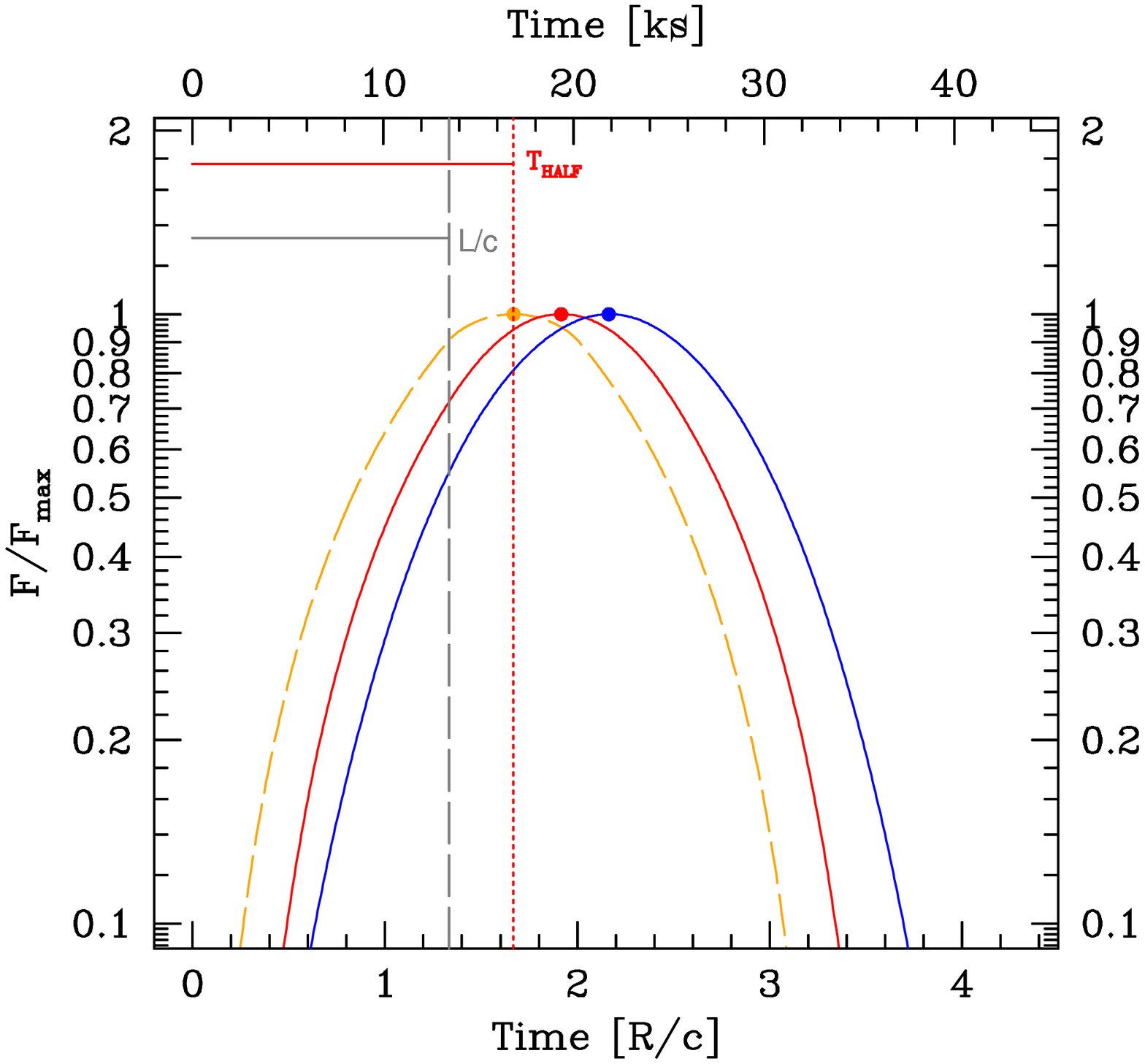}
\hfill
\includegraphics[height=0.35\linewidth,bb=112 184 550 692,clip=]{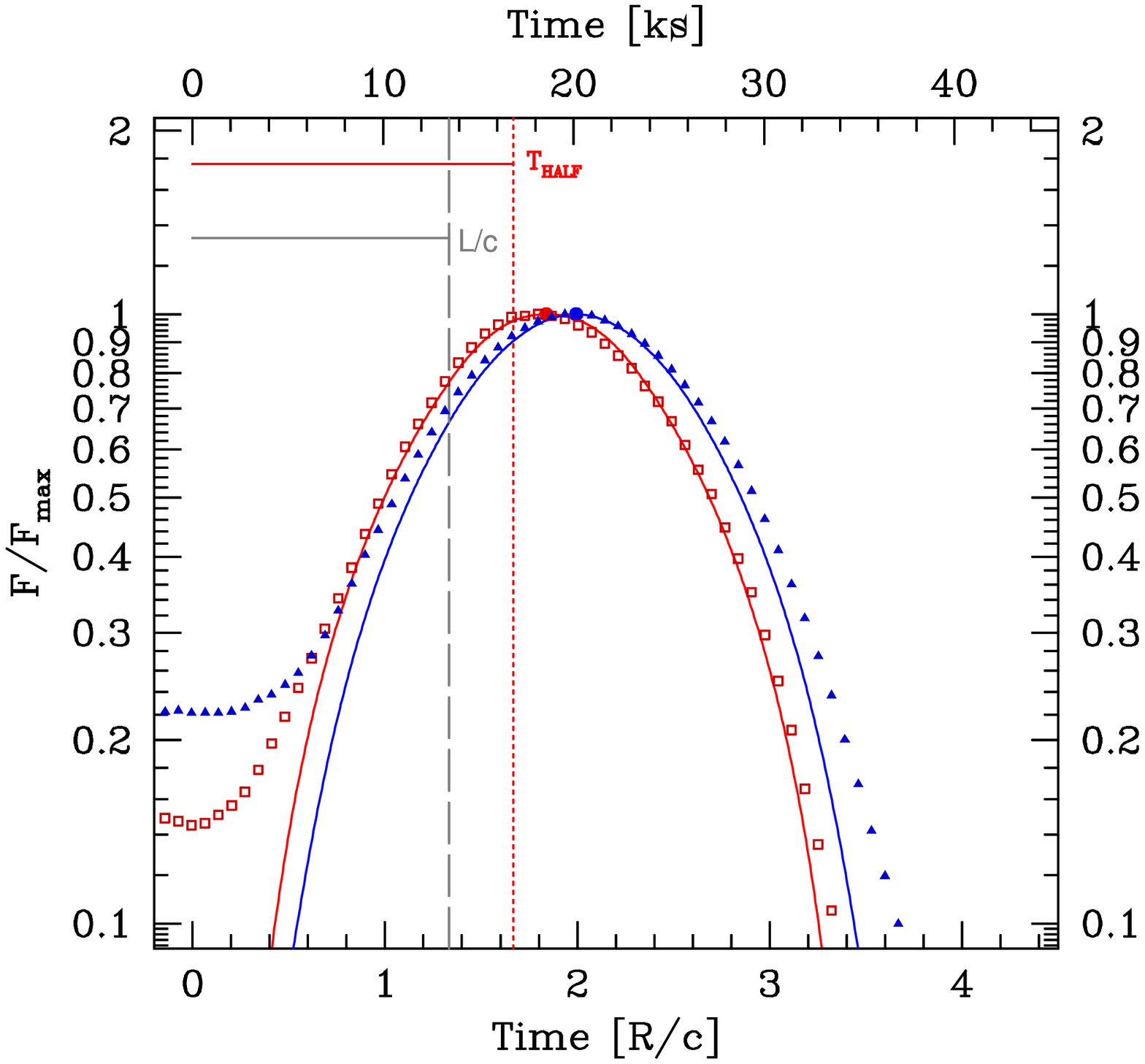}
\hfill
\includegraphics[height=0.35\linewidth,bb=112 184 586 692,clip=]{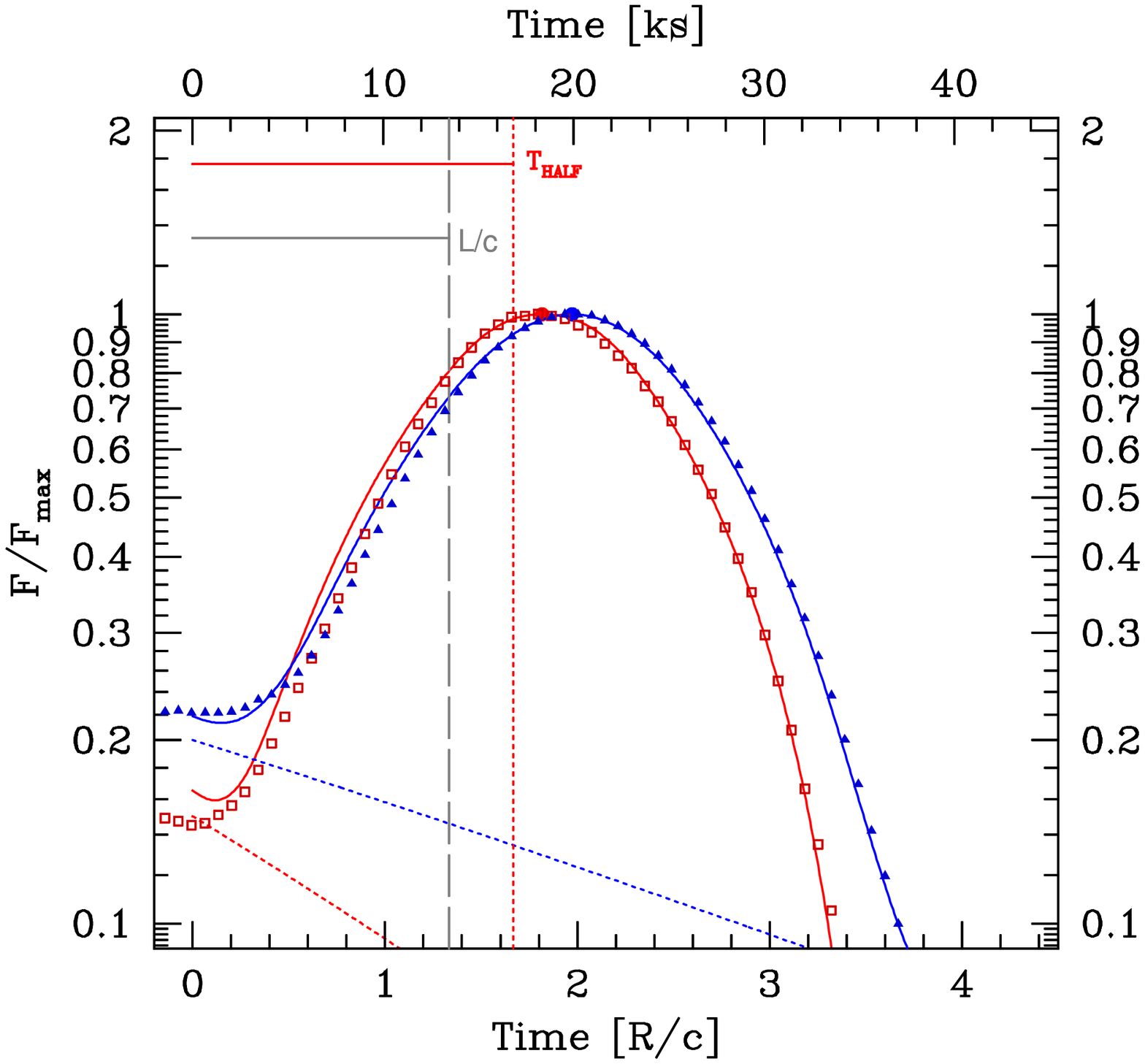}
\hfill
}
\caption{%
Light curves for a purely geometrical toy-model of the observations of
a cylinder seen at an angle $\varphi=90\degr$ with respect to its axis. 
The meaning of the vertical lines is the same as in
Fig.~\ref{fig:standard_lc}c,~\ref{fig:empty_fg_lc}c,~\ref{fig:bettersed_lc}c, 
except that since the `injection' begins at $t=0$, there is no first
grey dashed line.
(a) Three cases illustrating the time shift caused by the simple
variation of $\tau_\mathrm{on}$. 
The (light) curves are computed for $\tau_\mathrm{on}=10^{-3}$
(orange), 0.5 (red), 1.0 (blue), in units of $R/c$.
The orange case, with a $\tau_\mathrm{on}$ value yielding a negligible
slice thickness, peaks at the expected time.  
(b) Attempt at matching light curves and simulation data without any
baseline component contribution.  
The red curve is computed for $\tau_\mathrm{on}=0.35$, the blue for
$\tau_\mathrm{on}=0.65$.
The symbols are the simulated light curves of case~1, shown in
Fig.~\ref{fig:standard_lc}c: red empty squares for 9$-$15\,keV and blue
triangles for 2$-$4\,keV.
(c) Same flare light curves combined with a slowly decaying baseline,
with time-scale proportional to $\tau_\mathrm{on}$. 
\label{fig:toy_model_lc}
}
\end{figure*}
%-------------------------------------------------------

If the cylinder was moving with Lorentz factor $\Gamma$, because of
relativistic aberration at a viewing angle $\theta \simeq 1/\Gamma$ we
would be observing the radiation that in the comoving frame leaves the cylinder
`sideways', at 90$\degr$ from its axis.
The observed frequencies would be blue-shifted and times compressed, but that
would be simply a scaling factor applied uniformly to them and for convenience
we can chose to use observed frequencies and times.
Therefore observing the toy-model at $\varphi=90\degr$ is equivalent to
observing the relativistically moving blob at $\theta=1/\Gamma$,  and in turn
this purely geometrical analysis captures some of the features of the realistic
model studied in this paper.

The results are illustrated in Figure~\ref{fig:toy_model_lc}.
In the first panel, we show the general features of the light curves
obtained with this model, most importantly the effect of the change in
the duration $\tau_\mathrm{on}$.
In Figure~\ref{fig:toy_model_lc}a we plot the curves for three cases,
showing the shift of the flare peak to later time as $\tau_\mathrm{on}$
increases.
For very short $\tau_\mathrm{on}$ the maximum is reached at the
expected time, that is when size of the plane that is the locus of
points simultaneously seen by the observer (red lines in
Fig.~\ref{fig:toy_model_cartoon}) is the largest possible for the given
viewing angle.
In general, however, the light curve peak will be shifted by $\Delta t=
\tau_\mathrm{on}/2$.
There is also a widening of the light curve, though much less
noticeable than the peak shift.
It is worth noting that at this extreme level of simplification
geometric effects can not produce any asymmetry in the light curves.

We tried to reproduce with this toy-model the \xray light curves from
the simulations of the first scenario presented here. 
The results are shown in Figure~\ref{fig:toy_model_lc}b,c.
The central panel shows the curves obtained by adjusting
$\tau_\mathrm{on}$ to match the peak time of the two \xray light curves
(plotted with symbols), yielding a very satisfactory result for values
around 0.35 and 0.65 $R/c$.
However the overall shape of the flares can not be matched well without 
adding a baseline component, to mimic in some way the fact that in the
simulations the blob was already `on' at a low level prior to the
beginning of the injection.
We therefore added to the toy-model light curves a slowly decaying
component, with an initial level such that the combination of the two
components would match the data.  Since in the simulations the
pre-existing component is left to evolve (cool) starting at the
beginning of the active phase, here we let the baseline component decay too.
The visual matching is not critically sensitive to the exact values of
the decay slopes, and to make the model more constrained we forced them 
to be in a fixed ratio with respect to the chosen $\tau_\mathrm{on}$, by
considering that the cooling times of the electrons emitting the baseline
photons are also related to their energies.
For the results shown in Figure~\ref{fig:toy_model_lc}c the slope is equal
to five times $\tau_\mathrm{on}$ for both cases.

The synchrotron cooling time-scale for electrons emitting in the $2-4$\,keV
and $9-15$\,keV bands, following the approximate expression of
eq.~\ref{eq:tau_sync_parameterized} are 
$\tau_\mathrm{soft} \simeq 0.73-1.04\,R/c$, 
$\tau_\mathrm{hard} \simeq 0.38-0.5\,R/c$.
Given the steepness of the \xray spectrum the emission in each band
is dominated by the lower energy electrons, hence the longer $\tau$ is
probably a more appropriate estimate. On the other hand, the above
cooling time-scales only consider synchrotron cooling. Including some
additional loss due to IC would decrease the value of $\tau$.
In any case, the similarity between these crude estimates of cooling
time-scales and the values of $\tau_\mathrm{on}$ corroborates the success
of the geometrical toy-model at fitting the simulation light curves.

The ability of the purely geometrical toy-model to reproduce the two
\xray light curves is indeed remarkable.  For \xrays this is
facilitated by the fact that the synchrotron emission is independent on
the internal delays due to photon diffusion that affect the evolution
of the IC emission from the blob.  It is not possible to apply a
similar toy-model to the \gray light curves.

It is worth noting that although this test shows how dominant the effect of
the geometry can be in shaping the light curve, at the same time we need to
highlight that some geometry parameters, such as the `thickness'
of the visible slices, are in effect determined by the physical conditions
of the emission region.

In this respect it is interesting to note that, at least in the setup
of the scenarios presented in this work, despite the apparent dominance
of the source geometry the effect of the energy dependent
physics-induced geometrical factors is detectable.
Hence multiwavelength datasets and time-resolved spectroscopy have the
potential to disentangle them from the source geometry.

%%%%%%%%%%%%%%%%%%%%%%%%%%%%%%%%%%%%%%%%%%%%%%%%%%%%%%%%%%%%%%%%%%%%%%
\section{Discussion}
\label{sec:discussion}
\label{sec:conclusions}

We introduced a coupled Fokker--Planck and Monte Carlo code allowing us
to study blazar phenomenology in unprecedented detail with a time
dependent and multi-zone model properly taking into account all light
travel time effects.

We presented three test scenarios, aimed at modeling the variability
exhibited by \mrk during the 2001 March 19 flare, and based on a
relatively standard choice of parameters.
The results of these tests are summarized in Table~\ref{tab:scorecard},
side by side with the features observed in the actual multiwavelength
observations \citep{fossati_etal:2008:xray_tev}.
There are a few fundamental issues that we wanted to address, which are
common throughout the phenomenology of all well studied blue blazars.
\begin{enumerate}

\item 
The shape of the flares, often quasi-symmetric for a wide range of
observational bands where the intensity variations are large (the main
ones being \xray and \gray).

\item 
The characteristics of the correlation between \xray and \gray
fluxes. 
There has been great interest in the slope of their relationship, in
particular because of the observation of a quadratic, or higher order,
relationship holding throughout some well sampled outbursts, challenging our
understanding of the physical conditions and causes of the variability.

\item
The phase of the correlation between variations in different bands, namely
the existence of time lags and their duration.

\end{enumerate}
None of the three test scenarios was able to reproduce all the characteristics
of the 2001 March 19 flare. 
Two features have been particularly challenging to match: 
the relationship between \xray and TeV \gray fluxes on the decay phase of the flare,
and the intra-band \xray time lag.
Moreover, the shape (symmetry) of the flare light curves could be reproduced
only by one of the three scenarios, case~1.

These aspects of phenomenology are among those more affected by
the spatial extent and geometry of the source, whose influence varies 
with observed energy band because of the relative importance of
geometrical and physical time-scales.
The impact of the geometrical factor, both due to the source intrinsic
structure and to the stratification of properties due to the physical
processes, emphasizes the necessity of a code like the one we introduce
here for modeling the variable high energy emission from blazar jets.

The difficulty of producing a quadratic relationship between the fluxes
in the \xray and \gray during the declining phase of the flare may indicate
that radiative cooling cannot fully explain the electron cooling mechanism.
The delayed evolution of the seed photon field due to internal LCTE compounds
the problem.
One alternative possibility could be a process causing energy loss
over a wide range of electrons energies (such that the IC seed photons
are also affected) on very similar/same time-scale, such as
adiabatic cooling, which could be associated with expansion of the blob, or
particle escape.
They are often invoked in qualitative discussions and in the context of
simpler models, treated by means of some phenomenological prescription.
The addition of such mechanisms to the code in a proper astrophysical way is
not immediate, but we are working on its implementation.
The escape term present in the Fokker-Planck equation is actually neglected for
these set of simulations. 
In a follow up work including particle acceleration and escape, this latter
seems to be effective and we obtain a quadratic flux relationship and hard
lags \citep[\eg][]{chen_etal:2011:proceedings_guanzhou}. 
About the effect of adiabatic expansion of the emitting blob, based on the
simplified analysis of \citet{katarzynski_etal:2005:tev_x_correlation},
\citet{aharonian_etal:2009:pks2155}, argue against its viability once the
implications of this expansion on the magnetic field and particle cooling are
taken into account. 
Nevertheless, its effect should be assessed with actual time-dependent
simulations of a source of finite size.

As to the hard intra-band \xray lags, in
Section~\ref{sec:applications:geometric_effects} by means of a toy-model we
illustrated an important factor affecting observations of time-lags: source
`stratification' combined with LCTE induces a systematic soft lag.
The simulations presented in this work do not include an acceleration term,
hence the lack of hard lags is not a complete surprise.
Nevertheless, the induced systematic geometric soft lag introduces an
additional constraint on viable electron acceleration and injection scenarios.
In this respect, the results of the above mentioned study focused on \xray
lags \citep{chen_etal:2011:proceedings_guanzhou}
% (Chen \etal, in preparation) 
suggest that continuous acceleration,
spatially extended (\eg diffuse diffusive acceleration due to turbulence,
\citealp{katarzynski_etal:2006:stochastic}), may be necessary, possibly
accompanied by an achromatic energy loss mechanism.  
It is interesting that this type of scenario seems to be able to produce also a
quadratic \xray/TeV relationship throughout a flare.

One of the most interesting aspects of this analysis was the comparison
between two possible hypotheses for the presence of an additional
component contributing to the observed SED.
The need for two components, a flaring and a quasi-steady one, to interpret
some of the observations has become more evident with the improvement
of multiwavelength observations 
\citep[for an interesting decomposition of
a Suzaku spectrum of \mrk see][]{ushio_etal:2009:mrk421_in_2006_with_suzaku}.  
Disentangling these two components is necessary in order to understand
the nature of the transient activity whose properties need to be seen
more clearly.  
This decomposition might also yield information on the average properties
of the relativistic jet, for which the less variable component might be
more representative.

%===============================================================================
\begin{table}
\caption{Summary of Simulations Results}
\label{tab:scorecard}
\begin{tabular}{l c@{}c c@{~~}c c@{~~}l c@{~~}l }
%-------------------------------------------------------------------------------
\hline
Feature        & Obs.     && 
\multicolumn{2}{c}{Case 1} &
\multicolumn{2}{c}{Case 2} &
\multicolumn{2}{c}{Case 3} \\
\hline
%%%%%%%%%%%%%%%%%%%%%%%%%%%%%%%%%%%%%%%%
\multicolumn{2}{@{}l}{Flare symmetry} \\
~~soft \xray     & Y        && Y    & $+$      & N        & $-$ & N         & $-$ \\
~~hard \xray     & Y        && Y    & $+$      & Y        & $+$ & Y         & $+$ \\
~~TeV \gray      & Y        && Y    & $+$      & N        & $-$ & N         & $-$ \\[8pt]
%%%%%
\multicolumn{2}{@{}l}{Flux-Flux Correlation}  \\ 
~~trend up       & 2        && 2    & $+$      & 2        & $+$ & 2         & $+$ \\
~~trend down     & 2        && 1    & $-$      & 1        & $-$ & 1         & $-$ \\[8pt]
%%%%%
\multicolumn{2}{@{}l}{Time Lags}       \\
~~\xray$-$\xray  & hard (2\,ks)    && soft & $-$      & soft     & $-$ & soft      & $-$ \\
~~\xray$-$\gray  & \gray (2\,ks)   && Y    & $+$      & Y        & $+$ & Y         & $+$ \\
%%%%%%%%%%%%%%%%%%%%%%%%%%%%%%%%%%%%%%%%
\hline
\end{tabular}
\end{table}
%===============================================================================

We considered the two simplest possibilities: 
i) that the secondary component is due to a population of electrons
that exists in the same region that will become active, and that will
be affected by the flare and evolve with it.
This pre-existing component can be interpreted as due to the
remnants of a previous outburst.
ii) That the secondary component is completely independent from the
flare, and it contributes just a steady SED diluting the transient
component from the observer point of view. 
Our simulations offer some hints as to their viability, and they favor the
first type of scenario.  
One fundamental difference between the two alternatives concerns the production
of IC emission, \ie the TeV band.
If the observed SED consists of the sum of two independent
contributions, emitted by electrons at two different locations 
then the only seed photons for IC scattering will be those produced by
the injected electrons themselves.
Starting from an empty blob, the energy density of synchrotron seed
photons needs some time to build up, which naturally results in a delay
in the variation of the IC scattered \grays.
This delay is caused by internal LCTE and it turned out to be quite significant
as illustrated by the TeV light curve of case~2 (Fig.~\ref{fig:empty_fg_lc}c),
yielding a flat-top flare not seen in observations.  
That TeV light curve is in fact an excellent example of LCTEs at work and of
the importance of a more advanced modeling code.

\smallskip
Naturally, the cases presented in this paper only represent an initial
study aimed at investigating the importance of LCTE which for
the first time could be fully accounted for.
These results can not be considered conclusive.
Nonetheless, despite their limited scope they make a strong case for a
true time-dependent and multi-zone modeling.  

The three scenarios discussed can be generally regarded as homogeneous blob
scenarios. The magnetic field is the same throughout the simulation volume, and
isotropic.  For what concerns electrons, the initial setup is homogeneous and
the injection is identical in all zones.  It is, however, worth emphasizing
that during the evolution of the simulation electrons properties become
inhomogeneous because of the different radiative cooling they experience in
different zones.

These simulations represent a first order implementation of a class of
scenarios for blazar flares often discussed in the literature, envisaging a
shock acting on a discrete blob/shell within the jet.  
We adopted a volume with relatively symmetric aspect ratio, to not depart too
much from the sphere `implied' by one-zone models while making it possible to
appreciate the effects of geometry, and attributed the flare to the injection
of a fresh electron population.

As discussed in Sections~\ref{sec:applications:goals}, \ref{sec:applications:parameter_choice},
and \ref{sec:applications:parameter_estimates} the main physical parameters are
fairly well constrained and the results can be regarded as meaningful for what
concerns the time-varying components of the model, as well as the nature of the
secondary emission component.
The simulations presented in this paper suggest that a simple injection in
the radiating region of particles with a formed spectrum produced in a
separate acceleration region whose emission is not significant, does not
provide a satisfactory match with some basic observational facts.
Moreover, the comparison of the background and foreground component
scenarios, in particular with respect to the TeV \gray evolution, clearly
favoring the first one (case~1), suggests that if a flare is caused by a
change affecting the electron population it may be necessary for it to
happen on a relatively hot blob, acting on the same particles, re-accelerating them.  
This in turn would support a scenario where flares are not fully independent of
each other but rather occur in the same region.  

%-------------------------------------------------
\subsection{Outlook}
\label{sec:discussion:outlook}

We are working on a few directions of expansion of this investigation.
The main immediate focus is on the two unsolved issues of hard intra-band \xray
lags and \xray-TeV flux correlation and we have introduced particle
acceleration and escape \citep[\eg][]{chen_etal:2011:proceedings_guanzhou}.
% (Chen \etal, in preparation).

Further areas of study concern inhomogeneities and different geometries, 
which can be easily studied with this code.
We can introduce a spatial structure to the magnetic field, either static or
changing according to some prescription, which could be motivated as caused by
compression and amplification of the tangled field by a shock, and we plan to
expand the code to deal with anisotropic magnetic field.
We have started to simulate more `exotic' geometries, \eg elongated or
flattened blobs, and the effect of an energy release in a small sub-region
(bubble) embedded in the simulated volume.

A particularly timely line of investigation concerns red blazars like FSRQs.
Because the peak of their high energy component occurs in the less
accessible MeV--GeV band, they have received much more limited attention,
except at the time of \textit{EGRET}, that however lacked the sensitivity
to follow even bright sources throughout their full variability cycles.
\textit{Fermi}/LAT has changed the \textit{status quo} by providing
continuous coverage of several bright blazars, and even more importantly by
being able to detect them during their more quiescent phases.
Among the most remarkable examples there are 3C~454.3 and PKS 1510$-$089
that already provided new clues about the nature of their \gray
emission and the structure of the jet by the study of the correlated
variations between \grays and their synchrotron or
thermal emission \citep{abdo_etal:2009:3c454,
bonning_etal:2009:3c454,
marscher_etal:2010:pks1510}.
Our code allows us to simulate EC scenarios, by illuminating the active
region with an external radiation field with full treatment of the
relativistic aberrations, thus allowing us to model external components
with different spectral, spatial and temporal properties.

\bigskip
We would like to thank J.~D.~Finke for the invaluable help he provided
during the development of this work. 
XC thanks Guy Hilburn for helpful discussions.
We also acknowledge the anonymous referee for the pointed comments and
suggestions that lead to a better manuscript.
This research has been supported by NASA grants NAG5-11796 and NAG5--11853, 
Chandra AR9-0016X, and Fermi Guest Investigator award NNX10AO42G.
GF thanks the European Southern Observatory for the extended
hospitality in its Santiago offices.
MB acknowledges support by NASA through Fermi Guest
Investigator Grant NNX09AT82G and Astrophysics Theory
Program Grant NNX10AC79G.
This work was supported in part by the Shared University Grid at Rice
funded by NSF under Grant EIA-0216467, and a partnership between Rice
University, Sun Microsystems, and Sigma Solutions, Inc.
This research has made use of NASA's Astrophysics Data System and 
of the NASA/IPAC Extragalactic Database (NED) which is operated by the Jet
Propulsion Laboratory, California Institute of Technology, under contract
with the National Aeronautics and Space Administration.

\label{lastpage}

\begin{thebibliography}{}
\expandafter\ifx\csname natexlab\endcsname\relax\def\natexlab#1{#1}\fi

\bibitem[{{Abbott} \& {Lucy}(1985)}]{Abbott_Lucy:1985:MC}
{Abbott}, D.~C., \& {Lucy}, L.~B. 1985, \apj, 288, 679

\bibitem[{{Abdo} {et~al.}(2009){Abdo}, {Ackermann}, {Ajello}, {Atwood},
  {Axelsson}, {Baldini}, {Ballet}, {Barbiellini}, {Bastieri}, {Battelino},
  {Baughman}, {Bechtol}, {Bellazzini}, {Berenji}, {Blandford}, {Bloom},
  {Bonamente}, {Borgland}, {Bouvier}, {Bregeon}, {Brez}, {Brigida}, {Bruel},
  {Burnett}, {Caliandro}, {Cameron}, {Caraveo}, {Casandjian}, {Cavazzuti},
  {Cecchi}, {Charles}, {Chaty}, {Chekhtman}, {Cheung}, {Chiang}, {Ciprini},
  {Claus}, {Cohen-Tanugi}, {Cominsky}, {Conrad}, {Costamante}, {Cutini},
  {Dermer}, {de Angelis}, {de Palma}, {Digel}, {Silva}, {Donato}, {Drell},
  {Dubois}, {Dumora}, {Farnier}, {Favuzzi}, {Focke}, {Foschini}, {Frailis},
  {Fuhrmann}, {Fukazawa}, {Funk}, {Fusco}, {Gargano}, {Gasparrini}, {Gehrels},
  {Germani}, {Giebels}, {Giglietto}, {Giommi}, {Giordano}, {Glanzman},
  {Godfrey}, {Grenier}, {Grondin}, {Grove}, {Guillemot}, {Guiriec}, {Hanabata},
  {Harding}, {Hartman}, {Hayashida}, {Hays}, {Hughes}, {J{\'o}hannesson},
  {Johnson}, {Johnson}, {Johnson}, {Kamae}, {Katagiri}, {Kataoka}, {Kawai},
  {Kerr}, {Kn{\"o}dlseder}, {Kocian}, {Kuehn}, {Kuss}, {Latronico}, {Lee},
  {Lemoine-Goumard}, {Longo}, {Loparco}, {Lott}, {Lovellette}, {Lubrano},
  {Madejski}, {Makeev}, {Massaro}, {Mazziotta}, {McEnery}, {McGlynn}, {Meurer},
  {Michelson}, {Mitthumsiri}, {Mizuno}, {Moiseev}, {Monte}, {Monzani},
  {Morselli}, {Moskalenko}, {Murgia}, {Nolan}, {Norris}, {Nuss}, {Ohsugi},
  {Omodei}, {Orlando}, {Ormes}, {Paneque}, {Panetta}, {Parent}, {Pelassa},
  {Pepe}, {Pesce-Rollins}, {Piron}, {Porter}, {Rain{\`o}}, {Rando}, {Razzano},
  {Reimer}, {Reimer}, {Reposeur}, {Reyes}, {Ritz}, {Rochester}, {Rodriguez},
  {Rahoui}, {Ryde}, {Sadrozinski}, {Sambruna}, {Sanchez}, {Sander},
  {Parkinson}, {Sgr{\`o}}, {Shaw}, {Smith}, {Smith}, {Spandre}, {Spinelli},
  {Starck}, {Strickman}, {Suson}, {Tajima}, {Takahashi}, {Takahashi}, {Tanaka},
  {Thayer}, {Thayer}, {Thompson}, {Tibaldo}, {Torres}, {Tosti}, {Tramacere},
  {Uchiyama}, {Usher}, {Vilchez}, {Villata}, {Vitale}, {Waite}, {Winer},
  {Wood}, {Ylinen}, {Zensus}, \& {Ziegler}}]{abdo_etal:2009:3c454}
{Abdo}, A.~A. {et~al.} 2009, \apj, 699, 817

\bibitem[{{Aharonian} {et~al.}(2009){Aharonian}, {Akhperjanian}, {Anton},
  {Barres de Almeida}, {Bazer-Bachi}, {Becherini}, {Behera}, {Benbow},
  {Bernl{\"o}hr}, {Boisson}, {Bochow}, {Borrel}, {Brion}, {Brucker}, {Brun},
  {B{\"u}hler}, {Bulik}, {B{\"u}sching}, {Boutelier}, {Chadwick},
  {Charbonnier}, {Chaves}, {Cheesebrough}, {Chounet}, {Clapson}, {Coignet},
  {Costamante}, {Dalton}, {Daniel}, {Davids}, {Degrange}, {Deil}, {Dickinson},
  {Djannati-Ata{\"i}}, {Domainko}, {O'C.~Drury}, {Dubois}, {Dubus}, {Dyks},
  {Dyrda}, {Egberts}, {Emmanoulopoulos}, {Espigat}, {Farnier}, {Feinstein},
  {Fiasson}, {F{\"o}rster}, {Fontaine}, {F{\"u}{\ss}ling}, {Gabici}, {Gallant},
  {G{\'e}rard}, {Giebels}, {Glicenstein}, {Gl{\"u}ck}, {Goret}, {G{\"o}hring},
  {Hauser}, {Hauser}, {Heinz}, {Heinzelmann}, {Henri}, {Hermann}, {Hinton},
  {Hoffmann}, {Hofmann}, {Holleran}, {Hoppe}, {Horns}, {Jacholkowska}, {de
  Jager}, {Jahn}, {Jung}, {Katarzy{\'n}ski}, {Katz}, {Kaufmann}, {Kendziorra},
  {Kerschhaggl}, {Khangulyan}, {Kh{\'e}lifi}, {Keogh}, {Klu{\'z}niak},
  {Kneiske}, {Komin}, {Kosack}, {Lamanna}, {Lenain}, {Lohse}, {Marandon},
  {Martin}, {Martineau-Huynh}, {Marcowith}, {Maurin}, {McComb}, {Medina},
  {Moderski}, {Monard}, {Moulin}, {Naumann-Godo}, {de Naurois}, {Nedbal},
  {Nekrassov}, {Niemiec}, {Nolan}, {Ohm}, {Olive}, {de O{\~n}a Wilhelmi},
  {Orford}, {Ostrowski}, {Panter}, {Paz Arribas}, {Pedaletti}, {Pelletier},
  {Petrucci}, {Pita}, {P{\"u}hlhofer}, {Punch}, {Quirrenbach}, {Raubenheimer},
  {Raue}, {Rayner}, {Renaud}, {Rieger}, {Ripken}, {Rob}, {Rosier-Lees},
  {Rowell}, {Rudak}, {Rulten}, {Ruppel}, {Sahakian}, {Santangelo},
  {Schlickeiser}, {Sch{\"o}ck}, {Schr{\"o}der}, {Schwanke}, {Schwarzburg},
  {Schwemmer}, {Shalchi}, {Sikora}, {Skilton}, {Sol}, {Spangler}, {Stawarz},
  {Steenkamp}, {Stegmann}, {Superina}, {Szostek}, {Tam}, {Tavernet}, {Terrier},
  {Tibolla}, {Tluczykont}, {van Eldik}, {Vasileiadis}, {Venter}, {Venter},
  {Vialle}, {Vincent}, {Vivier}, {V{\"o}lk}, {Volpe}, {Wagner}, {Ward},
  {Zdziarski}, \& {Zech}}]{aharonian_etal:2009:pks2155}
{Aharonian}, F. {et~al.} 2009, \aap, 502, 749

\bibitem[{{Arbeiter} {et~al.}(2005){Arbeiter}, {Pohl}, \&
  {Schlickeiser}}]{arbeiter_etal:2005:ssc_and_pions}
{Arbeiter}, C., {Pohl}, M., \& {Schlickeiser}, R. 2005, \apj, 627, 62

\bibitem[{{Bednarek} \&
  {Protheroe}(1997)}]{bednarek_protheroe:1997:SSC_constraints}
{Bednarek}, W., \& {Protheroe}, R.~J. 1997, \mnras, 292, 646

\bibitem[{{Bednarek} \&
  {Protheroe}(1999)}]{bednarek_protheroe:1999:mrk501_constraints}
---. 1999, \mnras, 310, 577

\bibitem[{{Blandford} \& {Eichler}(1987)}]{blandford_eichler:1987}
{Blandford}, R., \& {Eichler}, D. 1987, \physrep, 154, 1

\bibitem[{{B{\l}a{\.z}ejowski} {et~al.}(2005){B{\l}a{\.z}ejowski}, {Blaylock},
  {Bond}, {Bradbury}, {Buckley}, {Carter-Lewis}, {Celik}, {Cogan}, {Cui},
  {Daniel}, {Duke}, {Falcone}, {Fegan}, {Fegan}, {Finley}, {Fortson},
  {Gammell}, {Gibbs}, {Gillanders}, {Grube}, {Gutierrez}, {Hall}, {Hanna},
  {Holder}, {Horan}, {Humensky}, {Kenny}, {Kertzman}, {Kieda}, {Kildea},
  {Knapp}, {Kosack}, {Krawczynski}, {Krennrich}, {Lang}, {LeBohec}, {Linton},
  {Lloyd-Evans}, {Maier}, {Mendoza}, {Milovanovic}, {Moriarty}, {Nagai}, {Ong},
  {Power-Mooney}, {Quinn}, {Quinn}, {Ragan}, {Reynolds}, {Rebillot}, {Rose},
  {Schroedter}, {Sembroski}, {Swordy}, {Syson}, {Valcarel}, {Vassiliev},
  {Wakely}, {Walker}, {Weekes}, {White}, {Zweerink}, {Mochejska}, {Smith},
  {Aller}, {Aller}, {Ter{\"a}sranta}, {Boltwood}, {Sadun}, {Stanek}, {Adams},
  {Foster}, {Hartman}, {Lai}, {B{\"o}ttcher}, {Reimer}, \&
  {Jung}}]{blazejowski_etal:2005:multiwavelength_mrk421}
{B{\l}a{\.z}ejowski}, M. {et~al.} 2005, \apj, 630, 130

\bibitem[{{B{\l}a{\.z}ejowski} {et~al.}(2000){B{\l}a{\.z}ejowski}, {Sikora},
  {Moderski}, \& {Madejski}}]{blazejowski_etal:2000:compton_on_infrared}
{B{\l}a{\.z}ejowski}, M., {Sikora}, M., {Moderski}, R., \& {Madejski}, G.~M.
  2000, \apj, 545, 107

\bibitem[{{Bonning} {et~al.}(2009){Bonning}, {Bailyn}, {Urry}, {Buxton},
  {Fossati}, {Maraschi}, {Coppi}, {Scalzo}, {Isler}, \&
  {Kaptur}}]{bonning_etal:2009:3c454}
{Bonning}, E.~W. {et~al.} 2009, \apjl, 697, L81

\bibitem[{{B{\"o}ttcher}(2007)}]{boettcher:2007:emission_processes_review}
{B{\"o}ttcher}, M. 2007, \apss, 309, 95

\bibitem[{{B{\"o}ttcher} \& {Chiang}(2002)}]{boettcher_chiang:2002}
{B{\"o}ttcher}, M., \& {Chiang}, J. 2002, \apj, 581, 127

\bibitem[{{B{\"o}ttcher} {et~al.}(2003){B{\"o}ttcher}, {Jackson}, \&
  {Liang}}]{boettcher_jackson_liang:2003:MC_code}
{B{\"o}ttcher}, M., {Jackson}, D.~R., \& {Liang}, E.~P. 2003, \apj, 586, 389

\bibitem[{{B{\"o}ttcher} \& {Liang}(2001)}]{boettcher_liang:2001:MC_code}
{B{\"o}ttcher}, M., \& {Liang}, E.~P. 2001, \apj, 552, 248

\bibitem[{{B{\"o}ttcher} {et~al.}(2009){B{\"o}ttcher}, {Reimer}, \&
  {Marscher}}]{boettcher_reimer_marscher:2009:VHE_3c279}
{B{\"o}ttcher}, M., {Reimer}, A., \& {Marscher}, A.~P. 2009, \apj, 703, 1168

\bibitem[{{Canfield} {et~al.}(1987){Canfield}, {Howard}, \&
  {Liang}}]{canfield_howard_liang:1987:MC_IC_relativistic_electrons}
{Canfield}, E., {Howard}, W.~M., \& {Liang}, E.~P. 1987, \apj, 323, 565

\bibitem[{{Chang} \& {Cooper}(1970)}]{chang_cooper:1970}
{Chang}, J.~S., \& {Cooper}, G. 1970, Journal of Computational Physics, 6, 1

\bibitem[{{Chen} {et~al.}(2011){Chen}, {Fossati}, {Liang}, \&
  {B{\"o}ttcher}}]{chen_etal:2011:proceedings_guanzhou}
{Chen}, X., {Fossati}, G., {Liang}, E., \& {B{\"o}ttcher}, M. 2011, {Journal of
  Astrophysics and Astronomy}, {10.1007/s12036-011-9045-0}

\bibitem[{{Chiaberge} \&
  {Ghisellini}(1999)}]{chiaberge_ghisellini:1999:timedep}
{Chiaberge}, M., \& {Ghisellini}, G. 1999, \mnras, 306, 551

\bibitem[{{Cohen} {et~al.}(2007){Cohen}, {Lister}, {Homan}, {Kadler},
  {Kellermann}, {Kovalev}, \&
  {Vermeulen}}]{cohen_etal:2007:beaming_and_intrinsic_properties}
{Cohen}, M.~H., {Lister}, M.~L., {Homan}, D.~C., {Kadler}, M., {Kellermann},
  K.~I., {Kovalev}, Y.~Y., \& {Vermeulen}, R.~C. 2007, \apj, 658, 232

\bibitem[{{Coppi} {et~al.}(1993){Coppi}, {Blandford}, \&
  {Rees}}]{coppi_blandford_rees:1993:anisotropic_induced_compton}
{Coppi}, P., {Blandford}, R.~D., \& {Rees}, M.~J. 1993, \mnras, 262, 603

\bibitem[{{Coppi}(1992)}]{coppi92_tdep}
{Coppi}, P.~S. 1992, \mnras, 258, 657

\bibitem[{{Coppi} \& {Blandford}(1990)}]{coppi_blandford:1990}
{Coppi}, P.~S., \& {Blandford}, R.~D. 1990, \mnras, 245, 453

\bibitem[{{Costamante} {et~al.}(2001){Costamante}, {Ghisellini}, {Giommi},
  {Tagliaferri}, {Celotti}, {Chiaberge}, {Fossati}, {Maraschi}, {Tavecchio},
  {Treves}, \& {Wolter}}]{costamante_etal:2001:stretching_the_sequence}
{Costamante}, L. {et~al.} 2001, \aap, 371, 512

\bibitem[{{Crusius} \&
  {Schlickeiser}(1986)}]{crusius_schlickeiser:1986:synchrotron}
{Crusius}, A., \& {Schlickeiser}, R. 1986, \aap, 164, L16

\bibitem[{{Dermer}(1998)}]{dermer:1998:spectral_variability}
{Dermer}, C.~D. 1998, \apjl, 501, L157

\bibitem[{{Dermer} {et~al.}(1992){Dermer}, {Schlickeiser}, \&
  {Mastichiadis}}]{dermer_etal92}
{Dermer}, C.~D., {Schlickeiser}, R., \& {Mastichiadis}, A. 1992, \aap, 256, L27

\bibitem[{{Donnarumma} {et~al.}(2009){Donnarumma}, {Vittorini}, {Vercellone},
  {del Monte}, {Feroci}, {D'Ammando}, {Pacciani}, {Chen}, {Tavani},
  {Bulgarelli}, {Giuliani}, {Longo}, {Pucella}, {Argan}, {Barbiellini},
  {Boffelli}, {Caraveo}, {Cattaneo}, {Cocco}, {Wissel}, {Wood}, \&
  {Zitzer}}]{donnarumma_etal:2009:mrk421_2008}
{Donnarumma}, I. {et~al.} 2009, \apjl, 691, L13

\bibitem[{{Drury}(1983)}]{drury:1983:diffusive_shock_acceleration}
{Drury}, L.~O. 1983, Reports on Progress in Physics, 46, 973

\bibitem[{{Edelson} \& {Krolik}(1988)}]{edelson_krolik:1988:DCF}
{Edelson}, R.~A., \& {Krolik}, J.~H. 1988, \apj, 333, 646

\bibitem[{{Fabian} {et~al.}(2001{\natexlab{a}}){Fabian}, {Celotti}, {Iwasawa},
  \& {Ghisellini}}]{fabian_etal:2001:gb1428}
{Fabian}, A.~C., {Celotti}, A., {Iwasawa}, K., \& {Ghisellini}, G.
  2001{\natexlab{a}}, \mnras, 324, 628

\bibitem[{{Fabian} {et~al.}(2001{\natexlab{b}}){Fabian}, {Celotti}, {Iwasawa},
  {McMahon}, {Carilli}, {Brandt}, {Ghisellini}, \&
  {Hook}}]{fabian_etal:2001:pmn_j0525}
{Fabian}, A.~C., {Celotti}, A., {Iwasawa}, K., {McMahon}, R.~G., {Carilli},
  C.~L., {Brandt}, W.~N., {Ghisellini}, G., \& {Hook}, I.~M.
  2001{\natexlab{b}}, \mnras, 323, 373

\bibitem[{{Fabian} {et~al.}(1986){Fabian}, {Guilbert}, {Blandford}, {Phinney},
  \& {Cuellar}}]{fabian_etal:1986:variability_in_compact_sources}
{Fabian}, A.~C., {Guilbert}, P.~W., {Blandford}, R.~D., {Phinney}, E.~S., \&
  {Cuellar}, L. 1986, \mnras, 221, 931

\bibitem[{{Finke}(2007)}]{finke:2007:phd_thesis}
{Finke}, J.~D. 2007, PhD thesis, Ohio University

\bibitem[{{Fossati} {et~al.}(2008){Fossati}, {Buckley}, {Bond}, {Bradbury},
  {Carter-Lewis}, {Chow}, {Cui}, {Falcone}, {Finley}, {Gaidos}, {Grube},
  {Holder}, {Horan}, {Horns}, {Jordan}, {Kieda}, {Kildea}, {Krawczynski},
  {Krennrich}, {Lang}, {LeBohec}, {Lee}, {Moriarty}, {Ong}, {Petry}, {Quinn},
  {Sembroski}, {Wakely}, \& {Weekes}}]{fossati_etal:2008:xray_tev}
{Fossati}, G. {et~al.} 2008, \apj, 677, 906

\bibitem[{{Fossati} {et~al.}(2000{\natexlab{a}}){Fossati}, {Celotti},
  {Chiaberge}, {Zhang}, {Chiappetti}, {Ghisellini}, {Maraschi}, {Tavecchio},
  {Pian}, \& {Treves}}]{fossati_etal:2000:mkn421_temporal}
---. 2000{\natexlab{a}}, \apj, 541, 153

\bibitem[{{Fossati} {et~al.}(2000{\natexlab{b}}){Fossati}, {Celotti},
  {Chiaberge}, {Zhang}, {Chiappetti}, {Ghisellini}, {Maraschi}, {Tavecchio},
  {Pian}, \& {Treves}}]{fossati_etal:2000:mkn421_spectral}
---. 2000{\natexlab{b}}, \apj, 541, 166

\bibitem[{{Fossati} {et~al.}(1998){Fossati}, {Celotti}, {Ghisellini},
  {Maraschi}, \& {Comastri}}]{fossati_etal:1998:sequence}
{Fossati}, G., {Celotti}, A., {Ghisellini}, G., {Maraschi}, L., \& {Comastri},
  A. 1998, \mnras, 299, 433

\bibitem[{{Gaisser}(1991)}]{gaisser:1990:BOOK:CR_and_particle_physics}
{Gaisser}, T.~K. 1991, Cosmic Rays and Particle Physics (Cambridge: Cambridge
  University Press)

\bibitem[{{Ghisellini} {et~al.}(1998){Ghisellini}, {Celotti}, {Fossati},
  {Maraschi}, \& {Comastri}}]{ghisellini_etal:1998:seds}
{Ghisellini}, G., {Celotti}, A., {Fossati}, G., {Maraschi}, L., \& {Comastri},
  A. 1998, \mnras, 301, 451

\bibitem[{{Ghisellini} {et~al.}(1988){Ghisellini}, {Guilbert}, \&
  {Svensson}}]{gg_guilbert_svensson:1988:boiler}
{Ghisellini}, G., {Guilbert}, P.~W., \& {Svensson}, R. 1988, \apjl, 334, L5

\bibitem[{{Ghisellini} \& {Madau}(1996)}]{gg_madau:1996}
{Ghisellini}, G., \& {Madau}, P. 1996, \mnras, 280, 67

\bibitem[{{Ghisellini} {et~al.}(2010){Ghisellini}, {Tavecchio}, {Foschini},
  {Ghirlanda}, {Maraschi}, \&
  {Celotti}}]{gg_etal:2009:physical_properties_of_fermi_blazars}
{Ghisellini}, G., {Tavecchio}, F., {Foschini}, L., {Ghirlanda}, G., {Maraschi},
  L., \& {Celotti}, A. 2010, \mnras, 402, 497

\bibitem[{{Giebels} {et~al.}(2007){Giebels}, {Dubus}, \&
  {Kh{\'e}lifi}}]{giebels_etal:2007:mkn421}
{Giebels}, B., {Dubus}, G., \& {Kh{\'e}lifi}, B. 2007, \aap, 462, 29

\bibitem[{{Graff} {et~al.}(2008){Graff}, {Georganopoulos}, {Perlman}, \&
  {Kazanas}}]{graff_etal:2008:pipe}
{Graff}, P.~B., {Georganopoulos}, M., {Perlman}, E.~S., \& {Kazanas}, D. 2008,
  \apj, 689, 68

\bibitem[{{Kataoka}(2000)}]{kataoka:2000:phd_thesis}
{Kataoka}, J. 2000, PhD thesis, Univ. of Tokyo

\bibitem[{{Kataoka} {et~al.}(2000){Kataoka}, {Takahashi}, {Makino}, {Inoue},
  {Madejski}, {Tashiro}, {Urry}, \&
  {Kubo}}]{kataoka_etal:2000:crossingtimes_model}
{Kataoka}, J., {Takahashi}, T., {Makino}, F., {Inoue}, S., {Madejski}, G.~M.,
  {Tashiro}, M., {Urry}, C.~M., \& {Kubo}, H. 2000, \apj, 528, 243

\bibitem[{{Katarzy{\'n}ski} {et~al.}(2006){Katarzy{\'n}ski}, {Ghisellini},
  {Mastichiadis}, {Tavecchio}, \&
  {Maraschi}}]{katarzynski_etal:2006:stochastic}
{Katarzy{\'n}ski}, K., {Ghisellini}, G., {Mastichiadis}, A., {Tavecchio}, F.,
  \& {Maraschi}, L. 2006, \aap, 453, 47

\bibitem[{{Katarzy{\'n}ski} {et~al.}(2005){Katarzy{\'n}ski}, {Ghisellini},
  {Tavecchio}, {Maraschi}, {Fossati}, \&
  {Mastichiadis}}]{katarzynski_etal:2005:tev_x_correlation}
{Katarzy{\'n}ski}, K., {Ghisellini}, G., {Tavecchio}, F., {Maraschi}, L.,
  {Fossati}, G., \& {Mastichiadis}, A. 2005, \aap, 433, 479

\bibitem[{{Katarzy{\'n}ski} {et~al.}(2008){Katarzy{\'n}ski}, {Lenain}, {Zech},
  {Boisson}, \& {Sol}}]{katarzynski_etal:2008:pks2155}
{Katarzy{\'n}ski}, K., {Lenain}, J., {Zech}, A., {Boisson}, C., \& {Sol}, H.
  2008, \mnras, 390, 371

\bibitem[{{Kellermann} \&
  {Pauliny-Toth}(1981)}]{kellerman_pauliny:1981:compact_radio_sources}
{Kellermann}, K.~I., \& {Pauliny-Toth}, I.~I.~K. 1981, \araa, 19, 373

\bibitem[{{Kirk} {et~al.}(1998){Kirk}, {Rieger}, \&
  {Mastichiadis}}]{kirk_rieger_mastichiadis:1998}
{Kirk}, J.~G., {Rieger}, F.~M., \& {Mastichiadis}, A. 1998, \aap, 333, 452

\bibitem[{{Krawczynski} {et~al.}(2002){Krawczynski}, {Coppi}, \&
  {Aharonian}}]{krawczynski_coppi_aharonian:2002:timedep}
{Krawczynski}, H., {Coppi}, P.~S., \& {Aharonian}, F. 2002, \mnras, 336, 721

\bibitem[{{Krawczynski} {et~al.}(2004){Krawczynski}, {Hughes}, {Horan},
  {Aharonian}, {Aller}, {Aller}, {Boltwood}, {Buckley}, {Coppi}, {Fossati},
  {G{\"o}tting}, {Holder}, {Horns}, {Kurtanidze}, {Marscher}, {Nikolashvili},
  {Remillard}, {Sadun}, \& {Schr{\"o}der}}]{krawczynski_etal:2004:1es1959}
{Krawczynski}, H. {et~al.} 2004, \apj, 601, 151

\bibitem[{{Krawczynski} {et~al.}(2001){Krawczynski}, {Sambruna}, {Kohnle},
  {Coppi}, {Aharonian}, {Akhperjanian}, {Barrio}, {Bernl{\" o}hr}, {B{\"
  o}rst}, {Bojahr}, {Bolz}, {Contreras}, {Cortina}, {Denninghoff}, {Fonseca},
  {Gonzalez}, {G{\" o}tting}, {Heinzelmann}, {Hermann}, {Heusler}, {Hofmann},
  {Horns}, {Ibarra}, {Jung}, {Kankanyan}, {Kestel}, {Kettler}, {Konopelko},
  {Kornmeyer}, {Kranich}, {Lampeitl}, {Lorenz}, {Lucarelli}, {Magnussen},
  {Mang}, {Meyer}, {Mirzoyan}, {Moralejo}, {Padilla}, {Panter}, {Plaga},
  {Plyasheshnikov}, {P{\" u}hlhofer}, {Rauterberg}, {R{\" o}hring}, {Rhode},
  {Rowell}, {Sahakian}, {Samorski}, {Schilling}, {Schr{\" o}der}, {Siems},
  {Stamm}, {Tluczykont}, {V{\" o}lk}, {Wiedner}, \&
  {Wittek}}]{krawczynski_etal:2001:mrk421_x_tev}
---. 2001, \apj, 559, 187

\bibitem[{{Levinson}(2006)}]{levinson:2006:review}
{Levinson}, A. 2006, International Journal of Modern Physics A, 21, 6015

\bibitem[{{Li} \& {Kusunose}(2000)}]{li_kusunose:2000}
{Li}, H., \& {Kusunose}, M. 2000, \apj, 536, 729

\bibitem[{{Makino}(1999)}]{makino:1998:turku}
{Makino}, F. 1999, in Astronomical Society of the Pacific Conference Series,
  Vol. 159, BL Lac Phenomenon, ed. {L.~O.~Takalo \& A.~Sillanp{\"a}{\"a}}, 190

\bibitem[{{Mannheim}(1998)}]{mannheim:1998:science_cosmic_rays}
{Mannheim}, K. 1998, Science, 279, 684

\bibitem[{{Maraschi} {et~al.}(1999){Maraschi}, {Fossati}, {Tavecchio},
  {Chiappetti}, {Celotti}, {Ghisellini}, {Grandi}, {Pian}, {Tagliaferri},
  {Treves}, {Breslin}, {Buckley}, {Carter-Lewis}, {Catanese}, {Cawley},
  {Fegan}, {Fegan}, {Finley}, {Gaidos}, {Hall}, {Hillas}, {Krennrich},
  {Lessard}, {Masterson}, {Moriarty}, {Quinn}, {Rose}, {Samuelson}, {Weekes},
  {Urry}, \& {Takahashi}}]{maraschi_etal:1999:letter_sax_mkn421}
{Maraschi}, L. {et~al.} 1999, \apjl, 526, L81

\bibitem[{{Maraschi} {et~al.}(1992){Maraschi}, {Ghisellini}, \&
  {Celotti}}]{mgc92_3c279}
{Maraschi}, L., {Ghisellini}, G., \& {Celotti}, A. 1992, \apjl, 397, L5

\bibitem[{{Marscher} {et~al.}(2010){Marscher}, {Jorstad}, {Larionov}, {Aller},
  {Aller}, {L{\"a}hteenm{\"a}ki}, {Agudo}, {Smith}, {Gurwell}, {Hagen-Thorn},
  {Konstantinova}, {Larionova}, {Larionova}, {Melnichuk}, {Blinov},
  {Kopatskaya}, {Troitsky}, {Tornikoski}, {Hovatta}, {Schmidt}, {D'Arcangelo},
  {Bhattarai}, {Taylor}, {Olmstead}, {Manne-Nicholas}, {Roca-Sogorb},
  {G{\'o}mez}, {McHardy}, {Kurtanidze}, {Nikolashvili}, {Kimeridze}, \&
  {Sigua}}]{marscher_etal:2010:pks1510}
{Marscher}, A.~P. {et~al.} 2010, \apjl, 710, L126

\bibitem[{{Marscher} \& {Travis}(1996)}]{marscher_travis:1996}
{Marscher}, A.~P., \& {Travis}, J.~P. 1996, \aaps, 120, 537

\bibitem[{{Mastichiadis} \&
  {Kirk}(1997)}]{mastichiadis_kirk:1997:SSC_variability}
{Mastichiadis}, A., \& {Kirk}, J.~G. 1997, \aap, 320, 19

\bibitem[{{McHardy} {et~al.}(2007){McHardy}, {Lawson}, {Newsam}, {Marscher},
  {Sokolov}, {Urry}, \& {Wehrle}}]{mchardy_etal:2007:3c273_variability}
{McHardy}, I., {Lawson}, A., {Newsam}, A., {Marscher}, A.~P., {Sokolov}, A.~S.,
  {Urry}, C.~M., \& {Wehrle}, A.~E. 2007, \mnras, 375, 1521

\bibitem[{{Nayakshin} \& {Melia}(1998)}]{nayakshin_melia:1998}
{Nayakshin}, S., \& {Melia}, F. 1998, \apjs, 114, 269

\bibitem[{{Press} {et~al.}(1992){Press}, {Teukolsky}, {Vetterling}, \&
  {Flannery}}]{numerical_recipes92}
{Press}, W., {Teukolsky}, S., {Vetterling}, W., \& {Flannery}, B. 1992,
  Numerical Recipes, the art of scientific computing {\rm second edition}
  (Cambridge University Press)

\bibitem[{{Protheroe}(1996)}]{protheroe:1996:cosmic_rays_acceleration}
{Protheroe}, R.~J. 1996, in Towards the Millennium in Astrophysics: Problems
  and Prospects, Erice 1996, {\rm eds. M.M. Shapiro and J.P. Wefel (World
  Scientific: Singapore)}

\bibitem[{{Protheroe}(2002)}]{protheroe:2002:factors_for_variability}
{Protheroe}, R.~J. 2002, \pasa, 19, 486

\bibitem[{{Punch} {et~al.}(1992){Punch}, {Akerlof}, {Cawley}, {Chantell},
  {Fegan}, {Fennell}, {Gaidos}, {Hagan}, {Hillas}, \&
  {Jiang}}]{punch_etal:1992:mkn421_tev_discovery}
{Punch}, M. {et~al.} 1992, \nat, 358, 477

\bibitem[{{Rachen}(2000)}]{rachen_correlated_x_tev}
{Rachen}, J.~P. 2000, in {\rm Proc. of GeV-TeV Gamma Ray Astrophysics Workshop:
  Towards a Major Atmospheric Cherenkov Detector VI, Snowbird, eds. B.L.
  Dingus, M.H. Salamon, D.B. Kieda, AIP Conf. Proc. 515 (New York: AIP).}, 41

\bibitem[{{Ravasio} {et~al.}(2002){Ravasio}, {Tagliaferri}, {Ghisellini},
  {Giommi}, {Nesci}, {Massaro}, {Chiappetti}, {Celotti}, {Costamante},
  {Maraschi}, {Tavecchio}, {Tosti}, {Treves}, {Wolter}, {Balonek}, {Carini},
  {Kato}, {Kurtanidze}, {Montagni}, {Nikolashvili}, {Noble}, {Nucciarelli},
  {Raiteri}, {Sclavi}, {Uemura}, \& {Villata}}]{ravasio_etal:2002:bllac}
{Ravasio}, M. {et~al.} 2002, \aap, 383, 763

\bibitem[{{Rebillot} {et~al.}(2006){Rebillot}, {Badran}, {Blaylock},
  {Bradbury}, {Buckley}, {Carter-Lewis}, {Celik}, {Chow}, {Cogan}, {Cui},
  {Daniel}, {Duke}, {Falcone}, {Fegan}, {Finley}, {Fortson}, {Gillanders},
  {Grube}, {Gutierrez}, {Gyuk}, {Hanna}, {Holder}, {Horan}, {Hughes}, {Kenny},
  {Kertzman}, {Kieda}, {Kildea}, {Kosack}, {Krawczynski}, {Krennrich}, {Lang},
  {Le Bohec}, {Linton}, {Maier}, {Moriarty}, {Perkins}, {Pohl}, {Quinn},
  {Ragan}, {Reynolds}, {Rose}, {Schroedter}, {Sembroski}, {Steele}, {Swordy},
  {Valcarcel}, {Vassiliev}, {Wakely}, {Weekes}, {Zweerink}, {VERITAS
  Collaboration)}, {Aller}, {Aller}, {Boltwood}, {Jung}, {Kranich}, {Nilsson},
  {Pasanen}, {Sadun}, \&
  {Sillanpaa}}]{rebillot_etal:2006:multiwavelength_mrk421}
{Rebillot}, P.~F. {et~al.} 2006, \apj, 641, 740

\bibitem[{{Rieger} \& {Duffy}(2004)}]{rieger_duffy:2004:shear_acceleration}
{Rieger}, F.~M., \& {Duffy}, P. 2004, \apj, 617, 155

\bibitem[{{Rieger} \& {Duffy}(2006)}]{rieger_duffy:2006:shear_acceleration}
---. 2006, \apj, 652, 1044

\bibitem[{{Rybicki} \& {Lightman}(1979)}]{rybicki_lightman}
{Rybicki}, G.~B., \& {Lightman}, A.~P. 1979, Radiative Processes in
  Astrophysics, {\rm John Wiley \& Sons} ((New York): John Wiley \& Sons)

\bibitem[{{Sambruna} {et~al.}(2000){Sambruna}, {Aharonian}, {Krawczynski},
  {Akhperjanian}, {Barrio}, {Bernl{\"o}hr}, {Bojahr}, {Calle}, {Contreras},
  {Cortina}, {Denninghoff}, {Fonseca}, {Gonzalez}, {G{\"o}tting},
  {Heinzelmann}, {Hemberger}, {Hermann}, {Heusler}, {Hofmann}, {Horns},
  {Ibarra}, {Kankanyan}, {Kestel}, {Kettler}, {K{\"o}hler}, {Kohnle},
  {Konopelko}, {Kornmeyer}, {Kranich}, {Lampeitl}, {Lindner}, {Lorenz},
  {Magnussen}, {Mang}, {Meyer}, {Mirzoyan}, {Moralejo}, {Padilla}, {Panter},
  {Plaga}, {Plyasheshnikov}, {Prahl}, {P{\"u}hlhofer}, {Rauterberg},
  {R{\"o}hring}, {Sahakian}, {Samorski}, {Schilling}, {Schmele},
  {Schr{\"o}der}, {Stamm}, {Tluczykont}, {V{\"o}lk}, {Wiebel-Sooth}, {Wiedner},
  {Willmer}, {Wittek}, {Chou}, {Coppi}, {Rothschild}, \&
  {Urry}}]{sambruna00_mrk501_xray_tev}
{Sambruna}, R.~M. {et~al.} 2000, \apj, 538, 127

\bibitem[{{Sikora} {et~al.}(1994){Sikora}, {Begelman}, \& {Rees}}]{sbr94_ec}
{Sikora}, M., {Begelman}, M.~C., \& {Rees}, M.~J. 1994, \apj, 421, 153

\bibitem[{{Sikora} {et~al.}(2001){Sikora}, {B{\l}a{\. z}ejowski}, {Begelman},
  \& {Moderski}}]{sikora_etal:2001:flares}
{Sikora}, M., {B{\l}a{\. z}ejowski}, M., {Begelman}, M.~C., \& {Moderski}, R.
  2001, \apj, 554, 1

\bibitem[{{Sikora} \& {Madejski}(2001)}]{sikora_madejski:2001:review}
{Sikora}, M., \& {Madejski}, G. 2001, in American Institute of Physics
  Conference Series, Vol. 558, , 275--288, booktitle was: American Institute of
  Physics Conference Series

\bibitem[{{Sikora} {et~al.}(2009){Sikora}, {Stawarz}, {Moderski}, {Nalewajko},
  \& {Madejski}}]{sikora_etal:2009:constraining_emission_models}
{Sikora}, M., {Stawarz}, {\L}., {Moderski}, R., {Nalewajko}, K., \& {Madejski},
  G.~M. 2009, \apj, 704, 38

\bibitem[{{Sokolov} \& {Marscher}(2005)}]{sokolov_marscher:2005:EC}
{Sokolov}, A., \& {Marscher}, A.~P. 2005, \apj, 629, 52

\bibitem[{{Sokolov} {et~al.}(2004){Sokolov}, {Marscher}, \&
  {McHardy}}]{sokolov_marscher_mchardy:2004:SSC}
{Sokolov}, A., {Marscher}, A.~P., \& {McHardy}, I.~M. 2004, \apj, 613, 725

\bibitem[{{Stawarz} \& {Ostrowski}(2002)}]{stawarz_ostrowski:2002:acceleration}
{Stawarz}, {\L}., \& {Ostrowski}, M. 2002, \apj, 578, 763

\bibitem[{{Stern} {et~al.}(1995){Stern}, {Begelman}, {Sikora}, \&
  {Svensson}}]{Stern_etal:1995:MC_large_particle}
{Stern}, B.~E., {Begelman}, M.~C., {Sikora}, M., \& {Svensson}, R. 1995,
  \mnras, 272, 291

\bibitem[{{Takahashi} {et~al.}(2000){Takahashi}, {Kataoka}, {Madejski},
  {Mattox}, {Urry}, {Wagner}, {Aharonian}, {Catanese}, {Chiappetti}, {Coppi},
  {Degrange}, {Fossati}, {Kubo}, {Krawczynski}, {Makino}, {Marshall},
  {Maraschi}, {Piron}, {Remillard}, {Takahara}, {Tashiro}, {Terasranta}, \&
  {Weekes}}]{takahashi_etal:2000:mkn421_1998}
{Takahashi}, T. {et~al.} 2000, \apjl, 542, L105

\bibitem[{{Takahashi} {et~al.}(1996){Takahashi}, {Tashiro}, {Madejski}, {Kubo},
  {Kamae}, {Kataoka}, {Kii}, {Makino}, {Makishima}, \&
  {Yamasaki}}]{takahashi_etal:1996:mkn421}
---. 1996, \apjl, 470, L89

\bibitem[{{Tanihata} {et~al.}(2004){Tanihata}, {Kataoka}, {Takahashi}, \&
  {Madejski}}]{tanihata_etal:2004:mrk421_evolution}
{Tanihata}, C., {Kataoka}, J., {Takahashi}, T., \& {Madejski}, G.~M. 2004,
  \apj, 601, 759

\bibitem[{{Tavecchio} {et~al.}(1998){Tavecchio}, {Maraschi}, \&
  {Ghisellini}}]{tavecchio_maraschi_gg:1998:tev}
{Tavecchio}, F., {Maraschi}, L., \& {Ghisellini}, G. 1998, \apj, 509, 608

\bibitem[{{Tavecchio} {et~al.}(2001){Tavecchio}, {Maraschi}, {Pian},
  {Chiappetti}, {Celotti}, {Fossati}, {Ghisellini}, {Palazzi}, {Raiteri},
  {Sambruna}, {Treves}, {Urry}, {Villata}, \& {Djannati-Ata{\"
  i}}}]{tavecchio_etal:2001:mkn501}
{Tavecchio}, F. {et~al.} 2001, \apj, 554, 725

\bibitem[{{Tramacere} {et~al.}(2007){Tramacere}, {Massaro}, \&
  {Cavaliere}}]{tramacere_massaro_cavaliere:2007:synchrotron_signatures}
{Tramacere}, A., {Massaro}, F., \& {Cavaliere}, A. 2007, \aap, 466, 521

\bibitem[{{Ulrich} {et~al.}(1997){Ulrich}, {Maraschi}, \&
  {Urry}}]{ulrich_maraschi_urry:1997:review}
{Ulrich}, M.-H., {Maraschi}, L., \& {Urry}, C.~M. 1997, \araa, 35, 445

\bibitem[{{Urry} \& {Padovani}(1995)}]{urry_padovani:1995:review}
{Urry}, C.~M., \& {Padovani}, P. 1995, \pasp, 107, 803

\bibitem[{{Ushio} {et~al.}(2009){Ushio}, {Tanaka}, {Madejski}, {Takahashi},
  {Hayashida}, {Kataoka}, {Mazin}, {R{\"u}gamer}, {Sato}, {Teshima}, {Wagner},
  \& {Yaji}}]{ushio_etal:2009:mrk421_in_2006_with_suzaku}
{Ushio}, M. {et~al.} 2009, \apj, 699, 1964

\bibitem[{{Virtanen} \&
  {Vainio}(2005)}]{virtanen_vainio:2005:particle_acceleration}
{Virtanen}, J.~J.~P., \& {Vainio}, R. 2005, \aap, 439, 461

\end{thebibliography}
\end{document}